# Thèse préparée en vue de l'habilitation à diriger les recherches

(refus notifié par l'Université de Bordeaux en Novembre 2006)

---

# Modélisation d'écoulements complexes pour des applications aérospatiales

Philippe Reynier

Septembre 2006

# Introduction

Depuis le début des années quatre-vingts, la mécanique des fluides numérique, appelée aussi CFD (acronyme anglais pour “Computational Fluid Dynamics”), a connu un formidable développement. Son utilisation, au début restreinte à des applications militaires et spatiales, a envahi en une génération l’ensemble des secteurs industriels, de l’industrie automobile au traitement de l’eau en passant par la sidérurgie et la microélectronique. De façon parallèle à cette expansion, la qualité des prédictions s’est affinée et a permis par exemple de réduire dans l’industrie aéronautique le nombre d’essais en soufflerie. Le calcul numérique est désormais utilisé non seulement dans un objectif de comparaison de résultats numériques et expérimentaux mais aussi en amont. Il sert ainsi à la préparation d’essais et à valider et calibrer de nouvelles techniques de mesures. Nous assistons depuis quelques années à l’apparition d’une véritable intégration entre le calcul numérique et l’expérience, avec cette évolution ces deux domaines ne sont plus concurrents mais complémentaires. Une dernière preuve, s’il en est nécessaire, de la maturité de la CFD est son utilisation pour l’optimisation de forme. L’exécution de cette tâche, effectuée auparavant de façon artisanale et empirique, est en train de déboucher sur la création d’outils couplant la CFD et l’optimisation. Les algorithmes qu’elle met en oeuvre peuvent être génétiques ou basés sur la méthode de l’adjoint, suivant l’objectif recherché. Ces derniers développements préfigurent l’apparition d’outils numériques multi-objectifs et multidisciplinaires.

Il existe une grande variété d’écoulements rencontrés dans le domaine aérospatial. Ils peuvent être caractérisés par leur nombre de Mach, un autre nombre adimensionnel ou par un facteur de forme. Ce dernier paramètre peut permettre de définir une complexité de type géométrique. La notion de complexité est en elle-même subjective et arbitraire. Suivant les travaux de la littérature, elle est relative à la géométrie ou aux phénomènes physiques caractérisant l’écoulement. Le but de ce travail est d’en dégager les aspects principaux et les moyens d’y répondre sur le plan de la modélisation et du calcul numérique. Généralement, la complexité de type géométrique mène à l’utilisation de gros maillages et à coût de calcul en conséquence. La présence de phénomènes physiques complexes, nécessite l'utilisation de modèles qui souvent ne représentent que partiellement la physique de l'écoulement. Du point de vue du calcul numérique, ces modèles sont pour la plupart non linéaires et nécessitent une attention particulière dans la résolution des équations de transport de l'écoulement. Ce point renvoie à une troisième complexité, moins apparente, qui est numérique. Elle est bien illustrée par les difficultés rencontrées pour la résolution d'écoulements compressibles à bas nombres de Mach, pour lesquels la formulation des schémas numériques usuellement basée sur la densité est instable. Un traitement particulier des équations est alors requis, avec soit le recours à un préconditionnement de la matrice associée, soit l'utilisation d'un schéma numérique spécifique à ce type de problème. La complexité du problème peut aussi être multiple, associant par exemple, une géométrie complexe à des phénomènes physiques non linéaires.

Dans ce travail, nous avons choisi d'aborder les différents aspects de la complexité des écoulements relatifs au secteur aérospatial que cela soit sur le plan numérique, géométrique ou physique. Les activités de recherche présentées incluent des travaux relatifs à la propulsion, à l'aérodynamique des missiles supersoniques et à la rentrée atmosphérique. Nous aborderons aussi d'autres problèmes, plus exotiques, rencontrés lors d'opérations visant à minimiser le nombre de débris spatiaux comme la passivation de réservoirs ou la rentrée de véhicules en fin de vie. Dans chaque cas, nous nous efforcerons de mettre en évidence, l'origine de la complexité de l'écoulement, dans le but de trouver la méthodologie la plus simple pour résoudre le problème posé. L'objectif n'est pas de trouver un schéma numérique universel, pierre philosophale de la CFD, mais de présenter, pour chaque application, la méthode retenue avec les résultats obtenus mais aussi les éventuelles limites qu'elles soient d'origine physique ou numérique.

Les applications et les méthodes présentées sont fortement dépendantes du contexte dans lesquelles ont été rencontrées ces activités de recherche. Mes travaux ont débuté dans un environnement académique à l'IMFT (Institut de Mécanique des Fluides de Toulouse). Cependant, les activités étaient déjà clairement tournées vers une application industrielle : l'étude des instabilités en sortie des injecteurs du moteur fusée Vulcain développé pour Ariane 5. L'étude a donc comporté un volet académique et une partie plus proche du problème industriel. La suite de mes activités de recherche s'est déroulée à l'ESTEC, établissement de l'Agence Spatiale Européenne situé aux Pays-Bas, et au DLR (Centre

Aérospatial Allemand) à Brunswick en Allemagne. Ces deux organisations, l'ESTEC et le DLR, se situent à la croisée de la recherche et de l'industrie, avec pour conséquence des activités de recherche et développement très proches des applications industrielles.

Le Chapitre 1 est consacré à la propulsion, et plus particulièrement aux problèmes associés à l'injection des ergols dans le moteur fusée Vulcain. Ce chapitre correspond aux travaux de recherche effectués, en coopération avec la SNECMA et le CNES, à l'Institut de Mécanique des Fluides de Toulouse. Nous présenterons une étude des instabilités évoluant dans la couche cisaillée des jets ronds compressibles et turbulents. Les effets du nombre de Mach mais aussi des conditions d'émission sur les structures cohérentes de la couche de mélange seront analysés. Les principales caractéristiques du schéma numérique retenu ainsi que celles de la modélisation semi-déterministe de la turbulence, méthode proche de la simulation aux grandes échelles, seront présentées. La deuxième partie de ce chapitre sera dévolue à l'application de cette méthode aux jets coaxiaux. Une comparaison avec des résultats expérimentaux sera montrée pour des jets homogènes et permettra de préciser le niveau de modélisation de turbulence nécessaire pour ce type d'écoulement. Ensuite, sera traité le calcul de jets coaxiaux inhomogènes en présence de larges gradients transversaux de densité et de vitesse. L'étape ultime sera l'approche monophasique des sprays, et notamment de leur région proche de l'injecteur, dans le but de préciser le processus d'atomisation et d'arrachage de filaments de fluide lourd par un jet annulaire de gaz léger. La conclusion fera le bilan de ces travaux et dressera les acquis et limitations de la méthodologie utilisée.

Le Chapitre 2 abordera le traitement numérique d'une géométrie complexe, celle de l'aile en nid d'abeilles. Ces travaux ont été effectués au DLR qui s'intéresse depuis plusieurs années à cette technologie. Elle est actuellement surtout maîtrisée par les Russes qui l'utilisent pour leurs missiles balistiques et comme système de secours du véhicule Soyouz. Ce type d'ailes présente l'avantage d'avoir de très bonnes performances aérodynamiques dans le domaine supersonique. Il est ici associé à un missile supersonique/hypersonique. D'un point de vue numérique, la résolution de la configuration est bien sur possible mais elle engendre un coût de calcul élevé. Celui-ci est accru d'un facteur cinq s'il est comparé à l'effort nécessaire pour calculer un cops seul. Cela est dû au fait que 80 % du maillage est localisé au niveau des ailes. Pour s'affranchir de cet inconvénient, un "disque d'action" a été développé. Ici, cette technique consiste à remplacer les ailes par une condition limite artificielle à l'intérieur de l'écoulement. Au travers de cette limite, les forces induites par la présence de l'aile sont intégrées aux équations de bilan. Ce disque d'action initialement intégré dans un code structuré a été appliqué et validé pour une configuration complète. Dans un premier temps, il a été couplé à une base de données permettant d'estimer les forces induites par l'aile, cela a permis de valider la technique mais cela a aussi montré ses limites. Finalement, incorporé dans un solveur non structuré et couplé à un module permettant le calcul des forces, cette méthode a démontré ses capacités pour le calcul des performances d'un missile avec un coût de calcul sensiblement égal à celui d'un corps seul. Ce travail illustre en outre l'intégration entre la CFD et l'expérience, à travers l'utilisation de bases de données expérimentales et d'un module semi empirique par un code de calcul.

Deux exemples d'écoulements complexes sur le plan de leur résolution numérique seront présentés dans le Chapitre 3. L'ensemble des travaux correspondant aux applications décrites a été effectué à l'ESTEC. La première application est la passivation, dans l'espace, des réservoirs d'hydrazine du SCA (Système de Contrôle d'Attitude) d'ARIANE 5. Lors de la passivation, la vidange des réservoirs entraîne la dépressurisation de l'hydrazine et éventuellement son changement de phase. Le but de l'étude était de valider le processus de passivation, d'estimer le risque d'obstruction de la conduite et la taille des débris dans l'espace. Sur le plan numérique, la première difficulté fut de simuler un écoulement compressible à très bas nombre de Mach, la seconde est liée au changement de phase qui est un problème raide modélisé de façon non linéaire. La deuxième étude a trait à la prédiction de l'explosion de l'ATV ("Automated Transfer Vehicle") lors de sa rentrée dans l'atmosphère terrestre à la fin de sa mission. Pour cela, les calculs de rentrée ont permis de prédire la localisation la plus plausible d'une fissure dans la structure. Ensuite, l'écoulement interne couplé à une analyse des températures et pressions critiques conduisant à l'explosion du véhicule ont permis de raffiner la prédiction de la destruction de l'ATV durant sa fin de mission. Dans ce chapitre, et ce pour chaque application, nous nous efforcerons de dégager la difficulté de sa résolution numérique et les solutions apportées.

Finalement, le Chapitre 4 sera consacré aux projets et perspectives de ces travaux avec notamment une présentation d'activités déjà initiées et de projets de recherche à plus long terme sur la rentrée atmosphérique et ses phénomènes associés. L'ensemble des travaux menés dans le cadre de ce chapitre a été effectué à l'ESTEC dans le cadre d'études et de projets de l'ESA. La première moitié du chapitre sera spécifique à la rentrée martienne avec tout d'abord l'estimation des capacités d'une soufflerie à haute enthalpie pour les atmosphères de $CO_2$. Ensuite, seront montrés des résultats de calculs de rentrée martiennes, effectués dans le but d'estimer l'influence de phénomènes tels que la catalycité et le rayonnement dus au plasma entourant la capsule. Ce dernier phénomène, fortement dépendant des espèces chimiques présentes dans l'écoulement et de la vitesse d'entrée, est susceptible d'engendrer des flux thermiques très élevés non seulement sur le bouclier de protection thermique mais aussi au niveau de l'arrière de la capsule. La seconde partie de ce chapitre sera dévolue à la rentrée terrestre. La première application concernera l'étude Euroreturn pour une mission de retour d'échantillons vers Vénus, avec des résultats de simulation numérique d'une rentrée terrestre rapide où l'ablation de la protection thermique est prise en compte. Les activités numériques effectuées pour l'analyse de mission et la préparation de la reconstruction du vol de l'IRDT ("Inflatable Re-entry and Descent Technology") seront montrées. Dans la dernière partie de ce chapitre, nous aborderons l'optimisation de forme pour les manœuvres d'aérocapture. Ici, il s'agit de résoudre un problème de nature multidisciplinaire dont la solution est dépendante des objectifs techniques recherchés. Un algorithme génétique a été retenu pour l'étude et les résultats obtenus sur une capsule montrent la dépendance de la forme par rapport aux critères choisis. Ce chapitre permettra de faire une revue des différents phénomènes physiques caractéristiques des rentrées atmosphériques et de la façon de les traiter.

Une conclusion générale sur l'ensemble des travaux présentés dans ce mémoire d'HDR suivra. Son but sera de dresser un état des lieux pour la modélisation des écoulements complexes, de dégager un bilan des différentes activités de recherche menées et de donner des directions possibles pour la poursuite de ces travaux.

# 1 Complexité physique : Modélisation des jets

## 1.1 Introduction

### 1.1.1 Le problème industriel

Le contrôle de l'instabilité des jets d'injection de moteur-fusée est un point fondamental pour assurer la sécurité de fonctionnement d'un lanceur. La modélisation et la simulation numérique des phénomènes physiques liés à l'injection de propergols dans un moteur-fusée cryotechnique sont essentielles pour améliorer la performance du moteur. En effet, la performance d'un foyer de combustion dépend essentiellement de son système d'injection. Plus le processus d'atomisation-vaporisation-combustion qui s'y déroule est efficace, plus les dimensions de la chambre de combustion sont réduites, ce qui induit un gain en masse et en performance du moteur. La modélisation de ce processus est très complexe puisqu'il met en jeu, de manière couplée, les phénomènes d'arrachage de gouttes sur une surface libre, de vaporisation d'un spray et de gouttes à haute pression et de mélange réactif diphasique. Les connaissances sur l'ensemble de ces phénomènes sont actuellement trop parcellaires pour assurer une prédiction satisfaisante de l'ensemble de l'écoulement. Cela est du principalement au fait que l'atomisation est un phénomène non linéaire dont la théorie reste peu développée [Lian, 1990].

L'atomisation, ou breakup primaire, est le processus par lequel le cône liquide est détruit pour former des ligaments et des gouttes. Ce phénomène est important car il détermine les propriétés initiales de la phase dispersée, qui influencent le taux de mélange et le breakup secondaire. Ce dernier correspond a la transformation des gouttes et ligaments en gouttelettes. Différents régimes de breakup [Reitz, 1978] existent, le cas présent correspond au régime d'atomisation caractérisé par un nombre de Weber élevé : $We > 40$ ; avec $We = \rho_g DU^2/\sigma$, où $\rho_g$ est la densité du gaz atomisant, D le diamètre du jet, U la vitesse d'éjection et $\sigma$ la tension de surface.

Dans le moteur-fusée cryotechnique Vulcain, développé pour Ariane 5, l'injection est assurée par un ensemble d'injecteurs coaxiaux (voir Figure 3). Dans la perspective du mélange et de la combustion, les atomiseurs de type coaxiaux sont plus efficaces que les atomiseurs à pression comportant un seul jet simple [Shavit, 1996]. Les injecteurs coaxiaux produisent des gouttelettes plus petites, une longueur de breakup plus courte, un angle de spray plus large et des sprays mieux contrôlés. Dans le cas du moteur Vulcain représenté sur la Figure 1, le jet central d'oxygène liquide à basse vitesse est entouré par un jet annulaire rapide d'hydrogène gazeux. Cela induit un fort gradient transversal de densité qui est équilibré par un gradient de vitesse de telle façon que le rapport de quantité de mouvement soit égal à un. Ces contrastes de densité et de vitesse ont un effet important sur l'instabilité et améliore l'atomisation.

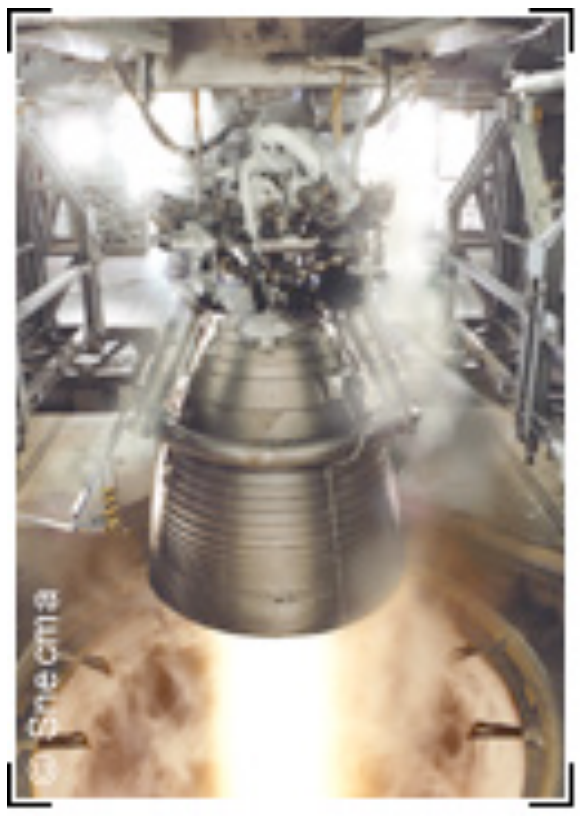

**Figure 1: Moteur Vulcain au banc d'essai.**

Les travaux expérimentaux de Gicquel et al [Gicquel, 1995] effectués sur le banc d'essai Mascotte de l'ONERA ont permis de définir trois régions, représentées sur la Figure 2, dans les écoulements de jets coaxiaux diphasiques pour des conditions proches de celles rencontrées dans le moteur Vulcain. La première est la région proche où se déroule l'atomisation qui s'étend jusqu'à une distance de 6D (où D est le diamètre de l'injecteur). Ensuite se trouvent une zone fortement ligamenteuse jusqu'à 18D et enfin la zone d'atomisation secondaire en aval.

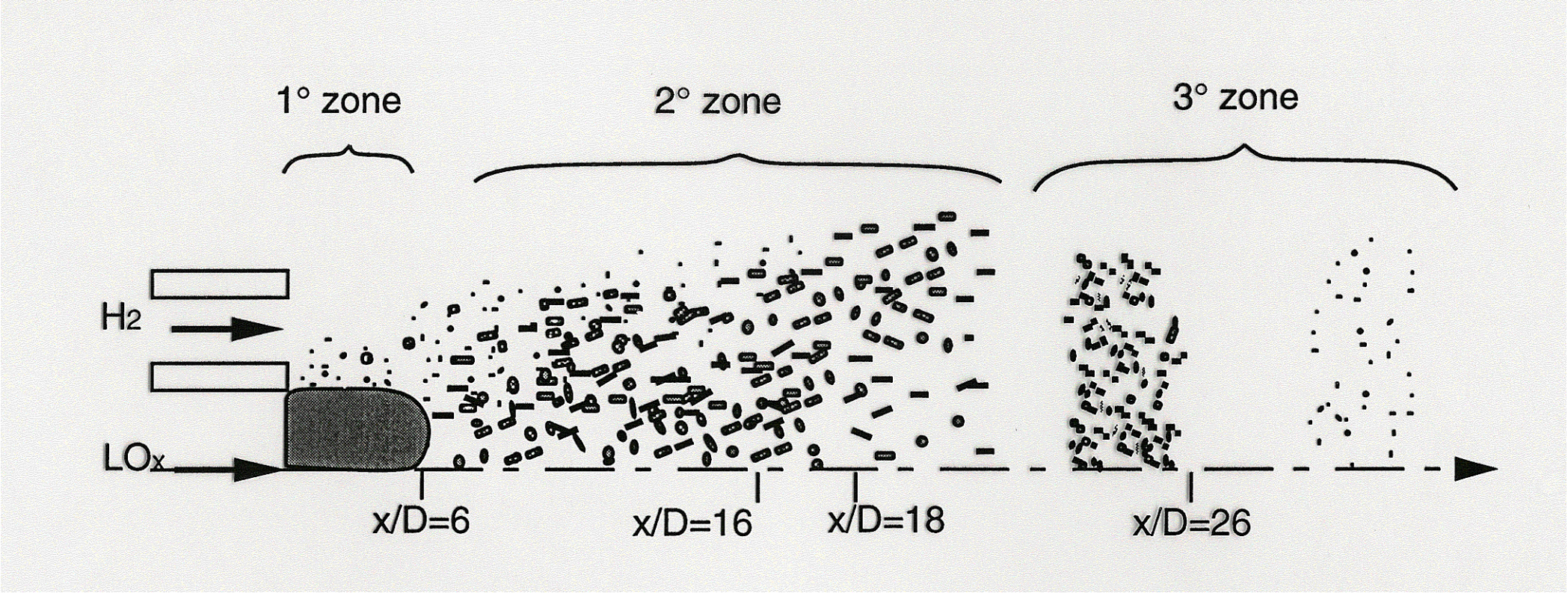

**Figure 2 : Régions du spray caractéristiques des injecteurs du moteur-fusée Vulcain.**

La difficulté du problème à résoudre et le manque d'études réalisées dans de telles configurations laissent en suspend de nombreuses questions quand au choix des paramètres physiques (rapports de vitesse, de densité, de pression et de température) et géométriques (rapport entre les diamètres des jets, forme de l'injecteur, retrait de jet central par rapport au jet annulaire) qui permettraient d'optimiser le mélange et ainsi d'améliorer l'efficacité et la sécurité de fonctionnement du moteur. Ainsi, un retrait du jet central est favorable au mélange, cependant s'il est trop important il risque de provoquer un accrochage de flamme. Dans le contexte de la combustion dans un moteur cryotechnique, c'est la zone proche qui nous intéresse le plus, en effet, c'est là qu'a lieu l'initiation du mélange. Cependant, la modélisation de cette région de l'écoulement est délicate du fait de l'intermittence des frontières, de la présence de parois et de l'instabilité de l'écoulement qui génère dans cette zone des structures instationnaires. Pour minimiser la longueur du cône liquide, dans les moteurs-fusées le débit total est reparti dans un grand nombre d'injecteurs : 516 pour la plaque d'injection, du moteur Vulcain, représentée sur la Figure 3.

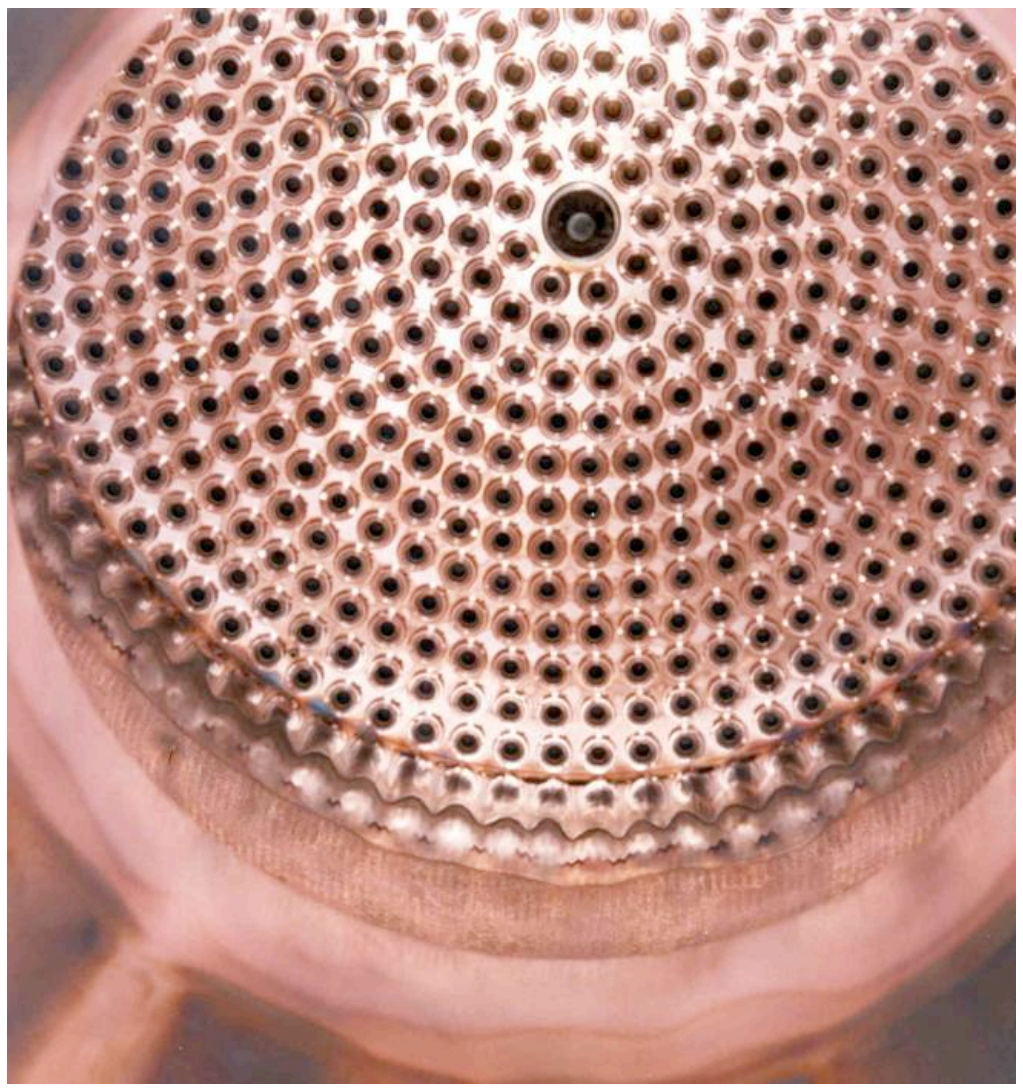

**Figure 3 : Plaque d'injection du moteur Vulcain avec 516 éléments d'injection [Vuillermoz, 2004].**

Il existe peu d'études détaillées sur l'influence de la turbulence sur le mécanisme de breakup. Cependant, plusieurs travaux [Reitz, 1982 ; Lefebvre, 1992; Issac, 1994] rapportent que la turbulence accélère le breakup (ce que l'on pressent intuitivement). Les propriétés du breakup turbulent mériteraient des études supplémentaires, en particulier pour les conditions où la formation des gouttes est contrôlée par les tourbillons de petites tailles. Peu d'efforts de recherche ont été effectués sur l'interaction entre les structures turbulentes et les gouttelettes (mais aussi les particules) et il peut être démontré [Sirignano, 1993] que dans un grand nombre d'écoulements à grands nombres de Reynolds, l'échelle de Kolmogorov est de l'ordre de la taille des gouttelettes.

D’après Herding [Herding, 1997], la turbulence du jet liquide diminue le niveau de vaporisation et favorise le transfert de chaleur dans l’oxygène froid.

### 1.1.2 Objectifs

Le principal but de cette étude est de préciser l’influence de la turbulence et plus précisément des conditions d’émission sur l’instabilité de la région proche des jets. Dans un premier temps la physique du spray et l’aspect diphasique de l’écoulement n’ont pas été considérés. Vu la difficulté du problème, tant sur le plan numérique que sur celui de la modélisation, une approche a d’abord été développée pour le jet rond simple et appliquée ensuite aux jets coaxiaux. Plusieurs modèles de fermetures des équations de Navier-Stokes moyennées ont été retenus dans le but de modéliser les écoulements de jets coaxiaux turbulents et notamment de prendre en compte leur anisotropie. Différentes conditions d’émission ont été utilisées pour estimer l’influence du nombre de Mach, mais aussi du niveau de turbulence initial, sur les structures instationnaires évoluant dans la couche cisaillée des jets ronds. Pour les jets coaxiaux, l’influence du rapport de vitesse sur la région proche a été analysée.

Le second objectif est de développer une approche monophasique des sprays. Le but est de préciser la nature du mécanisme de breakup dans la région proche et plus particulièrement d’estimer la taille des filaments arrachés au jet central par le jet annulaire ainsi que la longueur du cône liquide. Ici, le calcul numérique s’avère très utile pour explorer l’écoulement à cause du manque de mesures disponibles. Jusqu'à récemment, l’approche expérimentale ne permettait pas d’aboutir à des mesures fiables dans cette région d’un spray [Andrews, 1993]. Les problèmes principaux sont les effets engendrés par les perturbations du liquide ainsi que les faibles effets aérodynamiques des liquides à pression et température ambiantes où les techniques de mesures sont peu performantes. Enfin, la longueur du cône liquide est difficile à estimer du fait de la densité optique associée à l’écoulement. Quelques études expérimentales ont cependant été menées [Gicquel, 1995; Herding, 1997 ; Mayer, 1994 ; Barata et al, 2003] pour les chambres à combustion de moteurs-fusées à hydrogène et oxygène. Elles ont montré qu’une densité plus élevée du gaz augmente l’atomisation et mène à des gouttelettes plus fines. Jusqu'à ce que ces difficultés expérimentales soient surmontées, la compréhension du mécanisme d’atomisation restera limitée. L’application de nouvelles techniques de mesure, telles que l’holographie, permettra certainement d’améliorer notre connaissance de ce processus.

# 1.2 Aspects numériques

## 1.2.1 Codes de calcul

Pour les simulations numériques deux codes de calcul ont été utilisés : NS22 pour Navier-Stokes bidimensionnel avec un modèle de turbulence à deux équations et NS25 avec un modèle de turbulence à cinq équations. Ces deux codes provenant initialement de la NASA étaient disponibles à l'IMFT mais dans des versions limitées au calcul plan d'interaction onde de choc/couche limite [Vandromme, 1983, 1987]. Les deux solveurs étaient basés sur la méthode numérique aux volumes finis [MacCormack, 1981]. Elle résout les équations de Navier-Stokes instationnaire sous leur forme conservative. Le schéma, parabolique en temps et elliptique en espace, est explicite-implicite et la technique de prédiction-correction est utilisée.

L'application de ces deux codes au calcul d'écoulement de jets a nécessité un travail important avec dans les deux cas le passage en coordonnées axisymétriques. Les jets sont des écoulements délicats à simuler d'un point de vue des conditions limites : en conséquence, les conditions limites non réflectives proposées par Thompson [Thompson, 1987] et dérivées de la théorie des caractéristiques ont été incorporées dans chacun des codes et appliquées à la sortie du domaine. Pour les calculs de jets coaxiaux d'oxygène et d'hydrogène, une équation de transport de la concentration d'oxygène a été implantée. Les différents maillages utilisés pour les prédictions numériques ont été construits 'à la main'. Plusieurs simulations avec différents maillages et domaines de calcul ont permis de vérifier l'indépendance des résultats vis-à-vis de la discrétisation.

## 1.2.2 Calcul instationnaire

Avec des conditions appropriées d'utilisation, le schéma numérique utilisé par les deux codes de calcul a permis de prendre en compte les phénomènes instationnaires de l'écoulement. Dans la partie implicite du schéma, le traitement des termes de diffusion est approché, ainsi la méthode, considérée dans son ensemble, est mieux adaptée au calcul stationnaire. En conséquence, pour cette étude, seule la partie explicite a été utilisée dans le but de prendre en compte les instabilités évoluant dans les écoulements étudiés. Cela induit l'utilisation de pas de temps plus petits avec un nombre de CFL inférieur à un, pour assurer la stabilité du schéma. Ce dernier est précis au second ordre en temps et en espace et ne nécessite pas l'utilisation de dissipation numérique ajoutée pour sa stabilité. Ce dernier aspect avec la précision au second ordre en temps et l'utilisation de la seule partie explicite constituent les points fondamentaux qui permettent de prendre en compte l'instabilité de l'écoulement. En effet, un schéma plus diffusif avec la présence d'une dissipation numérique importante préviendrait la prédiction de structures instationnaires dans l'écoulement.

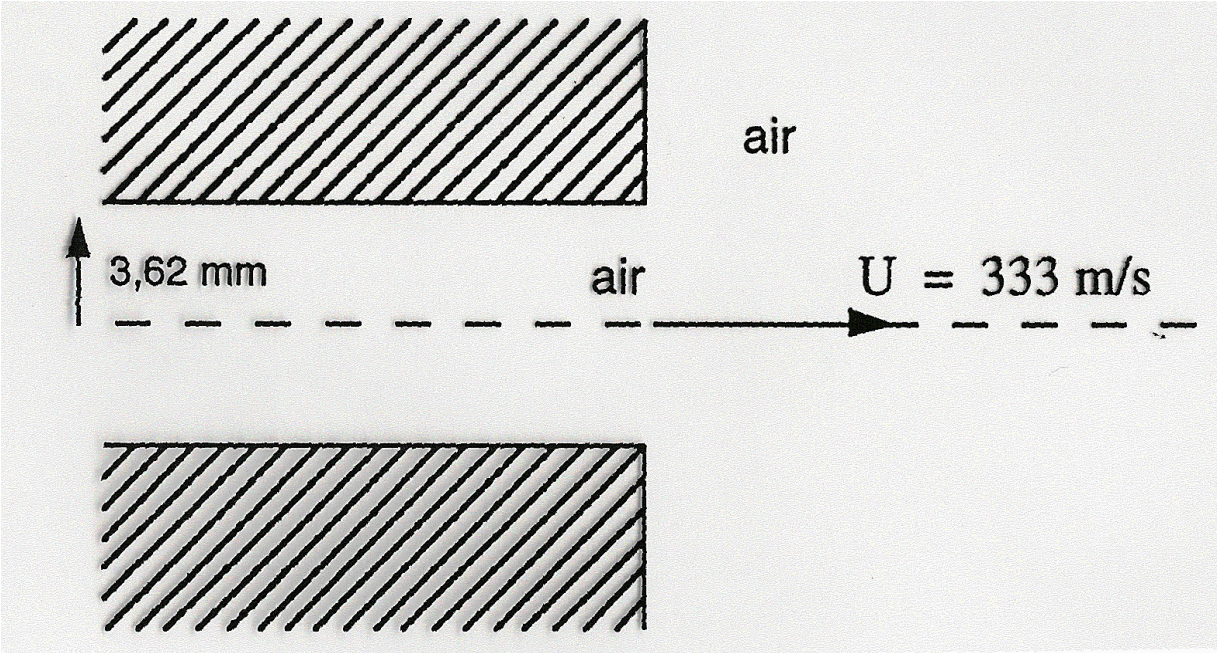

**Figure 4: Configuration simulée pour les jets ronds simples.**

La configuration de jet rond étudiée est représentée sur la Figure 4. Des conditions de symétrie sur l'axe et des conditions de parois sur la section d'entrée ont été appliquées. Au niveau de l'injection, les quantités ont été si possible dérivées de profils expérimentaux. Comme aucune donnée expérimentale n'est disponible pour des profils de dissipation d'énergie turbulente, cette variable a été extrapolée à partir de l'énergie turbulente en faisant l'hypothèse d'une turbulence en équilibre. À la sortie du domaine de calcul, les conditions limites non réflectives incorporées dans les codes ont été utilisées. Les mêmes conditions de calcul ont été appliquées pour les jets coaxiaux.

# 1.3 Modélisation

## 1.3.1 Modélisation semi-déterministe

La plupart des écoulements possèdent des caractères organisés et chaotiques. Les structures cohérentes, telles que Hussain [Hussain, 1883, 1986] les a définies, sont présentes dans le jet rond comme dans tout écoulement possédant une couche de mélange. Leur découverte dans la région proche du jet, où elles sont associées au mode axisymétrique du jet, est le fait de Crow et Champagne [Crow, 1971]. Plus tard, les travaux de Lau et Fisher [Lau, 1975] ont montré que la structure de base du jet rond turbulent consiste en une allée de tourbillons se déplaçant vers l'aval. La couche cisaillée du jet est en fait dominée par des structures cohérentes qui se comportent comme des instabilités de grandes échelles. Ces structures cohérentes jouent un rôle central dans l'écoulement, révélé par les travaux expérimentaux de Brown et Roshko [Brown, 1974] et Winant et Browand [Winant, 1974] sur la couche de mélange compressible. Elles ont été étudiées pour les raisons suivantes :

- Leur rôle dans la transition laminaire-turbulente ;
- Leur intérêt pour la modélisation théorique pour améliorer la prédiction de l'écoulement et du mélange.

Vu les nombres de Reynolds des écoulements étudiés, $Re_D > 50000$ pour les jets ronds, la simulation directe (DNS) et la simulation aux grandes échelles (LES) étaient difficilement applicables. Le choix s'est donc porté sur la modélisation semi- déterministe (SDM) développée par Ha Minh et Kourta [Ha Minh, 1993] qui est proche de la LES. Elle permet la simulation des instabilités naturelles sans excitation de l'écoulement.

Pendant longtemps, les simulations numériques ont été limitées à des problèmes approchés, tels que l'approximation de Prandtl, où l'aspect instationnaire de l'écoulement n'est pas considéré. Aussi, le développement des modèles de turbulence, au cours des années 70 a été fait pour la résolution de telles approximations. En conséquence de quoi, les modèles de turbulence statistiques donnent des résultats satisfaisants lorsque l'énergie turbulente est en équilibre spectral, elle est alors dite pleinement développée. Cependant, un grand nombre d'écoulements, particulièrement dans les applications industrielles, sont plutôt concernés par une turbulence jeune (i.e. proche de la transition). Deux traits importants la caractérisent :

- La turbulence peut être à bas nombre de Reynolds turbulent et les diffusions moléculaire et turbulente peuvent avoir le même ordre de grandeur ;
- Le spectre tridimensionnel de l'énergie turbulente n'est pas en équilibre, il est fortement altéré par la présence de structures organisées instationnaires.

Ces structures organisées constituent la partie cohérente de la turbulence. Elles sont fortement dépendantes des conditions initiales mais aussi des conditions limites et influencent fortement le niveau énergétique de la turbulence. Cela a été attesté dans la région proche des jets ou des expériences [Hussain, 1981 ; Sokolov, 1981], menées pour des jets avec un faible niveau initial de turbulence, ont révélé que près de la sortie la partie cohérente de l'énergie turbulente était plus importante que sa composante aléatoire. Par conséquent, pour ce type d'écoulement, la modélisation statistique classique [Favre, 1965] doit être reconsidérée pour tenir compte de l'instationnarité organisée. Reynolds et Hussain [Reynolds, 1972] ont démontré que chaque quantité physique instantanée, $f(x_k,t)$, peut être séparée en trois composantes:

$$f(x_k,t) = \underbrace{\overline{f}(x_k,t)}_{(a)} + \underbrace{f_c(x_k,t)}_{(b)} + \underbrace{f_r(x_k,t)}_{(c)} \qquad (1.1)$$

La partie (a) représente la quantité en moyenne temporelle, c'est généralement la seule accessible expérimentalement. La composante cohérente ou instationnaire organisée (b) possède un caractère déterministe et peut être prédite numériquement. La dernière partie (c) correspond aux fluctuations incohérentes ou aléatoires. Les tourbillons correspondants sont caractérisés par un spectre continu où des pics, traces des structures cohérentes, peuvent être surimposés. En théorie, la simulation directe de toutes les échelles de tourbillons (DNS) est possible. Cependant, si la DNS est la méthode la plus naturelle pour prédire un écoulement, son coût de calcul est trop élevé pour des écoulements à grands nombres de Reynolds. En fait, les parties dont la prédiction est raisonnablement envisageable sont le mouvement moyen et l'instabilité cohérente. Une nouvelle quantité, appelée moyenne de phase, est ainsi obtenue. Déterminée à partir des structures cohérentes au même moment de leur évolution, son expression est :

$$\langle f(x_k,t)\rangle = \overline{f}(x_k,t) + f_c(x_k,t) \qquad (1.2)$$

soit,

$$f(x_k,t) = \langle f(x_k,t)\rangle + f_r(x_k,t) \qquad (1.3)$$

Si $f_r = 0$ alors on retrouve la DNS. Le cas où la composante $f_r$ est très petite correspond à la LES. Finalement, si la quantité $\langle f(x_k,t)\rangle$ est minimisée et réduite à la moyenne temporelle, la modélisation classique est retrouvée. Ainsi la SDM apparaît comme une voie intermédiaire entre la modélisation classique et la DNS, et les équations de la modélisation classique sont conservées Néanmoins, si cette approche est très proche de la LES, elle en diffère par la fermeture des équations : modèle de sous maille pour la LES et modèle de turbulence pour la SDM. La principale différence concerne les structures instationnaires. La LES les sépare en fonction de leur taille : celles de grandes échelles à simuler et celles de petites échelles à modéliser. La SDM les classifie en fonction de leur nature : les structures cohérentes sont simulées et la turbulence aléatoire est modélisée. Si les structures cohérentes ont un caractère bidimensionnel, la SDM, contrairement à la LES, ne mène pas automatiquement à une approche tridimensionnelle.

La méthode décrite ci-dessus a été appliquée au calcul axisymétrique des écoulements de jets. Sur le plan de la simulation des instabilités, l'approche permet en principe de simuler le mode préféré du jet rond [Crow, 1971 ; Hussain 1981] qui est axisymétrique. Cependant, si ce mode domine l'instabilité initiale du jet, ce n'est pas le cas en aval de la région proche. La croissance des tourbillons évoluant dans la couche cisaillée est assurée par des interactions non linéaires au sein de l'écoulement. Ce phénomène caractéristique des couches cisaillées mène dans les jets, à l'apparition de structures azimutales en forme de lobes [Widnall, 1975] qui ont été étudiées pour des jets à basses vitesses, expérimentalement par Liepmann et Gharib [Liepmann, 1992] et numériquement par Verzicco et Orlandi [Verzicco, 1994]. Il existe peu d'études effectuées sur les ondes d'instabilité dans les jets compressibles. Cela est du au fait que les mesures de fluctuation de vitesse et de pression, surtout pour les jets supersoniques, ne donnent pas de résultats probants. Les simulations numériques, ici axisymétriques, ne permettent évidemment pas la simulation des structures tridimensionnelles qui évoluent dans les jets. Cette limitation a toutefois autorisé une représentation correcte de la région proche ce qui était le but principal de ces travaux.

### 1.3.2 Modèles de turbulence

Comme la partie à modéliser comporte toutes les composantes aléatoires, la situation est quasiment identique à la modélisation classique. Les équations instationnaires de Navier-Stokes écrites en moyenne pondérée par la masse volumique [Favre, 1965] restent valables. Les équations sont alors les mêmes et l'utilisation de la même méthode pour la fermeture des équations est tentante. Cependant, comme le rôle de la turbulence a été redéfini, quelques modifications peuvent s'avérer nécessaires dans le jeu des constantes des modèles à travers la recalibration de certaines d'entre elles.

Dans les calculs numériques, deux types de modèles ont été utilisés, deux sont issus de la modélisation classique tandis que le troisième a été adapté à la modélisation semi-déterministe. Pour les jets ronds, au vu de la faible anisotropie de la turbulence, une modélisation à deux échelles de type k-ε a été retenue. Deux modèles k-ε ont été appliqués au calcul des jets ronds, celui de Launder et Sharma [Launder, 1974] et celui proposé par Ha Minh et Kourta [Ha Minh 1993] où la constante $C_\mu$ a été recalibrée avec une valeur de 0.02 au lieu de 0.09 dans la modélisation standard. Il est à noter, que cette constante a été recalibrée pour une marche descendante et que la validité de cette recalibration pour des écoulements de jet n'a pas été démontrée.

Deux différents niveaux de fermeture des équations, tous deux issus de la modélisation classique, ont été retenus pour l'étude des jets coaxiaux. L'un est le modèle k-ε de Launder et Sharma [Launder, 1974], l'autre le modèle aux contraintes de Reynolds de Launder et al. [Launder, 1975]. L'application d'un modèle au second ordre permettra d'évaluer son influence sur les prédictions ainsi que le niveau d'anisotropie de la turbulence dans ce type d'écoulement.

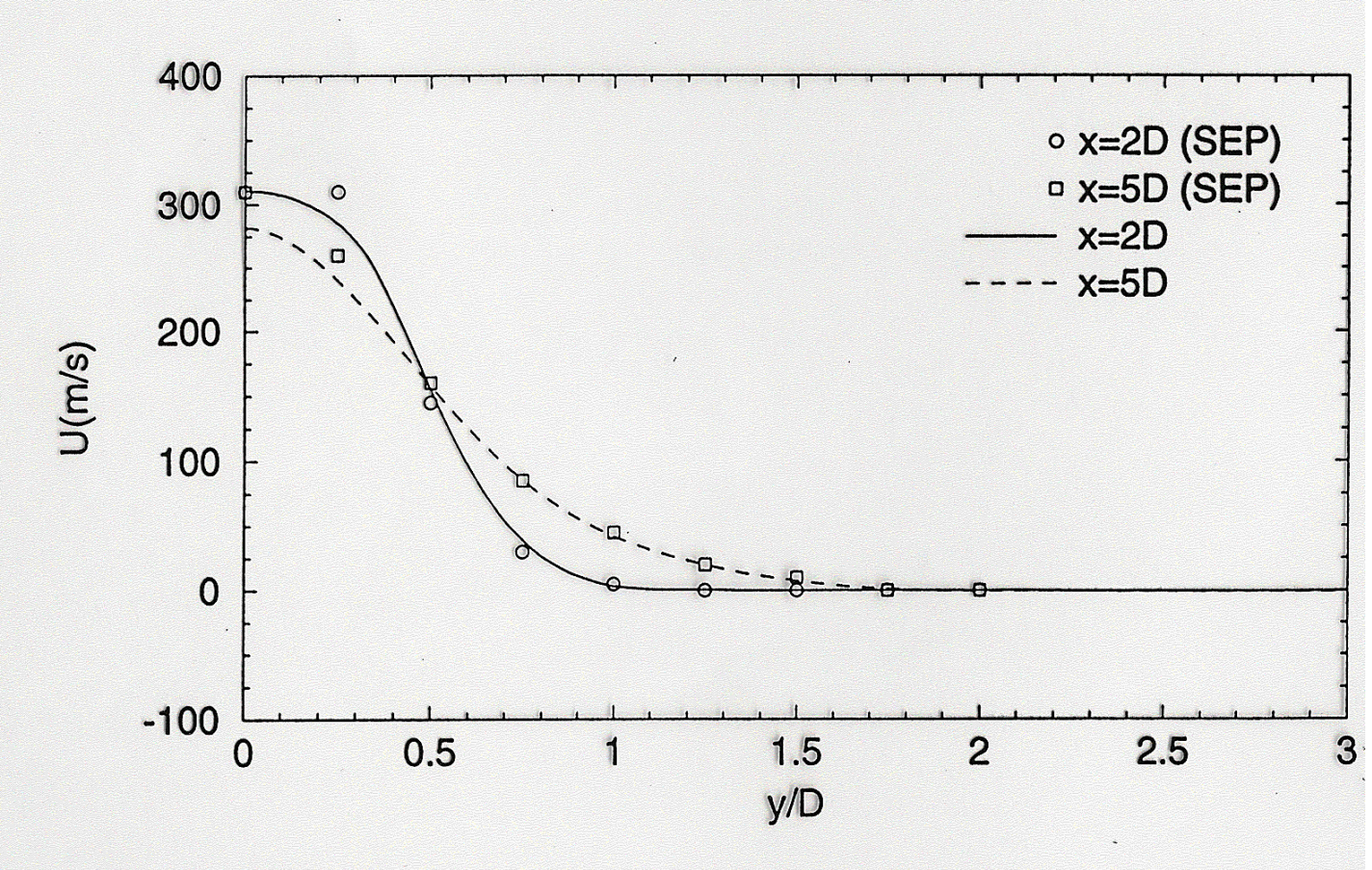

**Figure 5 : Profils de vitesse axiale dans la région proche (2D et 5D) comparés à ceux de la SEP [Delplanque, 1993].**

## 1.4 Un cas académique : le jet rond

Dans un premier temps, dans le but de mettre au point l'outil numérique, une configuration simple, celle du jet rond, pour laquelle de nombreux résultats sont accessibles dans la littérature, a été calculée. La configuration de la Figure 4 a été simulée pour différents nombres de Mach: 0.3, 0.96 et 1.5. Les profils initiaux de vitesse et d'énergie turbulente ont été dérivés de ceux mesurés expérimentalement par Durão [Durão, 1971] et Chassaing [Chassaing, 1979]. La première expérience correspond à une turbulence jeune, tandis que dans la seconde, le jet est pleinement développé.

### 1.4.1 Validation

Un premier calcul effectué pour un jet transsonique à Mach 0.96 avec des conditions initiales dérivées de celle de Durão [Durão, 1971] et le modèle k-ε de Launder et Sharma [Launder, 1974] a permis la validation de la méthode et du code de calcul. En l'absence de résultats expérimentaux obtenus pour un cas similaire, les résultats instationnaires obtenus ont été moyennés et confrontés à ceux obtenus à la SEP [Delpanque, 1993] avec le code Thésée et des conditions similaires. Les comparaisons obtenues sont représentées sur la Figure 5 pour la vitesse axiale et sur la Figure 6 pour la vitesse radiale. Les profils ont été comparés pour des distances de deux et cinq diamètres (2D et 5D). Un bon accord est observé entre les deux simulations. Une légère différence est observée sur le profil de vitesse axiale près de l'axe du jet, elle est due au fait que dans le calcul de la SNECMA un profil de vitesse carré a été imposé comme condition initiale du jet à l'entrée du domaine de calcul. Cette différence est absente de la Figure 6 qui représente la distribution de vitesse radiale où l'accord entre les deux calculs est quasiment parfait.

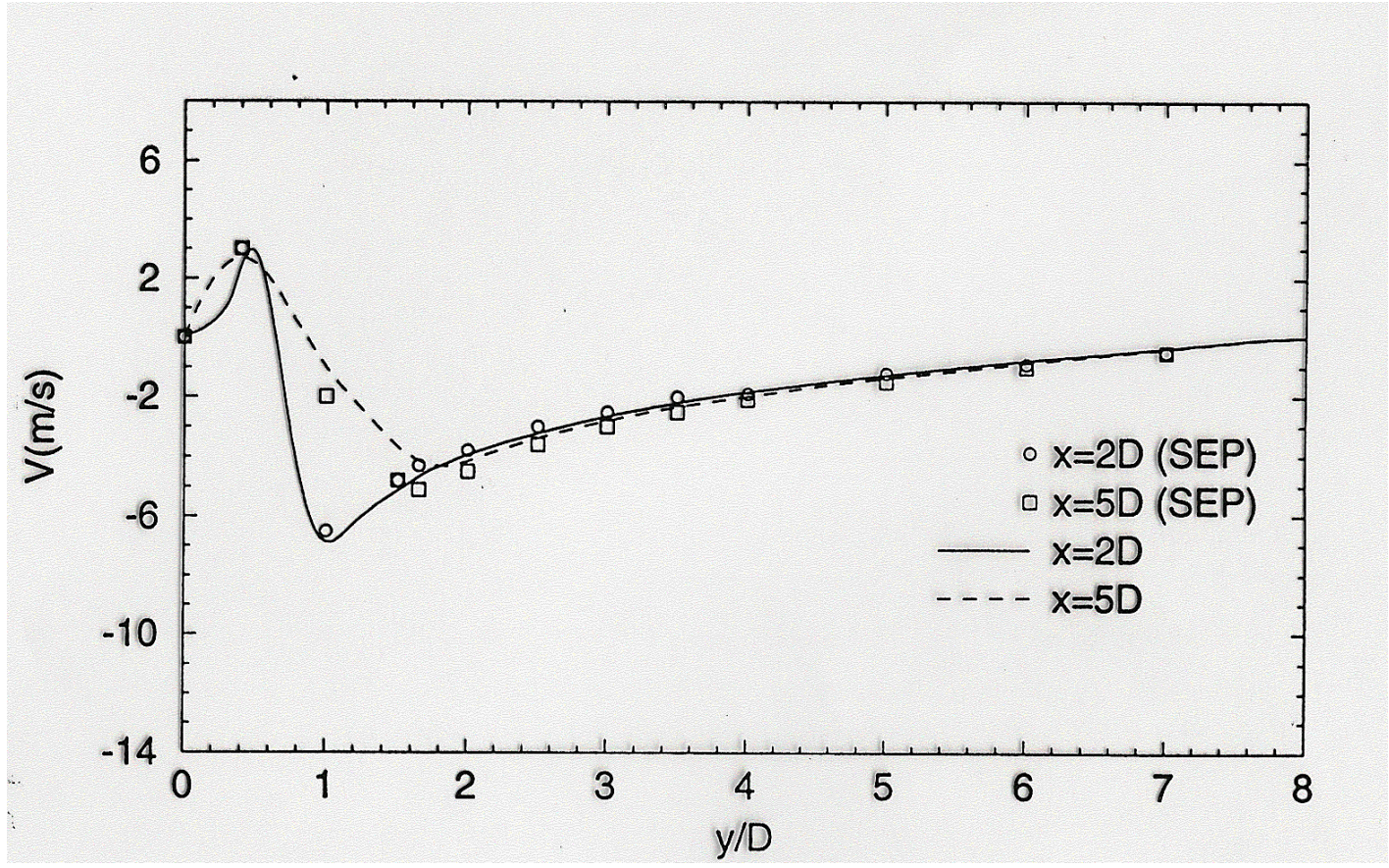

**Figure 6: Profils de vitesse radiale dans la région proche (2D et 5D) comparés à ceux de la SNECMA [Delplanque, 1993].**

| Modèle | Conditions d'émission | Écoulement |
|---|---|---|
| k-ε LS | Chassaing | Stationnaire |
| k-ε HK | Chassaing | Quasi-Stationnaire |
| k-ε LS | Durão | Instationnaire |
| k-ε HK | Durão | Instationnaire |

**Tableau 1 : Différents cas de calcul simulés à Mach 0.3 avec le modèle utilisé (HK = Ha Minh et Kourta [Ha Minh 1993]; LS = Launder et Sharma [Launder, 1974]), le type de conditions d'émission [Chassaing, 1979; Durão, 1971] et l'écoulement prédit.**

### 1.4.2 Influence des conditions d'émission et de la modélisation

Le jet à Mach 0.3 a été simulé pour deux types de conditions d'émission (turbulence pleinement développée ou non) et les deux modèles de turbulence retenus pour l'étude. Les différents cas de calcul et les résultats obtenus sont résumés dans le Tableau 1. Le but principal était d'évaluer l'influence des conditions initiales et de la modélisation sur la présence de structures instationnaires dans l'écoulement.

Les résultats montrent qu'avec une turbulence pleinement développée, dérivée des conditions de Chassaing [Chassaing, 1979], et le modèle LS [Launder, 1974], l'écoulement prédit est stationnaire. En effet, dans ce cas là, les structures cohérentes sont absentes de l'écoulement et on retrouve le contexte de la modélisation statistique classique. Le calcul pour les mêmes conditions initiales et le modèle HK [Ha Minh, 1993] met en évidence une très faible instationnarité de l'écoulement localisée dans la région initiale du jet. Ce résultat peut être interprété de deux façons : soit-il s'agit d'une instabilité d'origine numérique due au niveau plus faible de diffusion numérique dans la simulation associée à la valeur recalibrée de la constante $C_\mu$ ; soit-il s'agit de vestiges de structures organisées correspondant à une turbulence ‘vieille’. Ce résultat correspond à l’existence d’un faible niveau d’entraînement dans la région proche qui est en accord avec les résultats expérimentaux de Sahr et Gökalp [Sahr, 1991]. Cependant, l’absence d’une validation plus quantitative n’a pas permis de clore définitivement ce point.

Les simulations avec une turbulence plus près de la transition (Durão [Durão, 1971]) ont mené quel que soit le modèle à la simulation de structures organisées dans la région proche du jet. Ces instabilités sont plus marquées dans le cas du calcul avec le modèle HK. D’après Michalke [Michalke, 1984] leur présence est induite par la présence d’un pic de vorticité. C’est l’induction de la vorticité, présente dans la couche cisaillée, qui entraîne l’apparition des structures cohérentes. Ce pic de vorticité est absent quand la turbulence est pleinement développée. Les simulations effectuées avec le modèle LS et le jeu standard de constantes mettent en évidence un cône à potentiel plus court ainsi qu’une expansion plus importante du jet. Cela est dû à la différence de la valeur de la constante $C_\mu$ entre les deux modèles. Une valeur plus faible de cette constante induit une viscosité turbulente moindre et par conséquent un modèle moins diffusif.

| Mach | Modèle | Conditions d’émission |
|---|---|---|
| 0.3 | k-ε HK | Durão |
| 0.96 | k-ε HK | Durão |
| 1.5 | k-ε HK | Durão |

**Tableau 2 : Simulations effectuées pour l’étude des effets de compressibilité.**

### 1.4.3 Structures cohérentes

Dans le but d’étudier l’influence du nombre de Mach sur les structures organisées, la configuration de la Figure 4 a été simulée pour différents nombres de Mach en sortie (voir Tableau 2) avec des conditions d’émission de type Durão [Durão, 1971] et le modèle k-ε HK. Les résultats numériques ont mis en évidence un fort aspect instationnaire de la couche de mélange. Ceci est visible sur la Figure 7 qui représente le champ instantané de vitesse radiale du cas supersonique. Cette figure met en évidence l’allée de tourbillons présents dans la couche cisaillée. Ce résultat corrobore ceux de Lau et Fisher [Lau, 1975] pour lesquels le jet rond consiste essentiellement en une allée de tourbillons se déplaçant vers l’aval dans le couche cisaillée du jet. Cette présence de tourbillons dans la région proche du jet est fondamentale pour le processus d’expansion du jet [Verzicco, 1994].

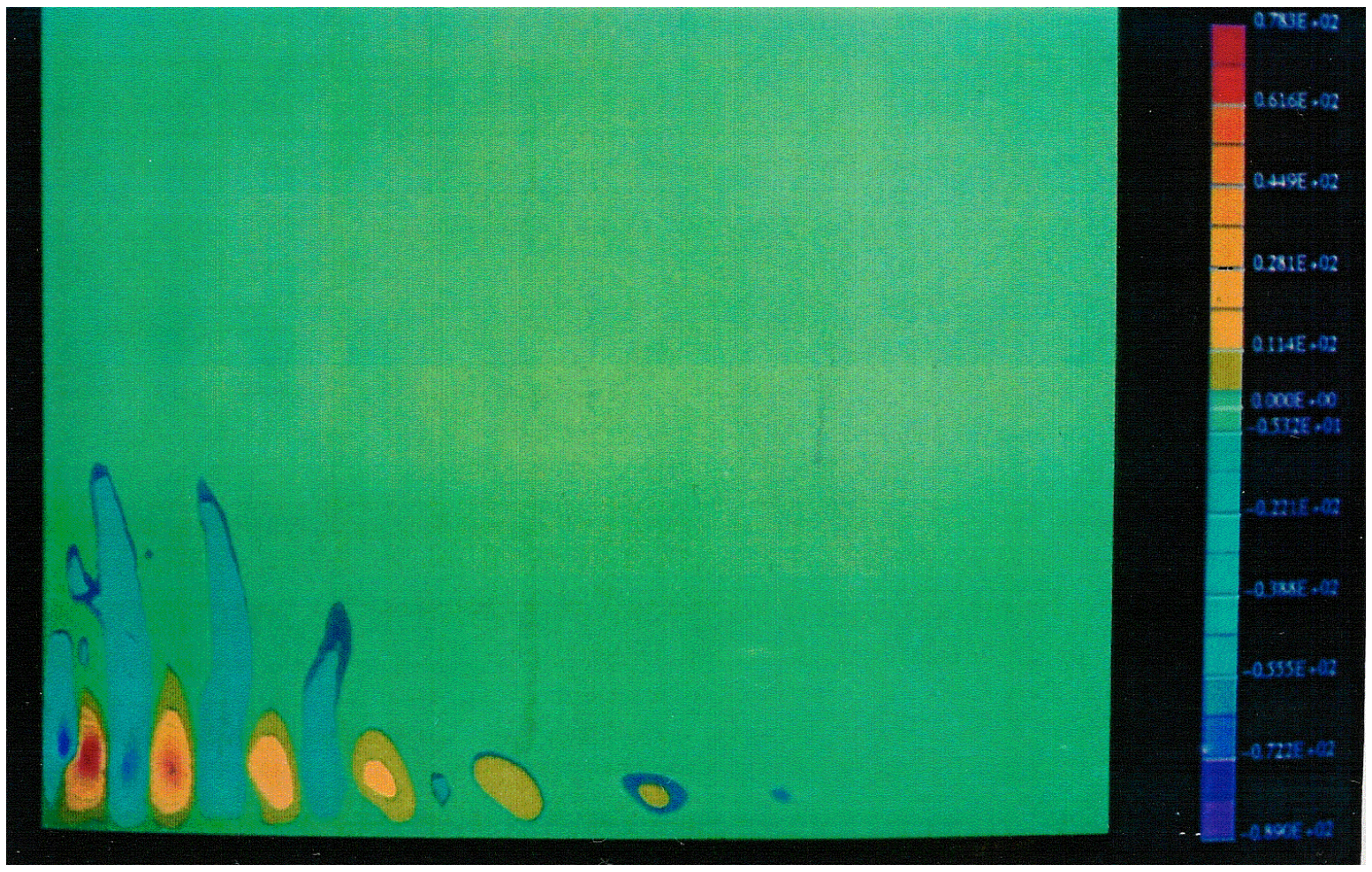

**Figure 7 : Champ instantané de vitesse radiale pour un jet rond à Mach 1.5.**

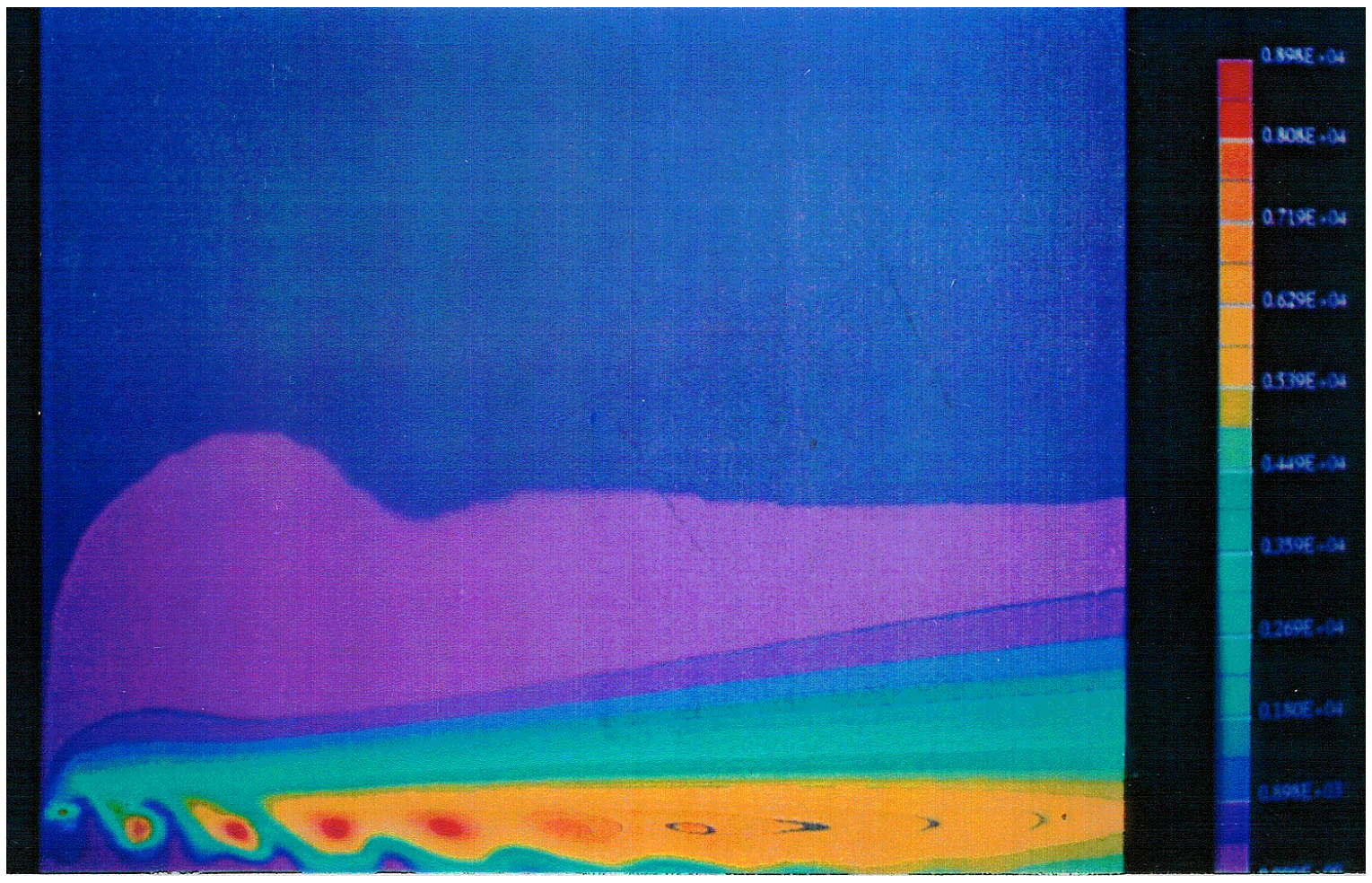

**Figure 8 : Champ instantané d'énergie cinétique turbulente pour un jet rond à Mach 1.5.**

Les régions localisées entre les tourbillons sont caractérisées par un cisaillement intense qui est la source d'un niveau élevé de turbulence dans la couche de mélange et mène à l'expansion du jet. Ceci est bien visible sur la Figure 8 où le champ instantané d'énergie cinétique de la turbulence est représenté. Ces tourbillons de grande échelle sont induits par l'instabilité de la couche cisaillée [Michalke, 1984]. Ici, quel que soit le nombre de Mach, l'instabilité est convective. Les travaux expérimentaux de Tam et Hu [Tam, 1989] ont démontré que seule l'instabilité de Kelvin-Helmholtz est active dans les jets subsoniques. Elle entraîne l'enroulement de la couche cisaillée et la formation d'anneaux tourbillonnaires, processus observé expérimentalement et numériquement [Liepmann, 1992 ; Verzicco, 1994]. D'après Ho et Huerre [Ho, 1984] ce processus est essentiellement un phénomène bidimensionnel. Il entraîne la présence de structures cohérentes qui contrôlent la dynamique et le mélange du jet. Ceci est vrai non seulement pour les jets subsoniques mais aussi pour des nombres de Mach supersoniques [Tam, 1989]. Ici, les tourbillons apparaissent rapidement (voir Figure 7) ce qui est en accord avec différentes expériences [Arnette, 1993 ; Liepmann, 1992] dans lesquelles ces structures apparaissent avant une distance de deux diamètres. Leur présence induit un processus de contraction et d'expansion du jet qui affecte le cône à potentiel [voir Figure 8] par le phénomène d'induction de Biot et Savart. Dans cette région, les zones les plus cisaillées sont situées entre les tourbillons, c'est donc là que l'énergie turbulente est produite [Panchapakesan, 1993] et que se situent les maxima d'intensité turbulente. Ceci est bien illustré par la Figure 8 qui met en évidence un fort niveau de turbulence dans la couche de mélange. Ce résultat est en accord avec la théorie sur les structures cohérentes [Hussain, 1983-1986]. Près de l'injection l'instationnarité est très forte, cela signifie que la partie cohérente de la turbulence est plus importante que la composante aléatoire. Ce résultat avec une domination des grandes structures par rapport aux petits tourbillons a déjà été mis en évidence expérimentalement [Hussain, 1981 ; Sokolov, 1981].

En se déplaçant vers l'aval, ces structures sont soumises à un processus de cisaillement et de contraction qui entraîne leur distorsion [Fourguette, 1991] puis leur destruction. Les structures organisées sont dégradées par les mécanismes non linéaires de l'écoulement, ici représentés par le modèle de turbulence. Les structures s'affaiblissent à partir de 5D et ont complètement disparu à 15D, la turbulence est alors dominée par les petites échelles. À ce mécanisme correspond le développement de la turbulence aléatoire, visible sur la Figure 8. Cette évolution des structures cohérentes est en accord avec la description faite par Sokolov et al [Sokolov, 1981]. La distance à laquelle ces structures disparaissent dépend des conditions initiales : 8D pour Hussain et Zaman [Hussain, 1981], 10D pour Liepmann et Gharib [1992], avant 11D dans le cas de cette étude.

Sur le plan de la modélisation, l'instationnarité simulée dans la région proche du jet rond supporte que l'égalité supposée entre la production et la dissipation qui est à la source de la modélisation classique n'est pas vérifiée dans l'ensemble de l'écoulement. Si cette hypothèse est vraie dans la région pleinement développée du jet, ce n'est pas le cas dans la zone proche [Rodi, 1972]. La présence de structures cohérentes altère fortement cette zone et le spectre de l'énergie n'y est plus en équilibre [Ha Minh, 1993]. Le niveau d'énergie turbulente est faible dans la région proche particulièrement au centre du jet. Un faible niveau de production explique cette dépression centrale aussi observée par Bogulawski et Popiel [Bogulawski, 1979]. Cette dépression est progressivement comblée vers l'aval par l'extension de la zone de mélange et la diffusion résultante.

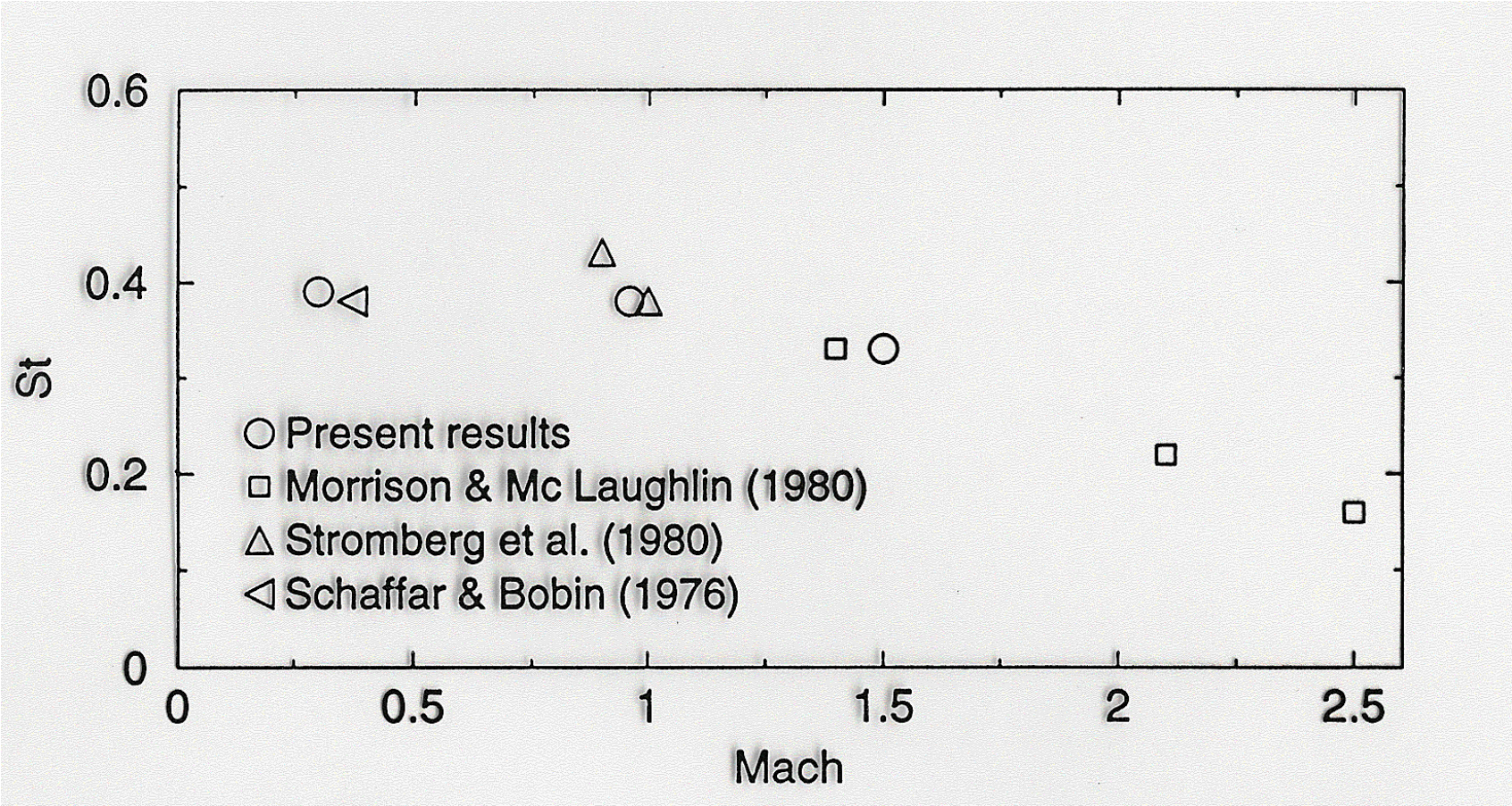

**Figure 9 : Évolution du nombre de Strouhal du mode préféré du jet rond avec le nombre de Mach.**

### 1.4.4 Influence du nombre de Mach

Dans le but d'étudier les effets de la compressibilité sur les jets ronds, la configuration de la Figure 4 a été calculée pour plusieurs nombres de Mach (voir Tableau 2). Pour chacun des cas de calcul, les instabilités simulées sont quasiment périodiques. Les nombres de Strouhal correspondants ainsi que ceux de plusieurs études expérimentales sont rapportés sur la Figure 9. Leur valeur est comprise entre 0.3 et 0.4 montrent qu'elles correspondent au mode préféré du jet rond [Crow, 1971]. La comparaison entre les différentes valeurs met en évidence un très bon accord entre le calcul et l'expérience. Entre Mach 0.3 et 0.96, le nombre de Strouhal reste quasi constant, cela n'est pas le cas pour le jet à Mach 1.5 : sa valeur passe de 0.38 (à Mach 0.96) à 0.33 (à Mach 1.5). Ceci est en accord qualitatif avec les études expérimentales de Morrison & McLaughlin [Morrison, 1980] et Lepicovski et al [Lepicovski, 1987] où le même phénomène a été remarqué. Cette diminution du nombre de Strouhal est à rapprocher de la théorie de Papamoschou et Roshko [Papamoschou, 1988] sur les couches de mélange compressibles. La décroissance du nombre de Strouhal correspond à une diminution du nombre de structures cohérentes. Ceci montre l'effet stabilisateur du nombre de Mach sur l'écoulement. Cet effet est dû à la diminution du taux de croissance de l'instabilité de Kelvin-Helmholtz [Miles, 1958].

Les fluctuations de l'énergie turbulente augmentent avec le carré du nombre de Mach. Cela correspond à une croissance de la force des structures cohérentes avec celui-ci, phénomène déjà observé par Arnette et al [Arnette, 1993]. En séparant les contraintes de Reynolds en deux composantes, Hussain & Zaman [Hussain, 1981] ont montré que la partie cohérente de la turbulence augmente pour les grands nombres de Reynolds. Ils ont aussi remarqué que les structures cohérentes devenaient plus énergétiques et le transport à travers la couche cisaillée plus grand. L'expansion du jet est alors plus rapide, elle augmente avec le nombre de Reynolds mais diminue avec le nombre de Mach. En effet, quand ce dernier augmente les structures sont plus énergétiques mais moins nombreuses en conséquence de quoi l'expansion du jet diminue quand les effets de compressibilité deviennent importants ce qui est le cas pour les jets supersoniques [Michalke, 1984].

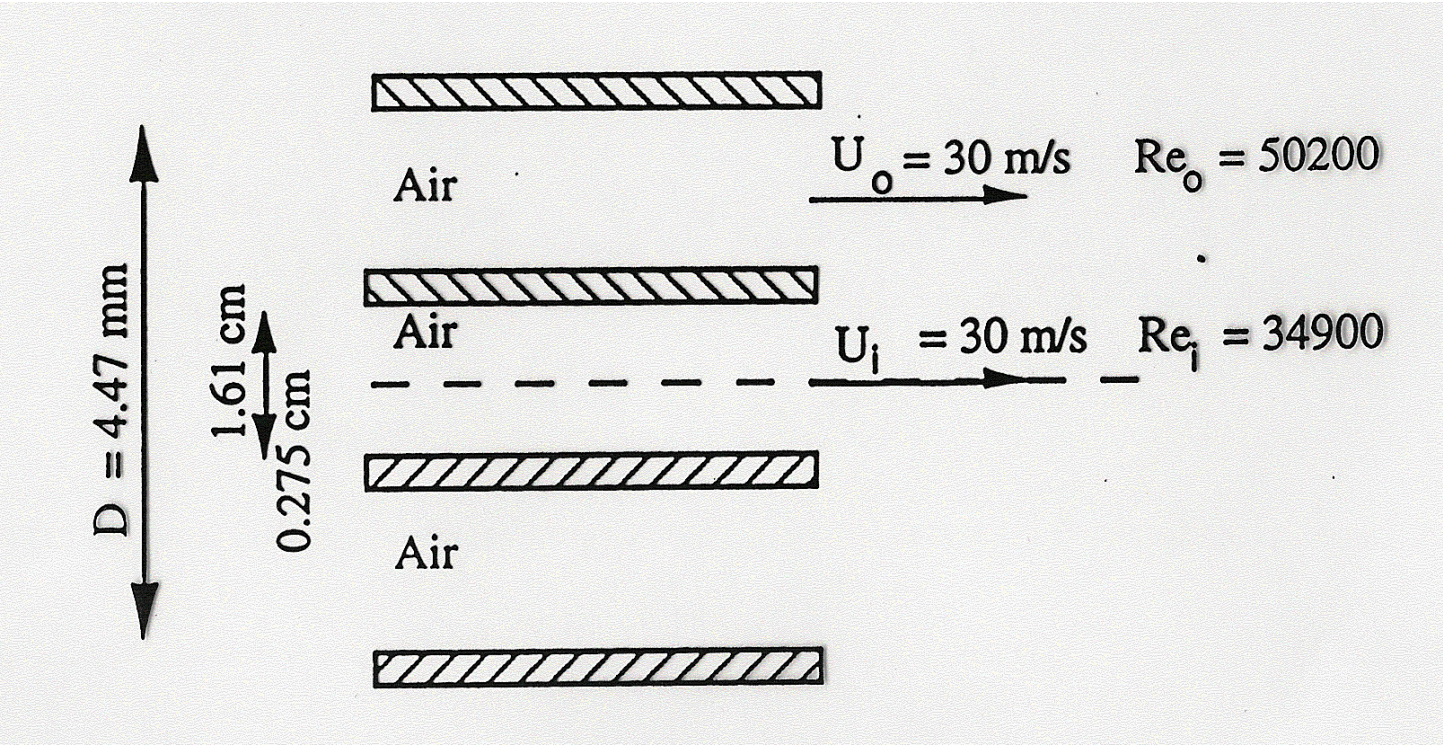

**Figure 10 : Configuration de jets coaxiaux étudiée par Ribeiro [Ribeiro, 1972].**

# 1.5 Application aux jets coaxiaux

La deuxième étape de ce travail a constitué à étudier les jets coaxiaux homogènes. Les écoulements de jets coaxiaux dans de l'air au repos ont été largement étudiés dans le passé [Ribeiro, 1980 ; Au, 1987 ; Dahm, 1992]. Ko et Au [Ko, 1981] ont démontré que l'écoulement peut être divisé en trois zones : la région initiale, la région intermédiaire et la région pleinement mélangée. La première s'étend jusqu'à la fin du cône à potentiel externe, la seconde jusqu'au point de rattachement, quand le maximum de la vitesse atteint l'axe et la dernière se situe en aval. Les jets coaxiaux sont caractérisés par la présence de deux couches de mélange, l'une confinée entre deux jets, l'autre située entre le jet annulaire et l'air au repos. L'aspect principal des jets coaxiaux est le mécanisme de mélange de deux écoulements initialement adjacents avec différents niveaux de turbulence et d'anisotropie, en fait deux histoires différentes de la turbulence. Un autre trait de cet écoulement, résultant du précèdent, est son instabilité. Pour des vitesses égales dans les deux jets, l'instabilité de type sillage prévaut, alors que l'instabilité de couche cisaillée domine pour les forts et faibles rapports de vitesse [Wicker, 1993; Dahm, 1992].

La première configuration étudiée a été celle de l'expérience de Ribeiro [Ribeiro, 1972]. Cette étude bien documentée a permis de comparer quantitativement les résultats numériques et expérimentaux. Ensuite, une seconde géométrie, plus proche de celles rencontrées dans les moteurs-fusées a été étudiée. Elle a permis d'évaluer l'impact du gradient de vitesse entre les jets.

Pour cette dernière, les conditions initiales ont été dérivées de l'étude expérimentale de Durão [Durão, 1971].

### 1.5.1 Validation

La géométrie, correspondante à l'expérience reconstruite numériquement [Ribeiro, 1972], est représentée sur la Figure 10. Les prédictions numériques ont été effectuées avec le modèle de turbulence de Launder et Sharma [Launder, 1974]. Les résultats mettent en évidence une forte instabilité dans la région proche. En l'absence de mesures expérimentales instationnaires, les résultats numériques ont été moyennés et comparés aux mesures [Ribeiro, 1972]. La Figure 11 représente la décroissance de la vitesse sur l'axe du jet. Un bon accord entre le calcul et l'expérience est observé, que cela soit dans la région proche ou dans la zone pleinement développée de l'écoulement.

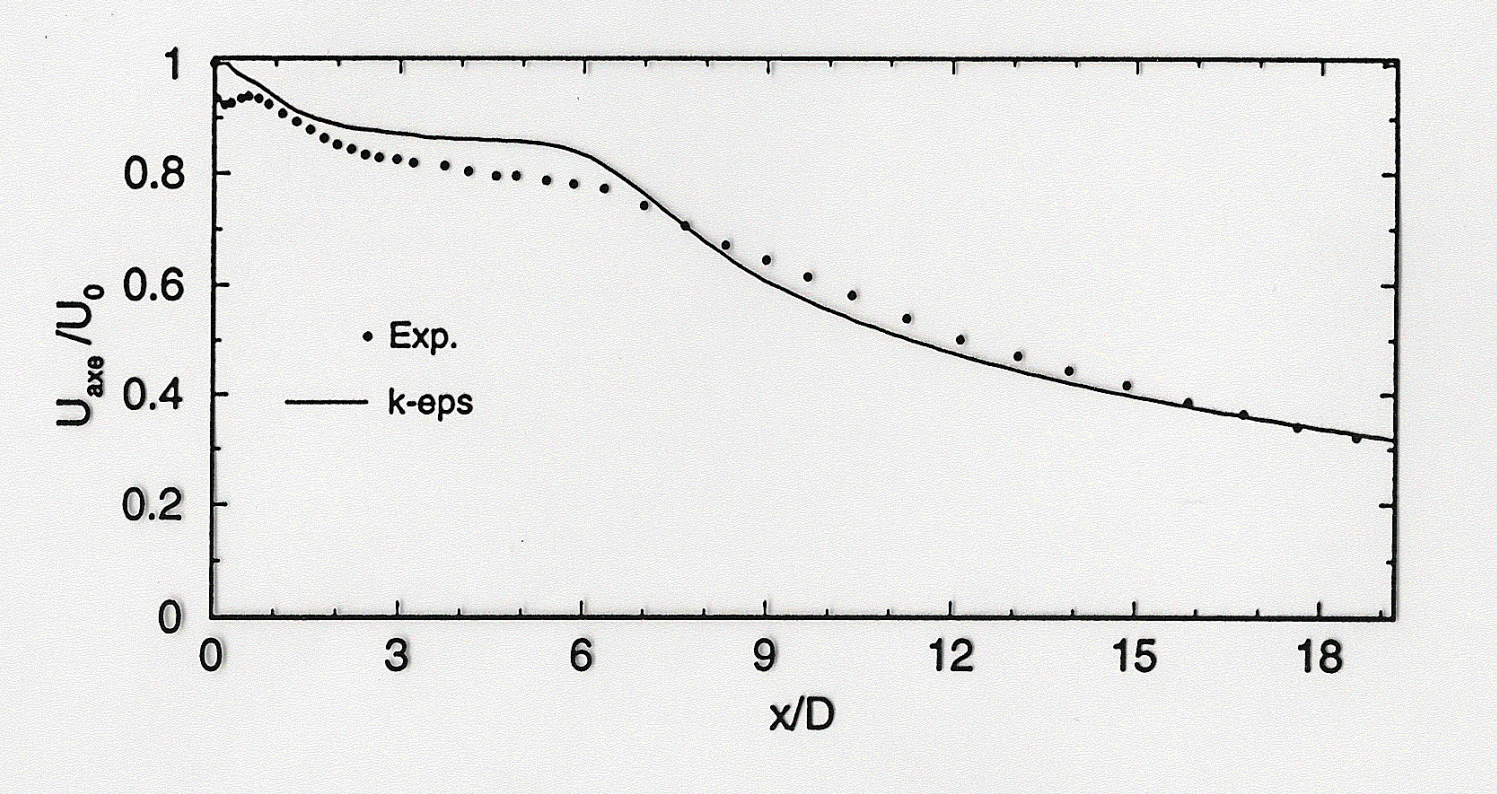

**Figure 11 : Distribution de la vitesse axiale adimensionnée.**

Les profils transversaux de vitesse dans la région proche sont représentés sur la Figure 12. Ils mettent aussi en évidence un bon accord entre les prédictions et l'expérience. Cet accord est cependant meilleur dans la zone pleinement développée de l'écoulement notamment en 6D et 10 D comme montré sur la Figure 13. Ce meilleur accord dans la région lointaine traduit la meilleure adaptation du modèle de LS dans la région où la turbulence est pleinement développée.

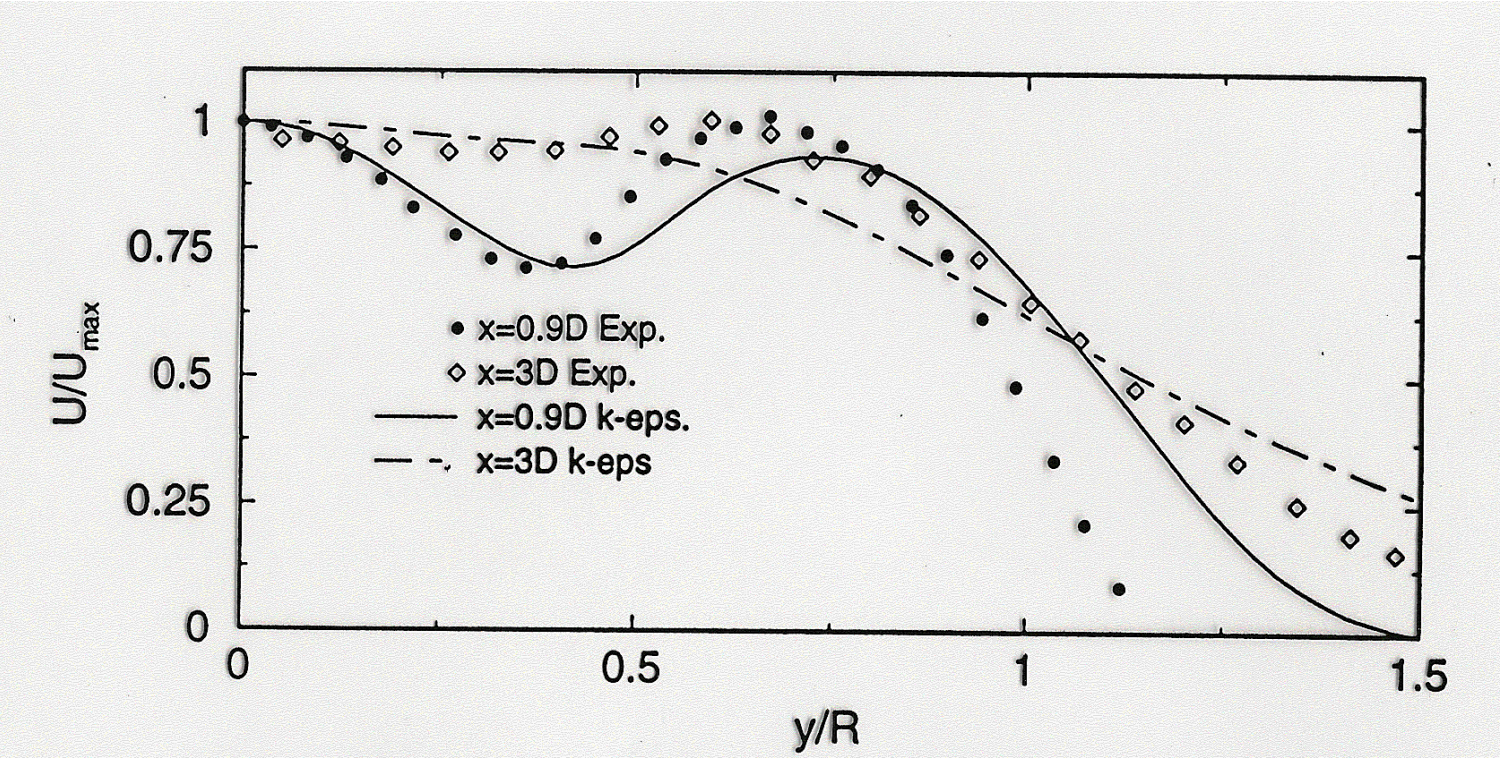

**Figure 12 : Profils transversaux de la vitesse adimensionnés par rapport à la vitesse sur l'axe en 0.9D et 3D.**

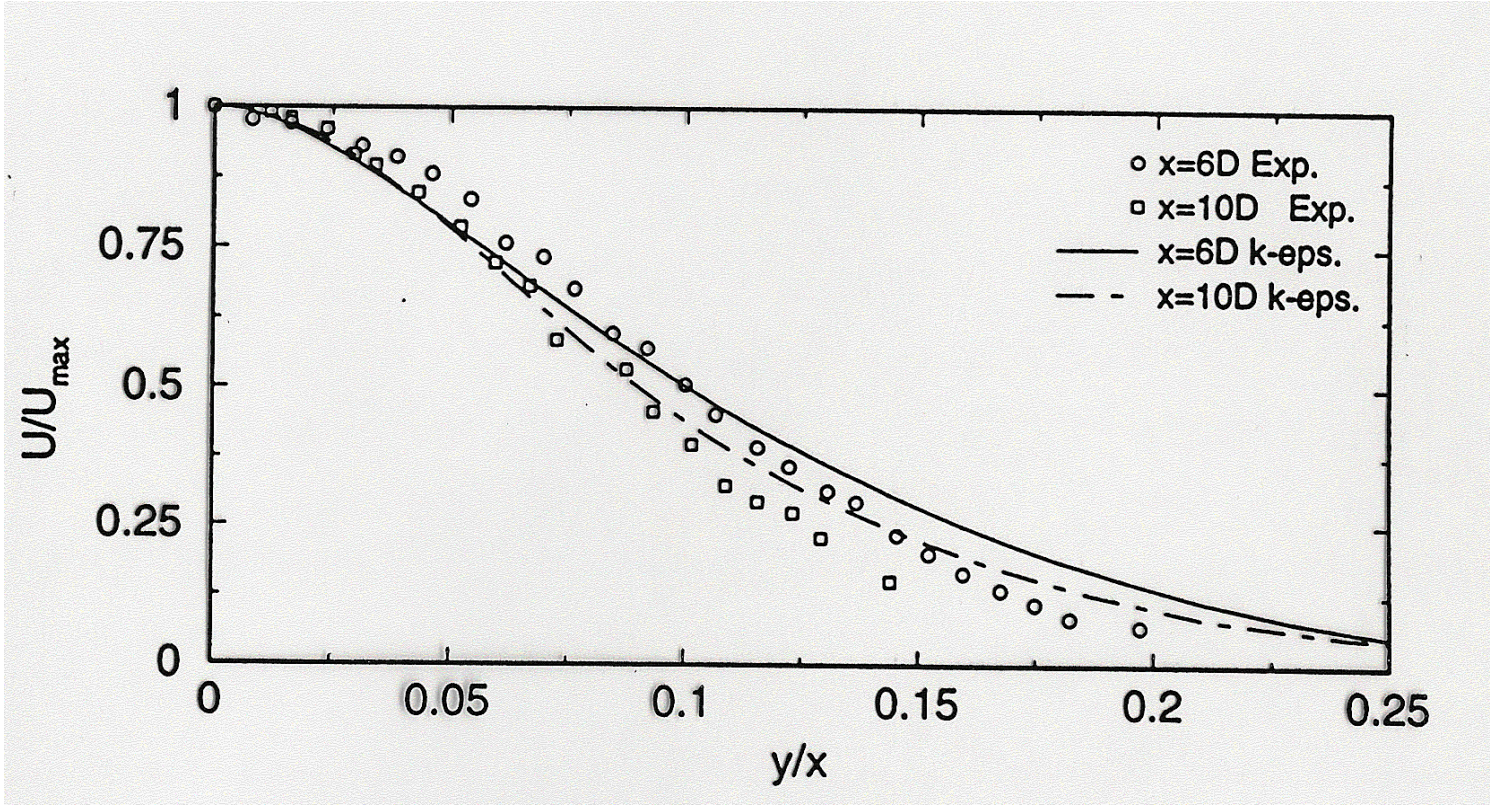

**Figure 13 : Profils transversaux de la vitesse adimensionnés par rapport à la vitesse sur l'axe en 6D et 10D.**

Un second calcul a été effectué avec un modèle de turbulence au second ordre [Launder, 1975]. L'utilisation de ce type de fermeture permet de prendre en compte l'anisotropie de la turbulence. Si cette dernière est faible dans les écoulements de jets ronds, il a paru pertinent d'évaluer son influence pour la prédiction des écoulements de jets coaxiaux. Les résultats obtenus n'ont pas mis en évidence une amélioration de la prédiction de l'écoulement lors de l'utilisation de ce modèle. Il en a été déduit que le niveau d'anisotropie de la turbulence est faible dans les écoulements de jets coaxiaux ce que laissait supposer l'expérience de Ribeiro [Ribeiro, 1972]. Cependant, ce niveau de fermeture des équations a été utilisé pour le calcul de la configuration de jets coaxiaux non homogènes (à densité variable).

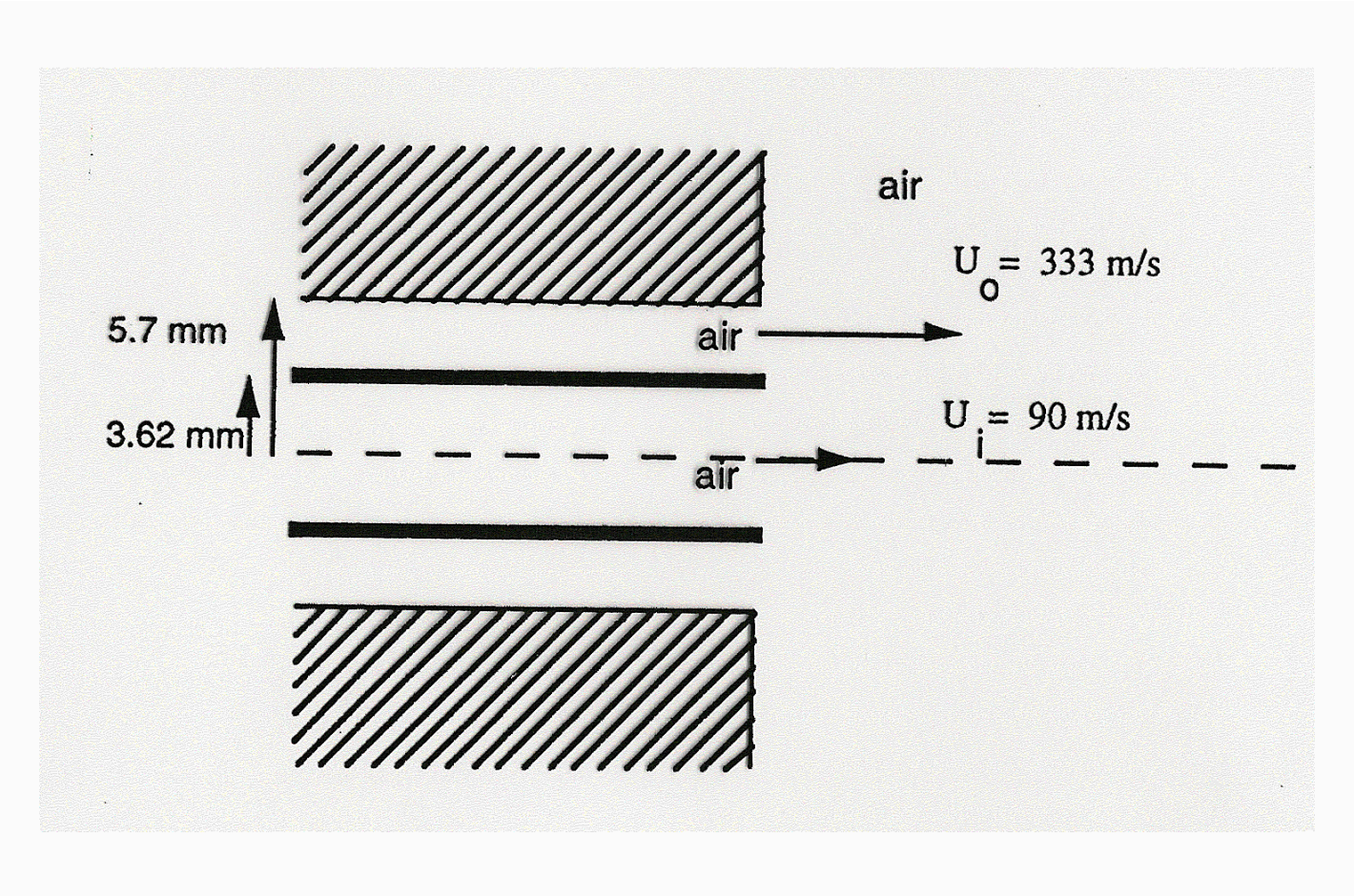

**Figure 14 : Configuration simulée pour les jets coaxiaux compressibles.**

### 1.5.2 Jets coaxiaux avec un fort rapport de vitesse

Dans le but d'observer numériquement l'effet du rapport de vitesse, une configuration avec un rapport de vitesse élevé a été calculée. Elle est représentée sur la Figure 14 et a été prédîtes pour des conditions initiales dérivées de l'expérience de Durão [Durão, 1971]. Le rapport de vitesse, $U_i/U_o$, entre les deux jets et égal à 0.27 avec un jet annulaire quasiment sonique à Mach 0.96.

Les résultats numériques ont mis en évidence une forte instabilité dans la région proche. Celle-ci est due à la présence de deux couches de mélange l'une confinée entre deux jets, l'autre située entre le jet annulaire et le fluide extérieur. Ces instabilités sont dues à la présence de structures cohérentes dans les deux couches cisaillées. Ces structures ont déjà été observées expérimentalement pour des jets coaxiaux homogènes [Dahm, 1993 ; Wicker, 1994] ou à densité variable [Gladnick, 1990].

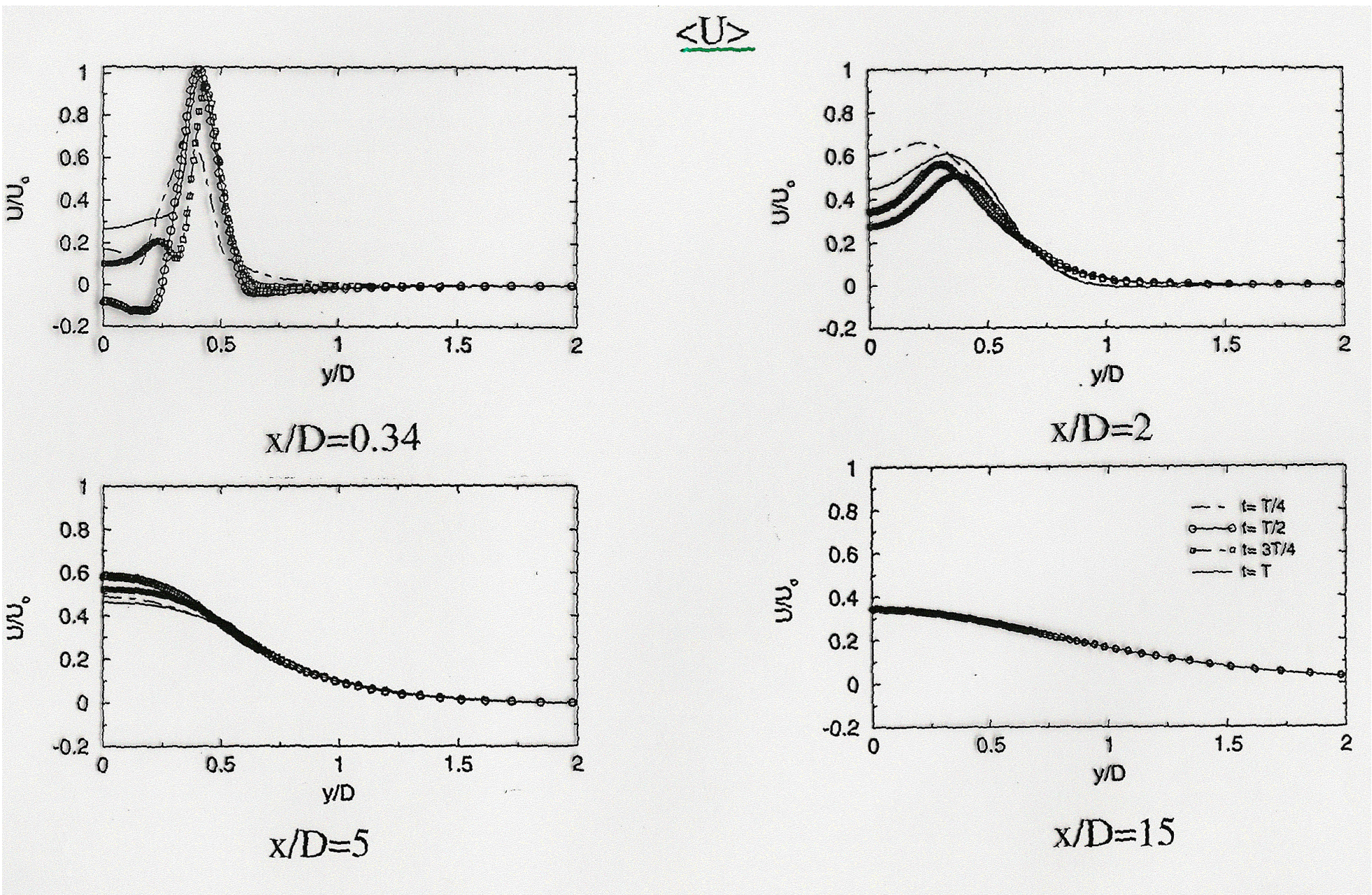

**Figure 15 : Profils transversaux de la vitesse axiale, adimensionnée par la vitesse en sortie du jet annulaire en 0.34D, 2D, 5D et 15D pour quatre instants d'une période.**

Les résultats numériques ont mis en évidence une forte instationnarité quasi périodique de l'écoulement dans la région proche des jets coaxiaux. Ceci est visible sur la Figure 15 qui représente les profils transversaux de la vitesse axiale en quatre instants d'une période, ce pour différentes distances depuis l'injection. La présence de structures cohérentes dans la région proche, traduit le mélange entre les deux jets ainsi que l'entraînement du fluide extérieur. Comme pour les jets ronds, ces structures tourbillonnaires, en forme de tores, sont induites par l'instabilité de Kelvin-Helmholtz. Dans la région initiale des structures azimutales peuvent être présentes [Dahm, 1992; Tang, 1994] mais ne peuvent être prédîtes ici par la simulation axisymétrique. Elles correspondent à une perte d'organisation de l'écoulement et mènent à la turbulence pleinement développée. Les instabilités simulées ici corroborent celles observées dans plusieurs expériences [Wicker, 1994 ; Gladnick, 1990] qui ont montré que la région proche est dominée par des structures instationnaires de grande échelle qui évoluent comme des instabilités dans les deux couches de mélange des jets coaxiaux. Ici, ces structures s'amortissent vers l'aval sous l'action de la diffusion et disparaissent dans la région développée entre 5D et 15D.

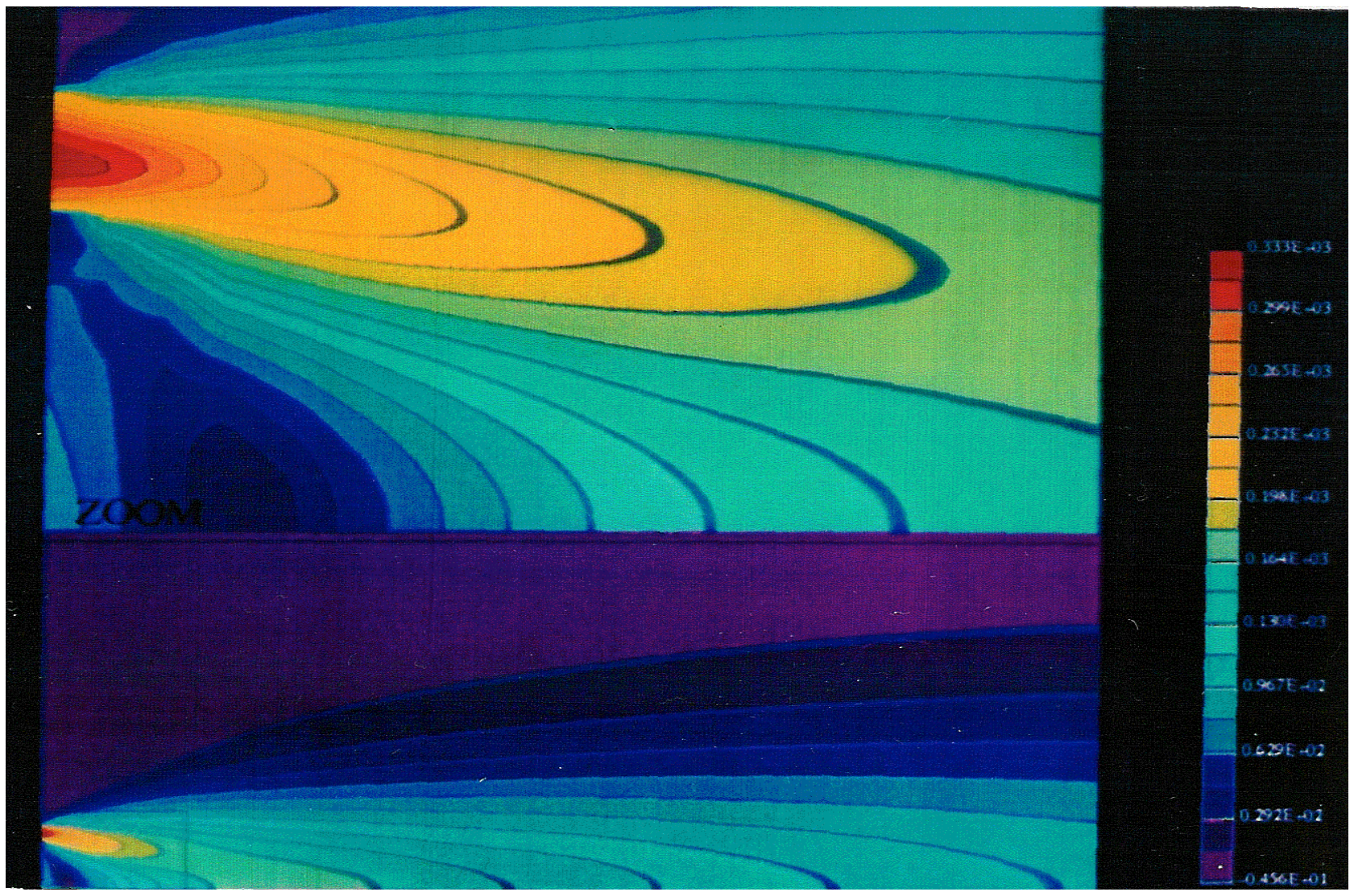

**Figure 16 : Champ de vitesse longitudinale en moyenne temporelle avec un zoom de la région initiale dans la partie supérieure.**

L'effet du rapport de vitesse est fondamental dans la région initiale de l'écoulement. Il influe notamment sur l'instabilité de la couche de mélange confinée entre les deux jets. Dans le cas présent, la Figure 16, qui représente le champ moyen de vitesse axiale, met en évidence l'existence d'une zone de recirculation située dans la partie centrale de la région initiale. Sous l'effet de la recirculation du jet annulaire, le cône à potentiel du jet central a pratiquement disparu comme dans l'expérience de Dixson [Dixson, 1983] pour des jets coaxiaux à effet de swirl. Dans ce calcul, le point de rattachement est situé à 4.3 D, les résultats d'Au et Ko [Au, 1987] prédisent un point de rattachement à 5.2D pour le même rapport de vitesse. Cette longueur de rattachement, inférieure ici de 25%, est influencée par la présence de la zone de recirculation qui a pour effet de supprimer le cône à potentiel du jet central.

Le phénomène de recirculation a déjà été observé expérimentalement pour des jets coaxiaux air-air et des rapports de vitesse inférieurs à 0.35 [Okajima, 1994], pour des jets avec effet de swirl [Dixson, 1983] et pour des jets non homogènes avec des forts rapports de quantités de mouvement [Camano, 1994]. Cependant d'autres études expérimentales pour des jets air-air [Au, 1987] n'ont pas mis en évidence cette recirculation et ce pour les mêmes rapports de vitesse. L'existence de cette région centrale de vitesse négative est due à la croissance des structures cohérentes qui pénètrent le jet central. Cette croissance dépend du rapport de vitesse entre les deux jets mais aussi des conditions initiales de l'instabilité, en termes d'intensité de la turbulence et d'épaisseur de la couche limite, qui influencent fortement la région initiale [Durão, 1971 ; Dahm, 1992]. Ainsi dans l'expérience d'Au et Ko [Au, 1987] le niveau initial de turbulence est faible tandis que dans celles d'Okajima [Okajima, 1994] le jet annulaire est pleinement développé. Il en est de même pour l'étude de Dixson [Dixson, 1983] où un effet de swirl, qui amplifie le niveau de turbulence à l'injection, est présent. En conclusion, un fort niveau initial de turbulence favorise la présence d'une zone de recirculation dans les jets coaxiaux à fort rapport de vitesse. Pour des jets hétérogènes gazeux la conclusion est la même mais c'est le rapport de quantité de mouvement et non de vitesse qui est à considérer. Cette présence d'une région de recirculation ainsi que les mêmes fréquences de l'instabilité illustre le couplage entre les deux couches de mélange et ce en accord avec l'expérience [Wicker, 1993, 1994].

# 1.6 Approche monophasique des sprays

## 1.6.1 Validité de l'approche

Comme nous l'avons vu en §1.1.2, l'objectif est d'étudier la région proche d'un spray qui est difficilement accessible par l'expérience. Les buts principaux sont de prédire la longueur du cône liquide et la taille des filaments arrachés au noyau liquide central par l'écoulement annulaire.

Le processus d'atomisation se déroule en deux phases : l'atomisation primaire ou breakup et l'atomisation secondaire. La première phase correspond à la destruction du cône liquide, la seconde à la production de gouttelettes à partir des filaments et gouttes issus de la première. Généralement lors du breakup, la destruction du noyau liquide est due à la génération d'instabilités qui initialement distordent puis cassent le cône liquide [Andrews, 1993 ; Issac, 1994 ; Mayer, 1994]. Pour le premier auteur les perturbations pourraient être générées par deux mécanismes : une onde hélicoïdale due à la relaxation du profil de vitesse dans le noyau liquide et une auto-induction cohérente des larges perturbations par les petites à travers une interaction non linéaire. Lors de l'atomisation secondaire [Wu, 1992], les gouttes sont produites par la destruction des filaments sous l'effet des ondes d'instabilité. Par conséquent, le diamètre des gouttes est directement relié à celui des filaments.

Dans les écoulements de jets d'injection, la vitesse du jet liquide est faible. Or dans ce cas là d'après Lefebvre [Lefebvre, 1992], l'atomisation devient presque seulement dépendante de la quantité de mouvement du gaz atomisant et du rapport de densité entre le liquide et le gaz. Une densité plus élevée du gaz accélère l'atomisation et produit des gouttelettes plus fines [Mayer, 1994]. Cette hypothèse est renforcée par plusieurs études expérimentales [Lin, 1987 ; Mayer, 1994 ; Yule, 1995]. Pour les jets d'injection de moteur-fusée, les résultats de Mayer [Mayer, 1994] ont montré que l'atomisation est largement indépendante des effets de haute température (du moins pour des pressions à l'intérieur de la chambre de combustion inférieures à 10 bars) et que par conséquent l'atomisation pour des injecteurs coaxiaux peut être traitée comme un écoulement froid.

En conséquence, les jets traités dans cette partie de l'étude seront considérés comme froids. Les effets de tension de surface seront eux aussi négligés. En effet, le processus d'atomisation, au moins dans la région proche, dépend essentiellement des effets aérodynamiques notamment des rapports de quantité de mouvement et de densité. Ce point est renforcé dans le cas de l'injection dans les moteurs cryotechniques. La pression à l'injection est supercritique et par conséquent les forces de tension de surface sont négligeables par rapport aux effets aérodynamiques avec un nombre de Weber supérieur à $10^4$ [Herding, 1997].

### 1.6.2 Calculs en présence de larges gradients de densité

Dans le but de se rapprocher de la configuration industrielle, avec un jet annulaire de gaz atomisant un jet central liquide, la géométrie de la Figure 14 a été simulée. L'objectif est d'approcher l'écoulement de spray à travers la simulation de jets coaxiaux gazeux mais non homogènes.

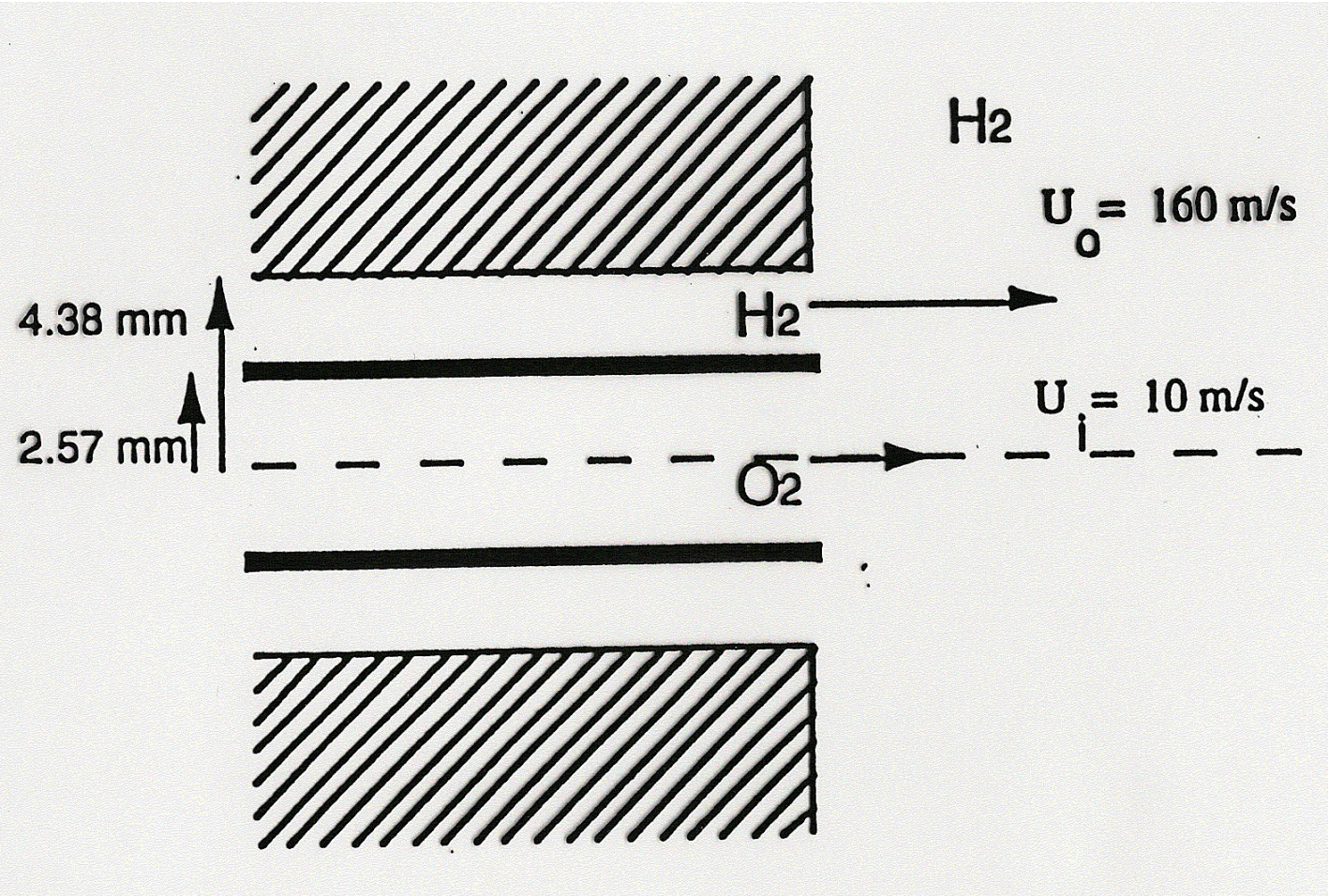

**Figure 17 : Configuration simulée pour les jets coaxiaux hétérogènes.**

Tout d'abord un écoulement d'hydrogène et d'oxygène gazeux avec des rapports de densité et de vitesse égaux à 16 (voir Figure 14) a été calculé. Les prédictions numériques ont été effectuées avec les deux modèles de turbulence utilisés en §1.5.1. Les conditions initiales ont été dérivées des données expérimentales [Durão, 1971] et les calculs ont été effectués avec un nombre de Schmidt (Sc = ν/d, où ν est la viscosité cinématique et d la diffusivité massique) égal à un [Chassaing, 1979]. Les résultats numériques visualisés sur la Figure 18 montrent le champ instantané de concentration d'oxygène et la distribution de densité en moyenne temporelle, obtenue avec le modèle aux contraintes de Reynolds [Launder, 1975] est visualisée sur la Figure 19.

Les résultats numériques montrent un écoulement beaucoup plus chaotique que pour les jets coaxiaux homogènes. La simulation ne prédit pas la présence d'une zone de recirculation, ici le gradient de vitesse est contrebalancé par celui de densité. Les calculs prédisent, en accord avec l'expérience de Mansour et Chigier [Mansour, 1994], un pic de turbulence localisé très près de l'injection ainsi qu'une décroissance rapide de la turbulence aléatoire. Il en est de même de l'anisotropie qui diminue très rapidement et ce toujours en accord avec cette étude [Mansour, 1994].

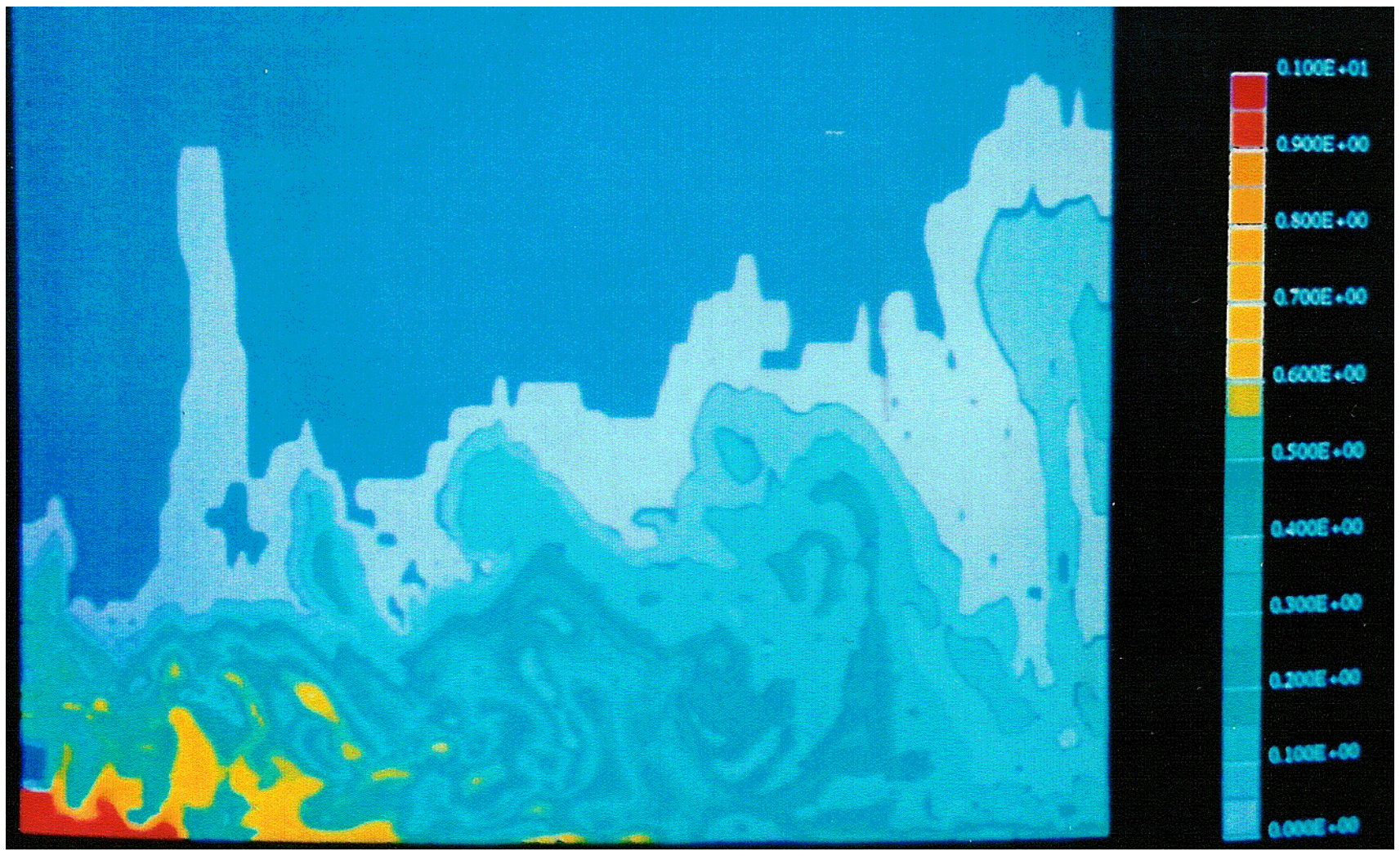

**Figure 18 : Champ instantané de concentration d'oxygène.**

La dispersion de l'oxygène est fortement affectée par la présence de structures organisées. Ceci est à rapprocher des résultats expérimentaux obtenus par Longmire et Eaton [Longmire, 1992] pour un jet ensemencé de particules qui montrent que la distribution de particules est fortement influencée par les structures cohérentes. On peut observer que la destruction du cône d'oxygène est assurée par les structures organisées comme dans le modèle d'atomisation par les structures cohérentes d'Andrews [Andrews, 1993]. La Figure 18 montre que l'arrachage de l'oxygène par l'hydrogène sous forme de paquets et de filaments commence dans la zone initiale aux environs de 2D. Ces filaments ont un diamètre largement inférieur à celui du jet central et sont étirés lors de leur convection vers l'aval. Le mélange déjà fort dans la zone initiale s'intensifie après 7D. La distribution de densité, considérée en moyenne temporelle, représentée sur la Figure 19, montre que la longueur du dard de gaz dense est de l'ordre de 7D. Cette valeur est très proche des mesures effectuées par Gicquel et al [Gicquel, 1995] dans des jets coaxiaux diphasiques oxygène/hydrogène pour lesquelles la longueur de la zone liquide (région avec plus de 50% de liquide) était de 6D.

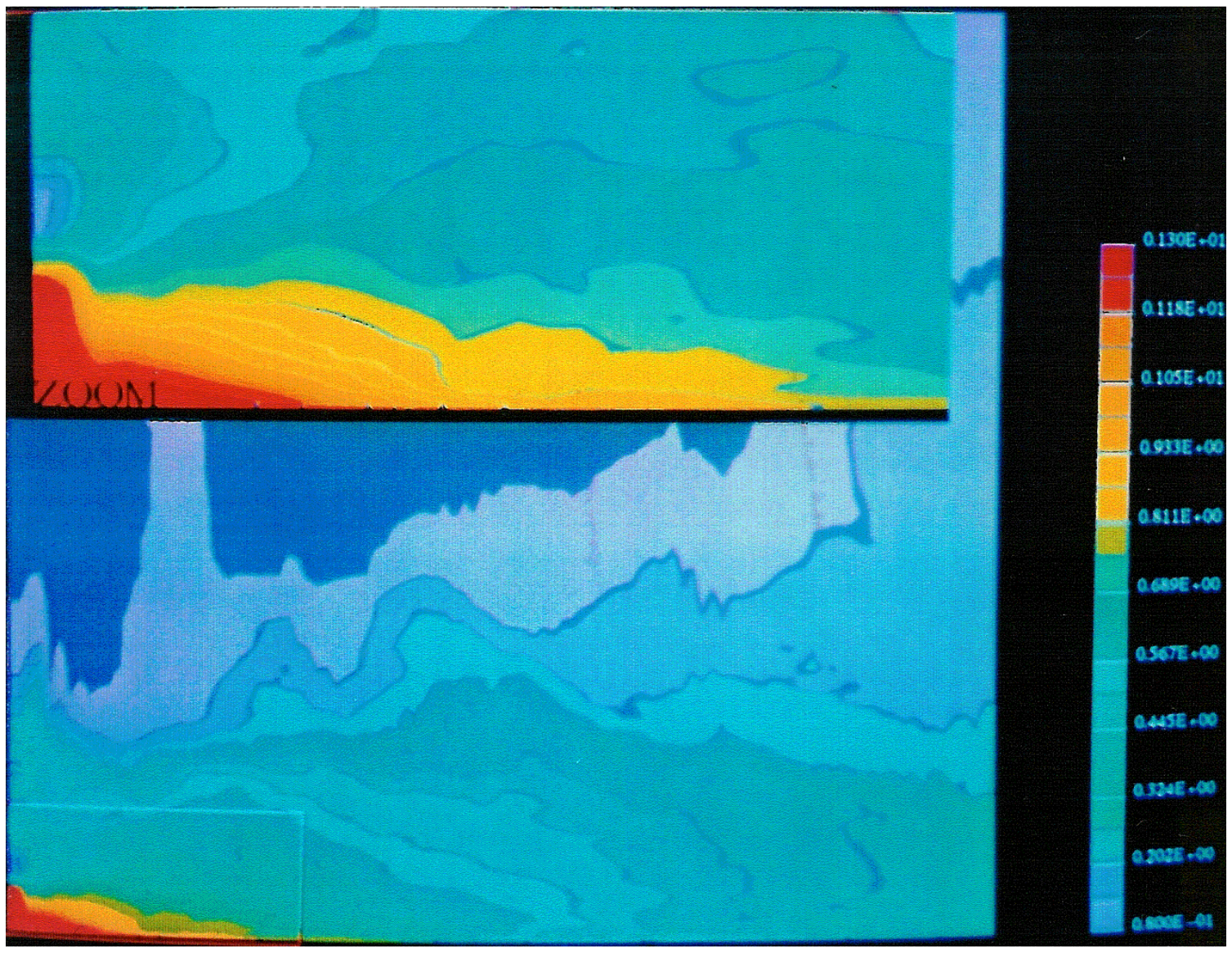

**Figure 19 : Distribution de la densité en moyenne temporelle.**

Dans un deuxième temps, plusieurs simulations numériques bidimensionnelles sans modèle de turbulence et avec des maillages plus raffinés ont été entreprises. Le même code de calcul que précédemment a été utilisé. Le but de l'exercice est d'effectuer de véritables simulations directes, cependant restreinte à des cas axisymétriques, pour différents rapports de densité. L'objectif était de préciser l'influence de ce rapport sur la destruction du jet central et d'approcher la configuration réelle, diphasique. Par conséquent, plusieurs calculs ont été entrepris pour des gradients de densité de plus en plus importants, de 1.05 à 32. Le calcul avec un rapport de densité de 16 est le plus intéressant car il peut être comparé aux simulations précédentes. La distribution instationnaire de la fraction massique d'oxygène est représentée sur la Figure 20. L'écoulement est plus chahuté que dans les simulations précédentes. Cependant, le même processus d'atomisation par les structures cohérentes est retrouvé avec une longueur assez proche du cône d'oxygène.

Le principal problème est apparu (être) d'ordre numérique. Le calcul devient de plus en plus instable quand le rapport de densité augmente et le cas pour un rapport de 32 n'a pas abouti à des résultats numériques exploitables. Il s'est avéré qu'en s'enroulant, les structures tourbillonnaires créent des zones de dépression avec des densités et des vitesses localement très faibles. Or, les schémas numériques basés sur la densité, c'est le cas ici, deviennent instables pour la simulation de tels écoulements. Nous reviendrons plus en détail sur ce problème et sur les techniques pour y remédier dans en §**Erreur ! Source du renvoi introuvable.**.

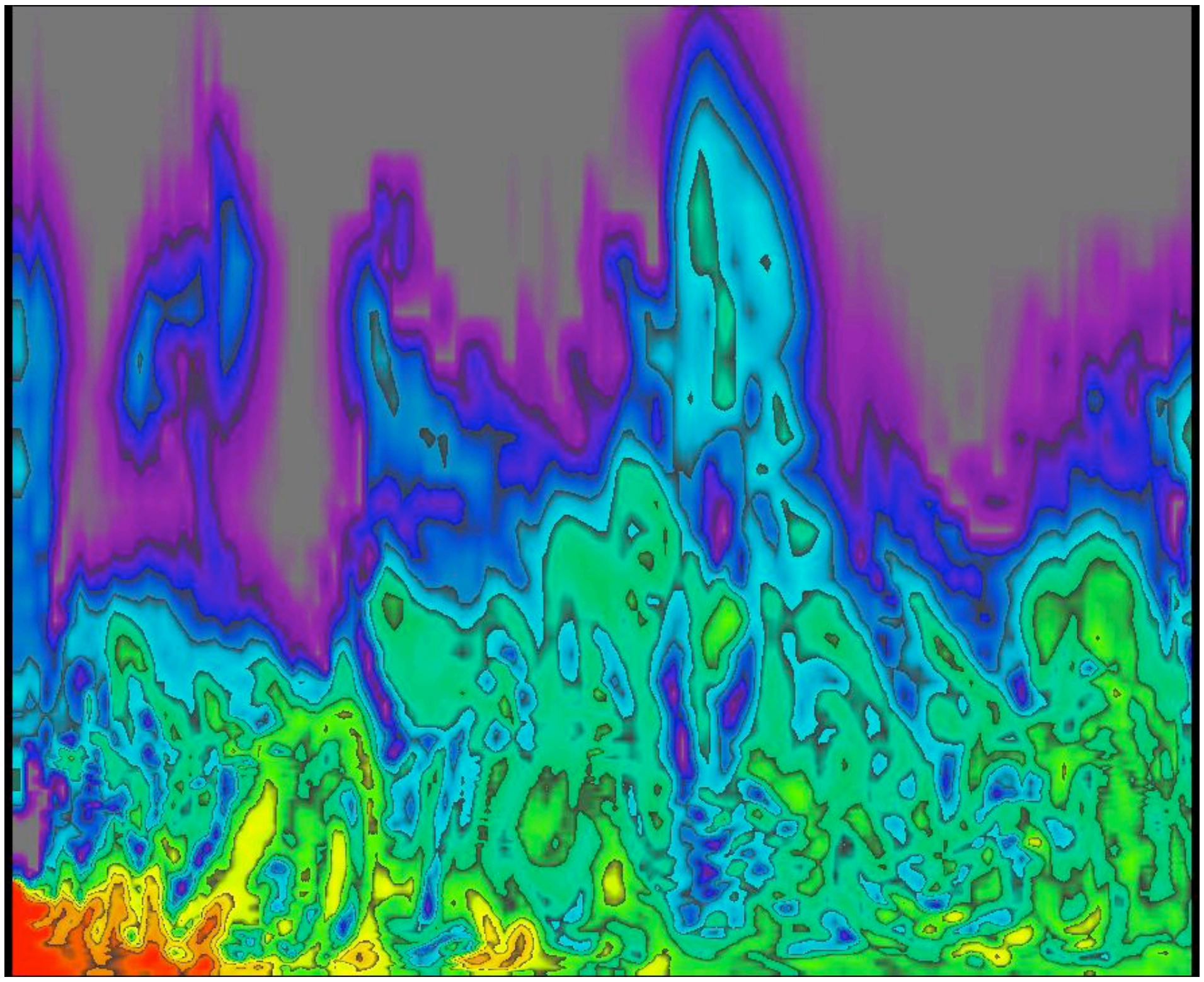

**Figure 20 : Champ instantané de la fraction massique d'oxygène.**

# 1.7 Conclusion

La modélisation semi-déterministe a été appliquée à la simulation des écoulements de jets. Son application a permis de simuler des instabilités dans la région proche des jets. Pour les jets ronds et coaxiaux homogènes, la comparaison des quantités considérées en moyenne temporelle par rapport aux résultats numériques ou expérimentaux d'autres études a mis en évidence un bon accord. Les nombres de Strouhal obtenus pour les jets ronds s'accordent eux aussi avec l'expérience. Les calculs pour différentes conditions initiales ont montré leur influence sur la présence de structures cohérentes. L'évolution des structures organisées, lorsqu'elles sont convectées vers l'aval, s'est avérée être en bon accord avec les observations de plusieurs études expérimentales rapportées dans la littérature. La présence d'une recirculation centrée sur l'axe du jet a été observée pour les jets coaxiaux à fort rapport de vitesse. La confrontation de ce résultat avec ceux de la littérature mène à la conclusion que le rapport de vitesse, ou de quantité de mouvement pour des jets a densité variable, est l'un des deux facteurs à l'origine de cette recirculation, l'autre étant l'état initial des jets en terme de niveau de turbulence et d'épaisseur de couche limite.

Si l'ensemble de ces résultats soutient la validité et l'intérêt de l'approche développée, il serait cependant particulièrement intéressant de confronter cette dernière à la LES. Cela permettrait d'une part une validation plus quantitative des résultats instationnaires mais aussi la possibilité de raffiner les modèles de turbulence utilisés à travers une recalibration de leurs constantes.

Les résultats obtenus pour l'approche monophasique des sprays mettent en évidence son intérêt pour prédire le processus d'atomisation initial des jets d'injection. Cette approche est originale et les résultats présentés ici soutiennent l'hypothèse que la longueur du dard liquide est accessible numériquement. Un second point est la mise en évidence du rôle joué par les structures cohérentes dans la destruction du cône liquide. Ce qui a limité l'application de l'approche, en terme de rapport de densité, est l'instabilité du schéma numérique qui s'accroît avec le rapport de densité. Ce type de problème sera abordé plus loin et nous verrons les solutions qui permettent de stabiliser les schémas numériques pour de très faibles nombres de Mach.

Ces travaux entièrement réalisés à l'Institut de Mécanique des Fluides de Toulouse ont fait l'objet de plusieurs communications référées en Annexes : [J1], [O1], [T], [C1-5], [A1-2], [R1] et [S1-2].

# 2 Complexité géométrique : L'aile en nid d'abeilles

## 2.1 Introduction

Les capacités des ailes en nid d'abeilles comme stabilisateurs sont connues depuis le début du XX$^{ème}$ siècle. Elles ont été d'abord utilisées pour le design de biplans (voir Figure 21) par des pionniers de l'aviation tels que O. et W. Wright en 1903 et A. Santos Dumont en 1906. Aujourd'hui, cette technique est utilisée dans les secteurs des lanceurs spatiaux, des missiles et des hydrofoils. Entre autres applications, elles servent de stabilisateurs arrière pour le système de secours du vaisseau Soyouz et pour les missiles balistiques russes. Depuis l'époque des pionniers de l'aviation, le design des ailes a évolué, un exemple en est le missile étudié dans ce chapitre, représenté sur la Figure 22.

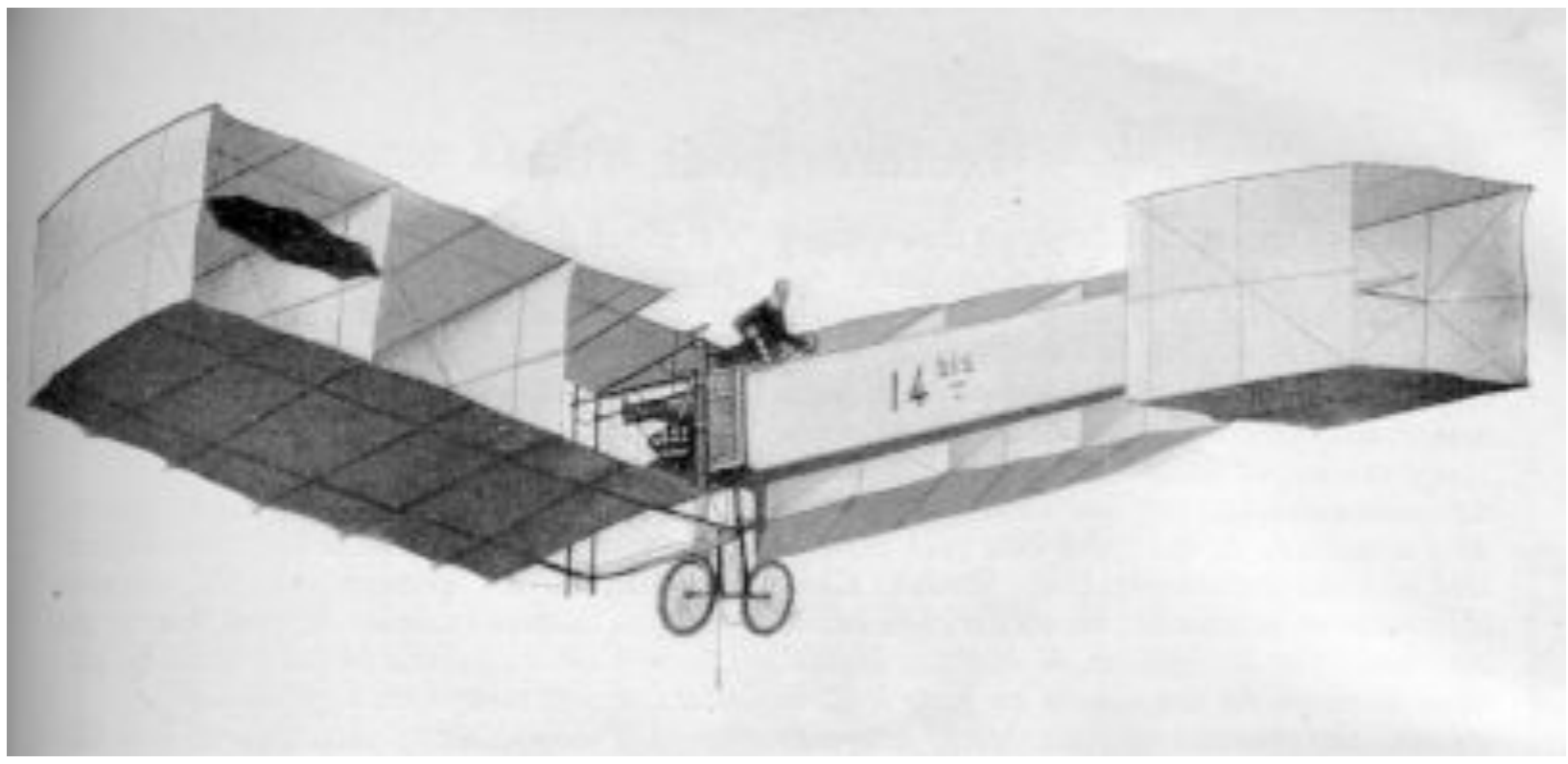

**Figure 21 : Le 14$^{bis}$ de A. Santos Dumont (1903).**

### 2.1.1 Caractéristiques aérodynamiques

Si les avantages des ailes en nid d'abeilles comme surfaces de contrôle aérodynamique sont bien connus, il existe cependant peu d'articles de la littérature qui leur sont consacrés. La plupart des études référencées, que ce soit sur les plans théorique, expérimental ou numérique datent pour les plus anciennes des années quatre-vingts [Belotserkovsky, 1985]. L'ensemble des travaux qui leur ont été consacrés, rapportent qu'elles sont complètement opérationnelles en régime supersonique, qu'elles possèdent des propriétés aérodynamiques importantes et qu'elles peuvent présenter plus d'avantages que les ailes planes. L'ensemble de ces avantages a été listé par Washington et Miller [Washington, 1998] :

- Excellentes caractéristiques de contrôle en régime supersonique ;
- Stockage compact et déploiement relativement facile ;
- Excellentes caractéristiques pour de grands angles d'attaque ;
- Faibles variations du centre de pression sur une grande plage de Mach ;
- Capacité à diminuer la traînée en modifiant la forme de l'aile ;
- Capacité à altérer la traînée et la portance par balayage de l'aile ;
- Rapports élevés entre la force et la masse de l'aile ;
- Faibles moments de charnière ce qui permet de réduire la taille de l'aile.

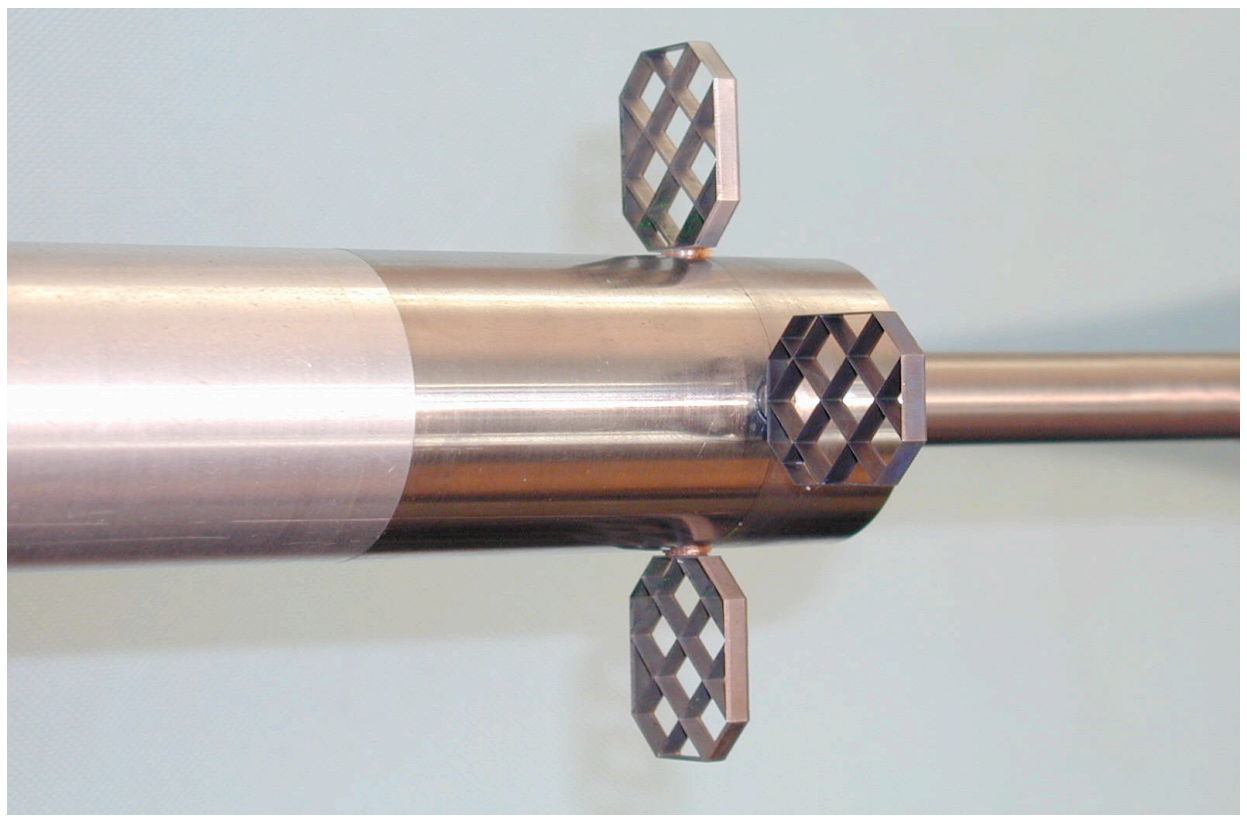

**Figure 22 : Missile équipé d'ailes en nid d'abeilles [Esch, 2000].**

Leur principal avantage est leur capacité à générer de la portance ce qui les rend idéales pour augmenter le moment de tangage [Khalid, 1998; DeSpirito, 2001]. Washington et Miller [Washington, 1998] ont mesuré que les ailes en nid d'abeilles produisaient la même force normale que les ailes planes à basse vitesse et 50 % de plus à Mach 2.5. Un autre point est les faibles moments de charnière générés par ce type d'aile, rapportés dans plusieurs campagnes expérimentales [Brooks, 1989; Simpson, 1998; Washington, 1993], ce qui permet de réduire la taille des systèmes d'action et de contrôle. Brooks et Burkhalter [Brooks, 1989] ont de plus montré que le moment de charnière n'augmentait que faiblement avec l'angle d'attaque. Leur efficacité dans le domaine supersonique ainsi que leur taille réduite et les faibles moments de charnière générés les rend très attractives pour des applications de type missile.

Elles présentent néanmoins deux désavantages importants : elles subissent une perte de stabilité en régime transsonique et ont des niveaux de traînée élevés. En régime transsonique, la force axiale engendrée par les ailes en nid d'abeilles n'est pas supérieure à celle produite par les ailes traditionnelles [Simpson, 1998]. Cette dernière étude a mis en évidence une perte de stabilité à Mach 1.45 indiquant que les effets transsoniques sont plus sévères que pour les ailes conventionnelles. Les calculs ont aussi mis en évidence l'apparition du choc pour les cellules individuelles des ailes en nid d'abeilles à Mach 0.8 ce qui explique leur faible performance en régime transsonique. En général, la combinaison d'une incidence élevée et d'une déflection importante réduit l'efficacité de ce type d'aile et ce indépendamment du nombre de Mach. Durant longtemps l'utilisation des ailes en nid d'abeilles a été restreinte par le niveau élevé de traînée engendré par cette technologie. En effet, à portance équivalente, la traînée induite par une aile en nid d'abeilles est plus importance que pour une aile conventionnelle. Cependant, dans la dernière décennie, plusieurs travaux expérimentaux [Miller, 1994; Washington, 1993] ont montré que la traînée pouvait être considérablement réduite en modifiant la forme de l'aile ainsi que l'épaisseur du cadre extérieur et des parois des cellules et ce avec un impact minimal sur la portance et les autres propriétés aérodynamiques.

### 2.1.2 Difficultés sur le plan numérique

Les ailes en nid d'abeilles ont aussi été étudiées théoriquement et numériquement. Plusieurs études [Kretxschmar, 1998 ; Brooks, 1989; Burkhalter, 1995] ont été effectuées pour l'analyse théorique de ces ailes à l'aide de méthode de type vortex. L'application des méthodes linéaires [Burkhalter, 1995] ne donne plus de résultats adéquats pour le calcul des coefficients aérodynamiques et ce pour des angles d'attaque supérieur à 5-8 degrés. Plus tard, Kretxschmar & Burkhalter [Kretxschmar, 1998] et Burkhalter et Franck [Burkhalter, 1996] ont montré que les formules empiriques développées pour étendre les capacités des prédictions aérodynamiques à des angles d'attaque élevés permettent d'avoir un bon accord avec les données expérimentales et ce pour une grande variété de configurations et une large plage de nombres de Mach.

Les premières études numériques consacrées aux ailes en nid d'abeilles et rapportées dans la littérature datent des années quatre-vingt dix [Sun, 1997 ; Khalid, 1998]. Les calculs effectués par Lesage [Lesage, 1998] pour une cellule isolée ont montré que l'approche Euler non visqueuse sous-estime les résultats expérimentaux. Les prédictions numériques pour une missile complet [Sun, 1997 ; Khalid, 1998 ; DeSpirito, 2001] ont mis en évidence un très bon accord avec l'expérience quand une approche Navier-Stokes est utilisée. Les solutions de type Euler ne permettent pas, comme prévu, de simuler les zones de séparation de l'écoulement : il en résulte un accord médiocre avec l'expérience. Si d'un point de vue numérique la résolution de la configuration est possible, elle augmente toutefois le coût de calcul par un facteur cinq du fait que 80 % du maillage est localisé au niveau des ailes [DeSpirito, 2001].

Depuis la fin des années quatre-vingt dix le DLR a mené plusieurs études sur les missiles supersoniques équipés d'ailes en nid d'abeilles. Plusieurs campagnes expérimentales sur des configurations complètes, ou pour des ailes isolées, ont été effectuées [Esch, 1997, 1999, 2000]. Sur le plan numérique, dans le but de s'affranchir de l'inconvénient du maillage, un "disque d'action" a été développé pour simplifier le problème (voir §2.2). Ce disque d'action a d'abord été intégré dans un code structuré, couplé à une base de données expérimentales, et testé sur une aile isolée [Reisch, 2000 ; C9]. Ensuite ce disque d'action a été appliqué et validé pour une configuration complète [C9, J3]. Lors de cette première phase de l'étude, le couplage entre le disque d'action et la base de données qui permet d'estimer les forces induites par l'aile, a montré ses limites. Finalement, cette technique a été incorporée dans un solveur non structuré et couplé à un module semi empirique et testée pour un missile complet.

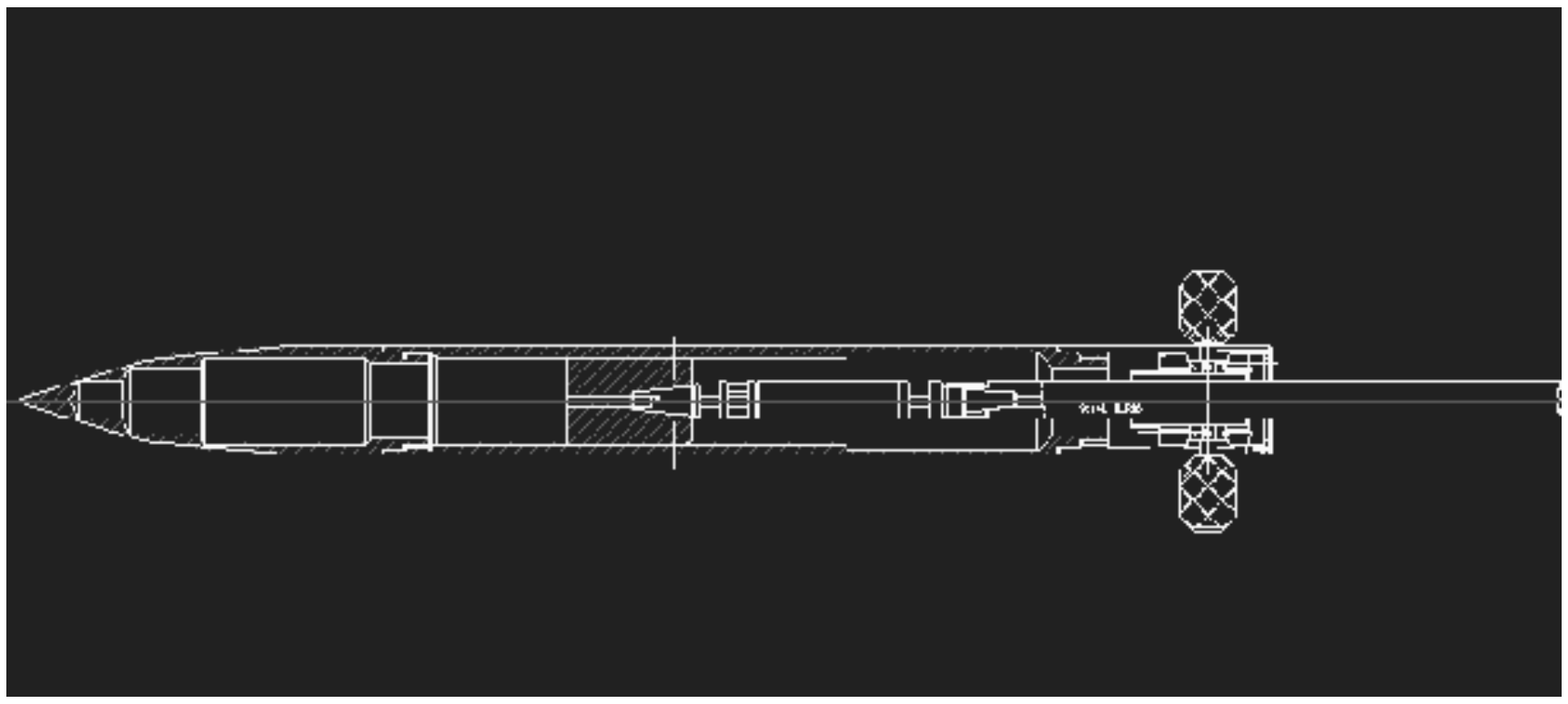

**Figure 23 : Missile équipé d'ailes en nid d'abeilles. Pour une meilleure visualisation, celles-ci sont parallèles et non perpendiculaires au missile.**

# 2.2 Disque d'action

## 2.2.1 Concept

Un disque d'action consiste en une condition limite artificielle au sein d'un écoulement. Sa théorie a été décrite en détail par Horlock [Horlock, 1978]. D'après cet auteur le concept de disque d'action est contemporain de la théorie de Rankine-Hugoniot relative aux écoulements à travers les hélices de navires. Depuis, la technique du disque d'action a été appliquée à un grand nombre de problèmes d'ingénierie tels que le rotor d'hélicoptère, le moulin à vent ou les turbomachines à étages multiples. Ici cette technique est utilisée pour modéliser les effets des ailes en nid d'abeilles dans le but de prédire, à un coût raisonnable, les performances aérodynamiques d'un missile. La configuration étudiée ici est représentée sur la Figure 23, les détails géométriques des ailes en nid d'abeilles utilisées sur la Figure 24. Avec la technique du disque d'action, chaque aile est remplacée dans l'écoulement par un jeu de conditions limites (voir Figure 25). A ces conditions limites, les forces induites par la présence de l'aile sont intégrées aux équations de bilan. Les avantages principaux de la technique sont de diminuer de façon importante le coût de calcul d'une telle configuration par rapport à un calcul complet Navier-Stokes et d'avoir un outil numérique adapté au design de véhicules munis de ce type de stabilisateur.

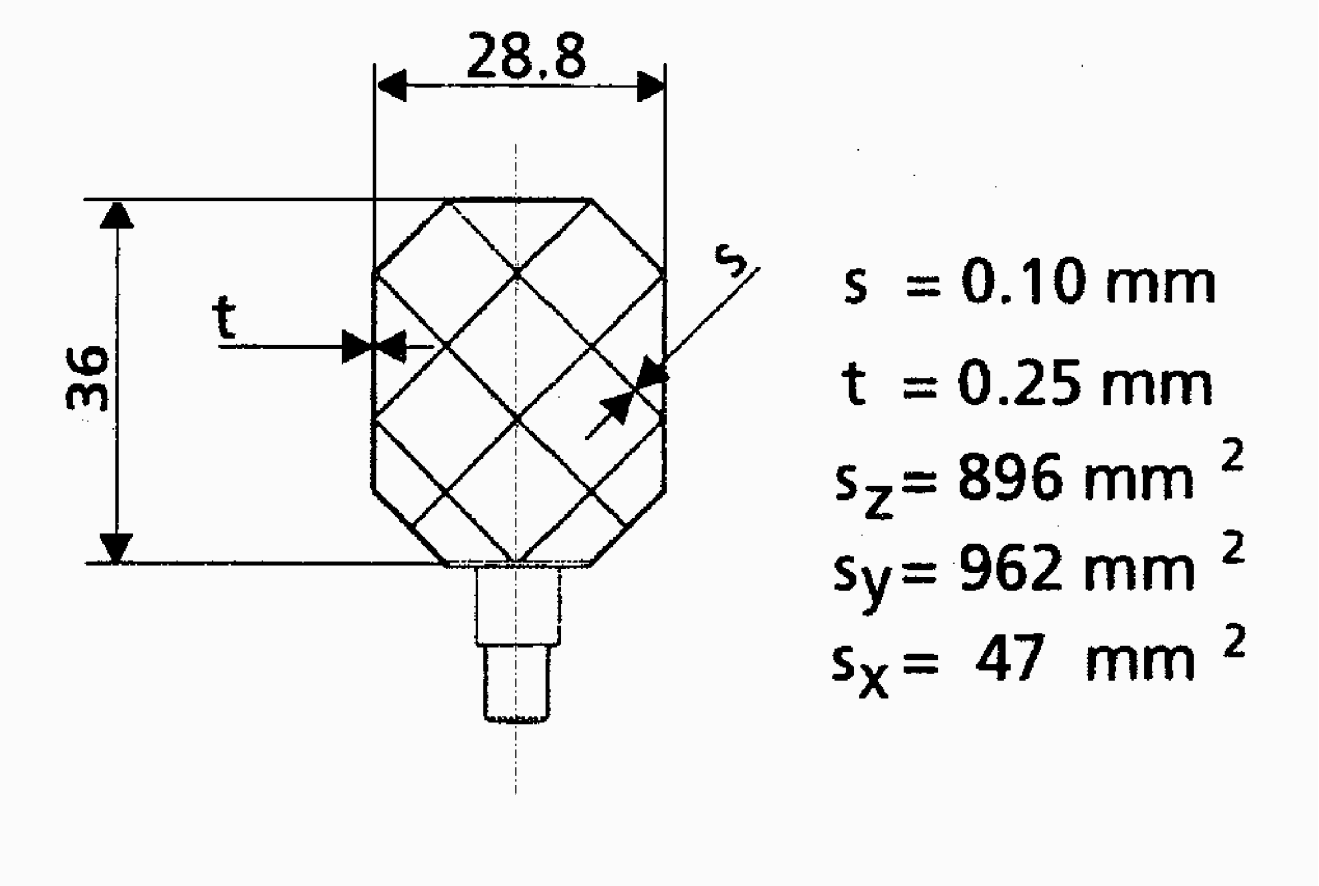

**Figure 24: Géométrie de l'aile en nid d'abeilles utilisée pour l'étude numérique et les mesures expérimentales [Esch, 2000].**

### 2.2.2 Equations du disque d'action

La théorie du disque d'action [Horlock, 1978] est principalement basée sur l'application des lois de conservation de la masse de la quantité de mouvement et de l'énergie. Les équations du disque d'action développé pour les ailes en nid d'abeilles sont brièvement décrites ci-après (voir [J3] pour plus de détails).

À une cellule du maillage située sur le disque d'action, la conservation de la quantité de mouvement s'écrit sous la forme suivante :

$$\frac{d\vec{J}}{dt} = \int_V \frac{\partial \rho \vec{q}}{\partial t} + \int_S \rho \overleftarrow{q} \cdot (\overleftarrow{q}\,\overleftarrow{n})\, dS = \sum \vec{F} \qquad (2.1),$$

où $\vec{J}$ est le vecteur de quantité de mouvement et $\vec{n}$ le vecteur normal à la surface *S*. Les forces extérieures appliquées sur la surface sont les forces de pression et la force de réaction de l'aile. La force de réaction suivant *x*, $F_x$, appliquée à une cellule du maillage de surface $S_x$ peut s'écrire :

$$F_x = -\frac{1}{2} c_x \gamma p Ma^2 \frac{S_{ref,LW}}{n} \frac{S_x}{t_v t_h} \qquad (2.2)$$

où $c_x$ est le coefficient aérodynamique de l'aile dans la direction *x*, γ le rapport des chaleurs spécifiques, *Ma* le nombre de Mach, $S_{ref,\ LW}$ la surface de référence de l'aile, *n* le nombre d'alvéoles de l'aile et $t_v$ et $t_h$ les dimensions d'une alvéole. Dans les équations de quantité de mouvement la force pariétale agissant sur le fluide est requise. Cette force suivant *x*, $D_x$, est déduite de l'équation (2.2) comme étant :

$$D_x = -\frac{1}{2} c_x \gamma p Ma^2 \frac{S_{ref,LW}}{n t_v t_h} \qquad (2.3)$$

Les autres composantes de la force de réaction de l'aile en nid d'abeilles à l'écoulement. $D_y$ et $D_z$ sont calculées de la même façon.

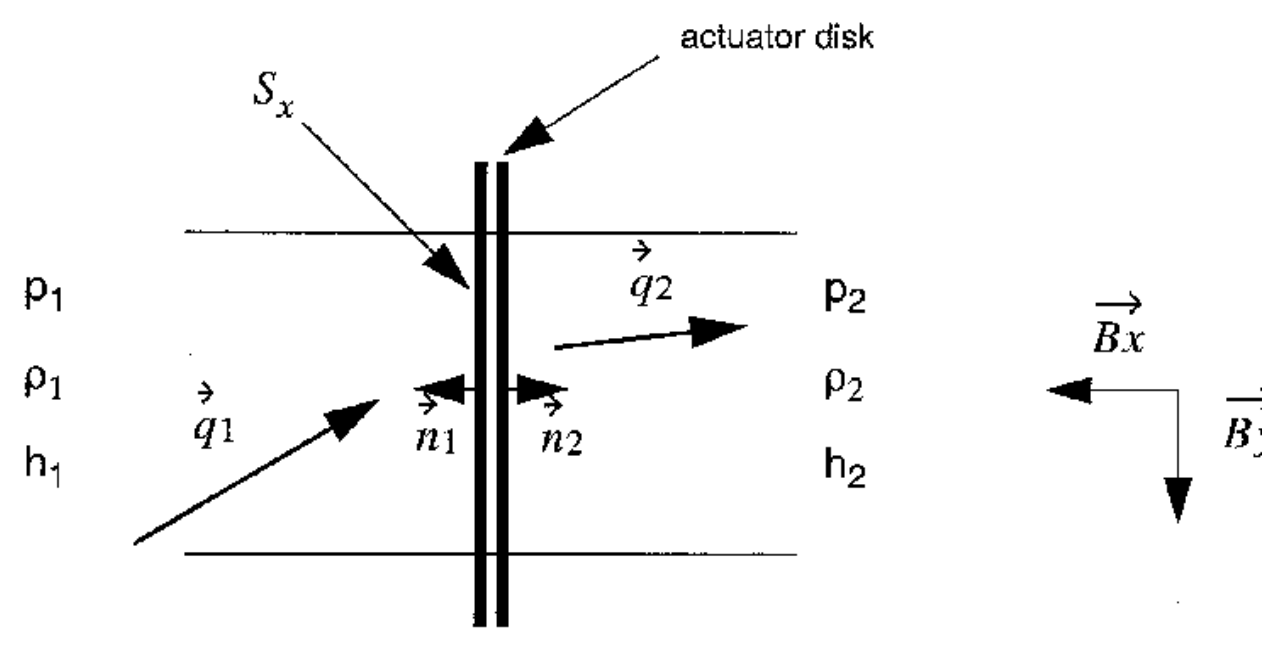

**Figure 25: Conditions de l'écoulement au niveau du disque d'action. Les indices 1 et 2 correspondent respectivement aux variables en amont et aval du disque, *h* est l'enthalpie, *p* la pression, *ρ* la densité, $\vec{q}$ le vecteur vitesse et $\vec{n}$ le vecteur normal au disque d'action.**

Dans la base de coordonnées associée à l'aile en nid d'abeilles $\vec{B}$ (voir Figure 25), l'équation de conservation de la quantité de mouvement (2.1) mène au système d'équations suivant :

$$\rho_1 u_1^2 - \rho_2 u_2^2 = p_2 - p_1 + D_x \qquad (2.4)$$

$$\rho_1 u_1 v_1 - \rho_2 u_2 v_2 = D_y \qquad (2.5)$$

$$\rho_1 u_1 w_1 \quad - \quad \rho_2 u_2 w_2 \quad = \quad Dz \tag{2.6}$$

où $u$, $v$ et $w$ sont les composantes de la vitesse dans le système de coordonnées associé à l'aile en nid d'abeilles. Deux autres équations sont nécessaires pour fermer le système, elles sont données par les lois de conservation de la masse et de l'énergie :

$$\rho_1 u_1 \quad = \quad \rho_2 u_2 \tag{2.7}$$

$$h_1 \quad + \quad \frac{1}{2} q_1^2 \quad = \quad h_2 \quad + \quad \frac{1}{2} q_2^2 \tag{2.8}$$

L'équation (2.8) est valide pour les écoulements visqueux tant qu'ils sont adiabatiques donc sans échange de chaleur avec les parois [Shapiro, 1953].

### 2.2.3 Conditions limites

Le disque d'action utilise les équations (2.4-8). À l'intérieur du maillage, une aile en nid d'abeilles est remplacée par un jeu de conditions limites. La partie amont du disque d'action correspond à des conditions limites de sortie du fluide tandis que sa partie aval correspond à des conditions d'entrée. Les composantes de la force de réaction de l'aile à l'écoulement sont incorporées aux équations de bilan au niveau de la partie aval du disque d'action.

Les conditions limites de sortie de la partie amont sont dérivées de la théorie des caractéristiques [Thompson, 1987]. Si l'écoulement est subsonique, une des variables doit être imposée pour obtenir un jeu de conditions limites numériquement stable. Les quatre autres variables sont extrapolées depuis l'intérieur de l'écoulement. Ici, le flux massique est la quantité qui est impose dans le cas d'un écoulement subsonique [J3]. Pour un écoulement supersonique toutes les variables sont extrapolées.

Pour la partie aval du disque d'action, si l'écoulement est supersonique, toutes les variables doivent être imposées. Elles sont calculées à partir de leurs valeurs respectives sur la frontière amont du disque d'action. Dans le cas d'un écoulement subsoniques, quatre variables doivent être imposées tandis que la cinquième est extrapolée depuis la frontière amont du disque d'action. Ici, c'est le carré de la vitesse que est extrapolé, plusieurs calculs ont en effet montré que cette technique menait à des distributions de pression plus réalistes que celles obtenues en extrapolant la pression [Reisch, 2000].

# 2.3 Aspects numériques

## 2.3.1 Codes de calcul

Le disque d'action a été incorporé dans le code de calcul FLOWer [Aumann, 2000] dans un premier temps puis dans le code TAU [Gerhold, 1997]. Ces deux codes sont développés par le DLR, le premier est basé sur une approche structurée tandis que le deuxième peut utiliser des maillages structurés, hybrides ou non structurés construit à partir de prismes, de pyramides, de tétraèdres ou d'hexaèdres. Les deux codes sont aux volumes finis et résolvent les équations de Navier-Stokes sous leur forme tridimensionnelle. Pour la discrétisation des flux, le schéma AUSM a été utilisé pour les calculs avec FLOWer et sa variante AUSM-DV pour les simulations effectuées avec TAU. Dans les deux codes l'intégration en temps est assurée par un schéma multipas, explicite de Runge-Kutta et des termes de dissipation numérique sont ajoutés pour atténuer les oscillations numériques de haute fréquence. Les schémas finaux sont précis au second ordre en espace. Dans le but d'économiser le temps de calcul, la technique multigrille a été utilisée pour l'ensemble des calculs. Ces deux codes sont vectorisés et parallélisés.

## 2.3.2 Intégration du disque d'action

Le disque d'action décrit en §2.2 a été intégré dans les deux codes de calcul [Reisch, 2000 ; R4]. Pour un missile tel que celui de la Figure 22, chaque aile en nid d'abeilles est remplacée par un disque d'action et par conséquent un jeu de conditions limites. Chaque disque d'action a exactement les mêmes dimensions qu'une aile.

Dans le code structuré, le disque d'action est situe à l'interface entre deux blocs [Reisch, 2000], la partie amont correspondant à la condition limite aval d'un bloc tandis que la partie aval du disque correspond à la condition d'entrée du bloc suivant.

L'intégration dans le solveur non structuré [R4] a été différente. Dans le maillage, il n'est plus possible d'utiliser une interface et il a été nécessaire de conserver l'épaisseur de l'aile, le disque d'action n'est plus alors strictement bidimensionnel. Sur les cotés de l'aile des conditions limites de sortie supersonique ont été appliquées. Des simulations avec une paroi non visqueuse ont mis en évidence un choc quand l'écoulement devient transsonique. La présence de ce choc induit une traînée et une portance additionnelles, or la contribution de l'aile est déjà prise en compte par le disque d'action. L'utilisation de cette condition limite pour les cotés de l'aile et l'ensemble du disque d'action ont été valides pour une aile isolée [O2]. Une seconde différence avec l'intégration dans le code structuré est que le disque d'action est constitué de deux surfaces avec a priori des maillages différents.  Or, le jeu de conditions limites est fait de telle manière que pour être appliqué, chaque cotés du disque a besoin de la valeur des variables de l'autre partie. En conséquence, pour chaque point d'un des cotés du disque d'action, un point miroir a été créé. À ce point miroir les valeurs des différentes variables de l'écoulement aux points adjacents sont interpolées.

# 2.4 Configuration

## 2.4.1 Géométrie et maillage

La configuration calculée, un missile muni de quatre ailes en nid d'abeilles, est représentée sur la Figure 22., à noter que le corps seul du missile (sans ailes) a été aussi calculé. Pour les prédictions avec le code FLOWer le maillage est constitué de quatre blocs. Le maillage est représenté sur la Figure 26, le nombre de mailles est de $128 \times 96 \times 64$ pour le missile proprement dit et $16 \times 128 \times 64$ pour le sillage. L'utilisation de la méthode multigrille a permis de vérifier l'indépendance des résultats par rapport au maillage. Le maillage en aval du culot s'étend sur une distance égale à un tiers de la longueur du missile. Le même maillage a été utilisé pour le missile complet et pour le corps isolé. Sur les frontières extérieures du maillage les conditions limites appliquées sont dérivées des relations caractéristiques [Thompson, 1987]. Les mêmes conditions ont été utilisées pour les calculs avec TAU avec un maillage hybride.

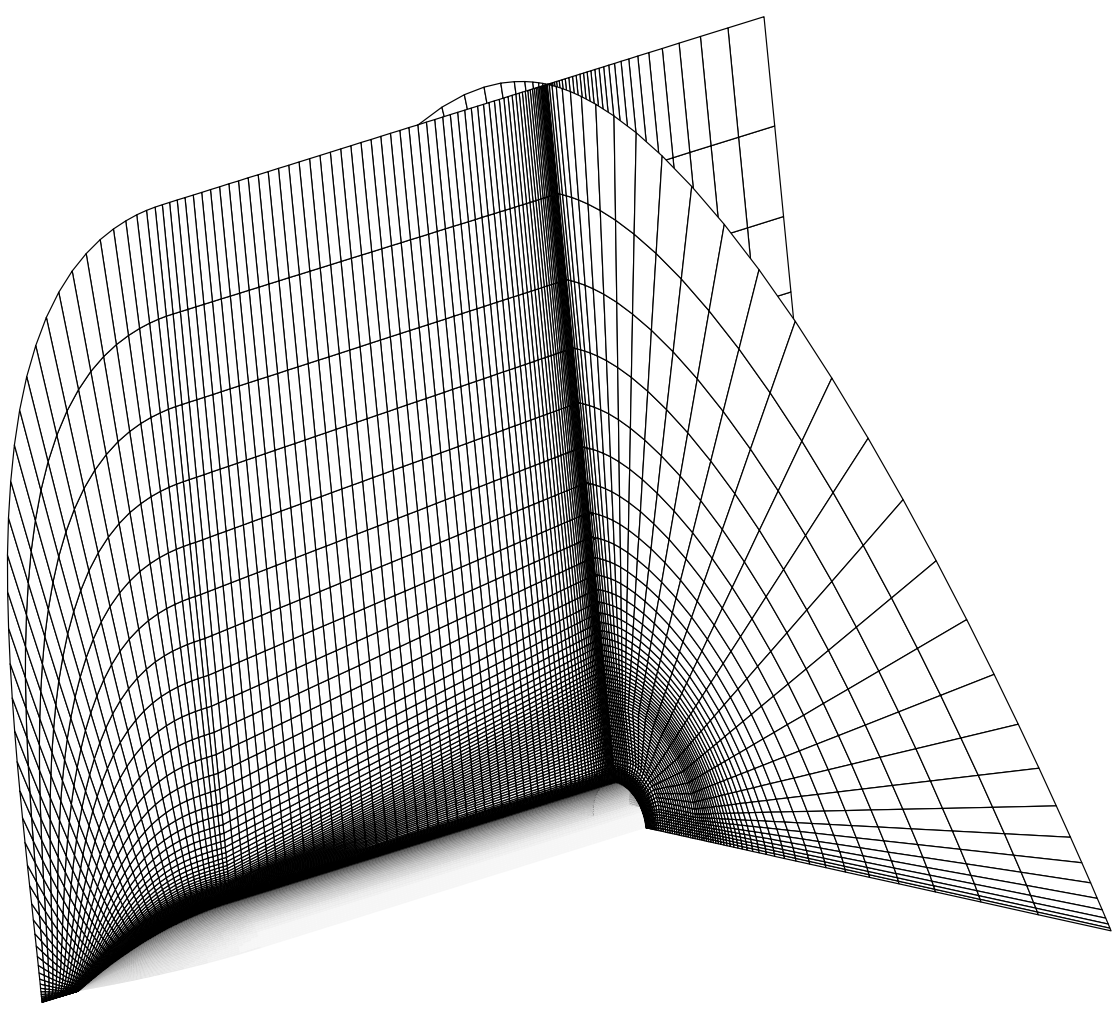

**Figure 26 : Maillage utilisé pour les calculs effectués avec le code FLOWer.**

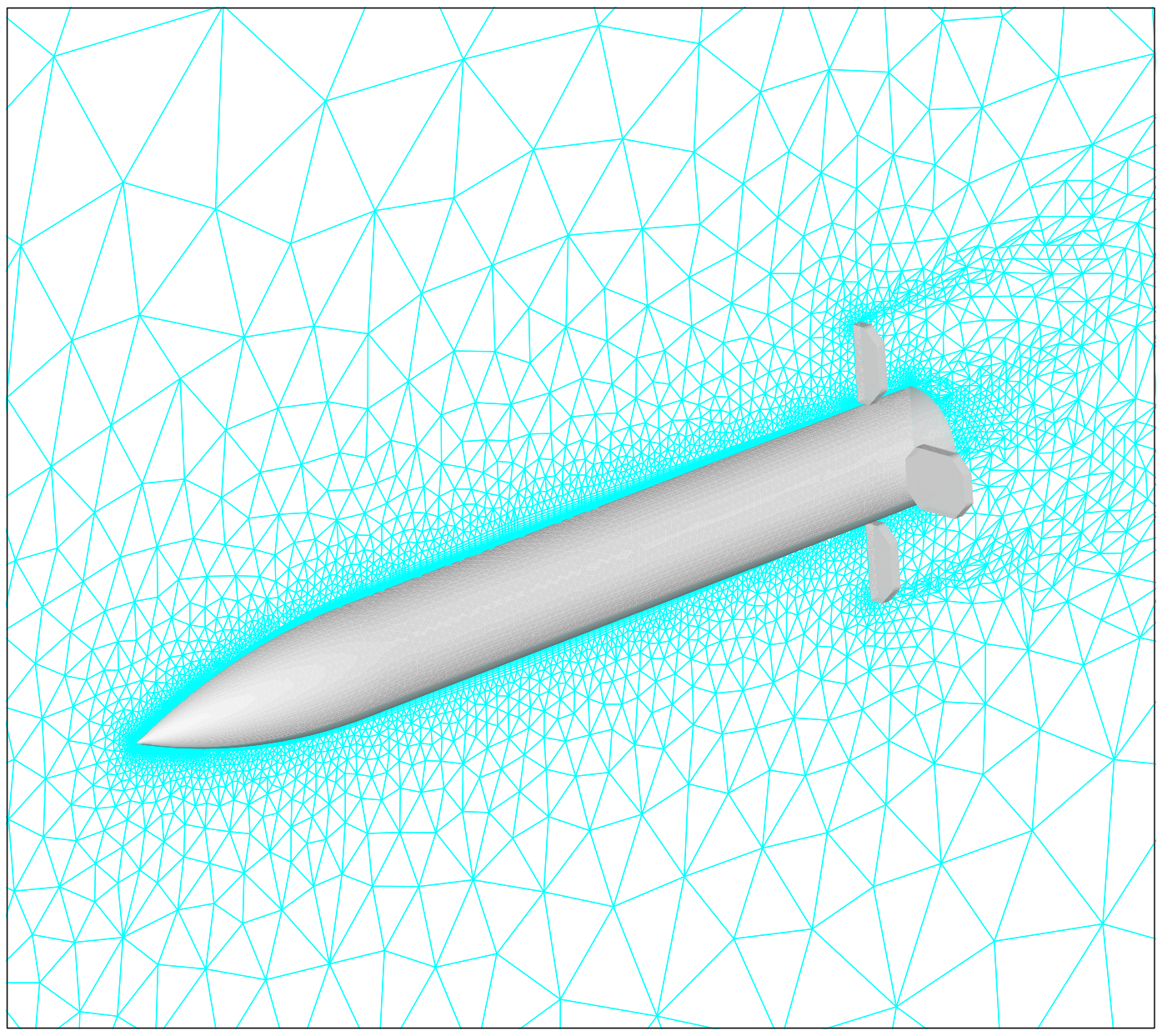

**Figure 27 : Plan de symétrie du maillage hybride après adaptation utilisé par TAU.**

Pour les calculs avec TAU un maillage assez grossier est utilisé au départ. Après convergence du calcul, ce maillage est adapté à l'aide d'une méthode de type gradient et environ 10 % de mailles supplémentaires sont ajoutées. Ce processus est répété plusieurs fois jusqu'à ce que l'indépendance des résultats soit atteinte. En pratique, le maillage est d'abord adapté au niveau du choc puis sur le reste du véhicule, la dernière étape est l'optimisation du maillage dans la couche limite. Le maillage final, représenté sur la Figure 27, comprend environ 400000 tétraèdres pour la partie non structurée et 560000 prismes localisées près des parois. Le maillage du corps isolé a nécessité à peu près le même nombre de cellules.

### 2.4.2 Cas de calcul

La configuration de la Figure 22 a été calculée à l'aide des maillages décrit en §2.4.1 pour une plage de nombres de Mach allant de 1.8 à 4. L'ensemble des calculs effectués est résume dans le Tableau 3. Le disque d'action a aussi été testé pour différents angles d'attaque : 5, 10 et 20 degrés. Le but initial était de développer une technique valable pour les grands nombres de Mach (jusqu'à 6). Les calculs avec FLOWer ont cependant mis en évidence une limitation de la base de données pour les nombres de Mach élevés, aussi les calculs avec angle d'attaque ont été effectués pour un nombre de Mach de 1.8. Pour les calculs avec angle d'attaque faits avec TAU, le couplage avec une méthode semi empirique a permis de s'affranchir des limitations de la base de données et l'écoulement a été prédit pour un nombre de Mach de 4. Dans l'objectif d'établir l'effet du disque d'action pour chaque cas de calcul, le corps seul (sans ailes) a aussi été prédit.

| Mach | Angle d’attaque | Codes |
|---|---|---|
| 1.8 | 0 | FLOWer + TAU |
| 2 | 0 | TAU |
| 3 | 0 | FLOWer + TAU |
| 4 | 0 | FLOWer + TAU |
| 1.8 | 5 | FLOWer |
| 1.8 | 10 | FLOWer |
| 1.8 | 20 | FLOWer |
| 4 | 5 | TAU |
| 4 | 10 | TAU |
| 4 | 20 | TAU |

**Tableau 3 : Cas de calcul simulés avec les codes TAU et FLOWer.**

L’étude expérimentale menée pour la même géométrie par Esch [Esch, 2000] a montré que l’écoulement devenait turbulent au niveau de la partie conique située à l’avant du missile. Les simulations numériques, résumées dans le Tableau 3, effectuées avec le code FLOWer pour des écoulements laminaires et turbulents [J3] ont mis en évidence un meilleur accord calcul/expérience pour les cas turbulents. En conséquence, les résultats montrés ici ont été obtenus à l’aide d’une modélisation de la turbulence. Le modèle retenu pour les simulations effectuées avec FLOWer est celui de Baldwin-Lomax, pour les prédictions avec TAU le modèle k-ω de Wilcox a été choisi.

Pour l’ensemble des calculs la méthode multigrille a été utilisée. Lors de chaque simulation, le calcul a d’abord été convergé sur un maillage grossier puis sur le maillage fin. Le maillage grossier est obtenu à partir du maillage fin en fusionnant deux cellules de ce dernier et ce dans chaque direction. La convergence a été obtenue après environ 50000 itérations pour les calculs avec le maillage structuré et un maximum de 30000 itérations pour le maillage hybride. Le nombre de CFL utilisé avec FLOWer varie de 1.5, pour une simulation laminaire à Mach 1.8, à 0.01 pour une prédiction turbulente à Mach 4. Pour les calculs avec TAU, le nombre de CFL utilisé a varié de 2 pour un calcul laminaire sans angle d’attaque à 0.5 pour un calcul turbulent avec 20 degrés d’angle d’attaque. La convergence des résultats par rapport au maillage a été vérifiée pour l’ensemble des simulations numériques.

## 2.5 Intégration CFD/Expérience

Dans l’objectif de prédire l’écoulement, en terme de forces et de moments, autour d’un missile muni d’ailes en nid d’abeilles, le disque d’action développé en §2.2 a été couplé à une base de données aérodynamiques puis à une méthode semi empirique basée sur la théorie des ailes en nid d’abeilles [Belotserkovsky, 1987]. Les coefficients aérodynamiques d’une aile, mesurés expérimentalement, servent d’entrées pour le disque d’action qui intégré dans un code de calcul permet la prédiction des coefficients de forces et de moments d’un missile complet. La technique considérée dans son ensemble repose sur l’intégration de la CFD et de l’expérience.

### 2.5.1 Couplage avec une base de données

Dans une première phase, le disque d’action, intégré dans le code FLOWer, a été couplé à une base de données expérimentales. Les coefficients de la base de données ont été obtenus pour la même aile isolée que celle utilisée pour le missile complet. Des calculs pour une aile isolée ont [C9] permis de valider l’outil numérique pour des écoulements subsoniques et supersoniques. Finalement, plusieurs simulations numériques (voir Tableau 3) pour différents nombres de Mach et angles d’attaque ont été effectuées pour évaluer les capacités de la technique développée pour prédire les coefficients aérodynamiques d’un missile équipé d’ailes en nid d’abeilles.

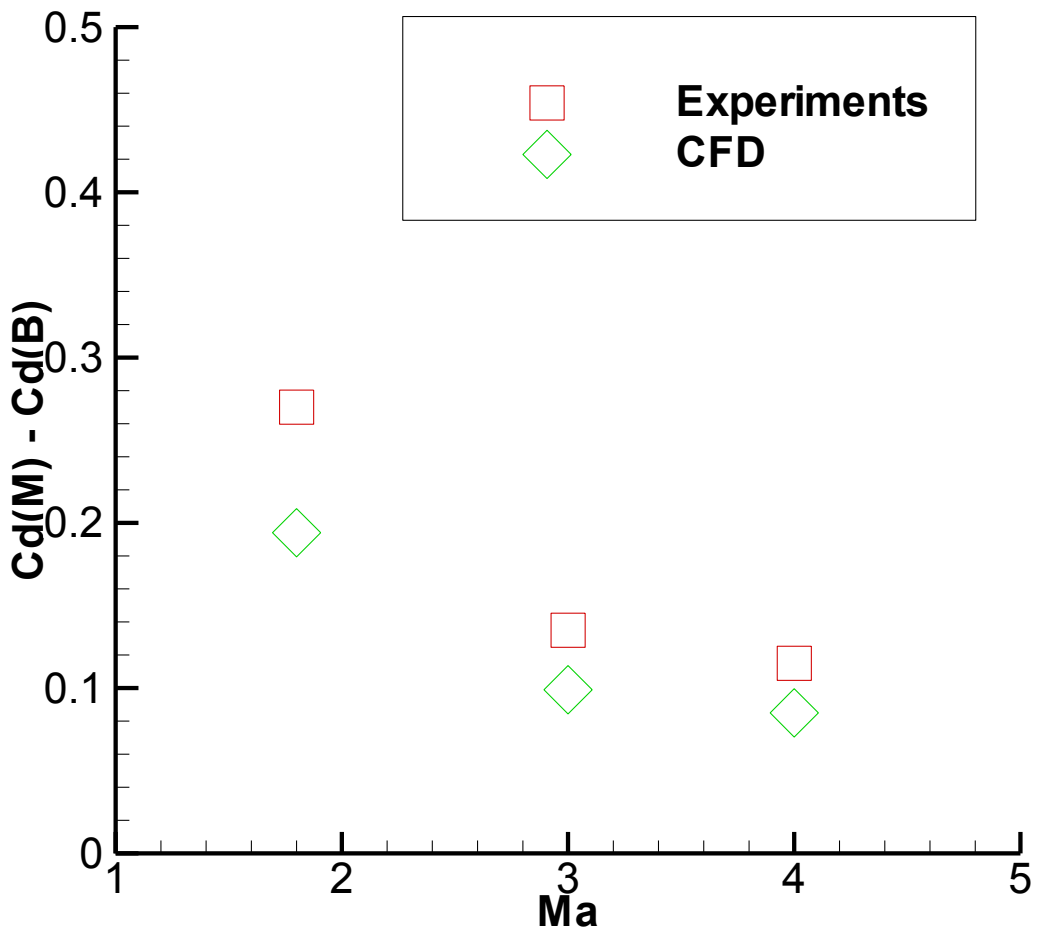

**Figure 28 : Différences entre les traînées du missile et du corps seul obtenues numériquement et expérimentalement pour les nombres de Mach 1.8, 3 et 4.**

### 2.5.2 Analyse des résultats

Le missile complet et le corps seul ont été calculés sans angle d’attaque pour la plage de nombres de Mach du Tableau 3. Pour évaluer l’efficacité du disque d’action, la différence entre les traînées du missile et celle du corps seul, obtenue numériquement a été comparée à celle issue de l’expérience [Esch, 2000]. Les résultats de cette comparaison sont représentés sur la Figure 28. L’intérêt de cette comparaison est d’isoler la différence induite par la présence des ailes en nid d’abeilles et de s’affranchir de sources potentielles d’erreur telle que celle entraînée par la base du missile. Celui-ci est maintenue par un support durant l’expérience, ce qui n’est pas pris en compte par les calculs. La Figure 28 montre que les prédictions sont capables de recouvrir plus de 75% de l'écart de traînée due aux ailes. Cependant, elles sous-estiment cet écart et ce quel que soit le nombre de Mach. Ceci est du aux interactions entre les ailes en nid d'abeilles et le missile [J3]. Ces interactions ont lieu d'une part entre les sillages des ailes et celui du missile avec une forte incidence sur ce dernier. D'autre part, au niveau des ailes la proximité du corps du missile entraîne des variations locales du nombre de Mach mais aussi des angles d'attaque et de roulis. Ces variations sont faibles mais elles sont susceptibles d'entraîner des erreurs d'interpolation dues aux limitations de la base de données. Ces différents points expliquent l'écart observé entre le calcul et l'expérience. Une base de données plus complète, ou une technique plus avancée, devrait clarifier ce point et améliorer les capacités du disque d'action. Toutefois, ce dernier est capable de recouvrir la majeure partie de l'écart entre les traînées mesuré expérimentalement.

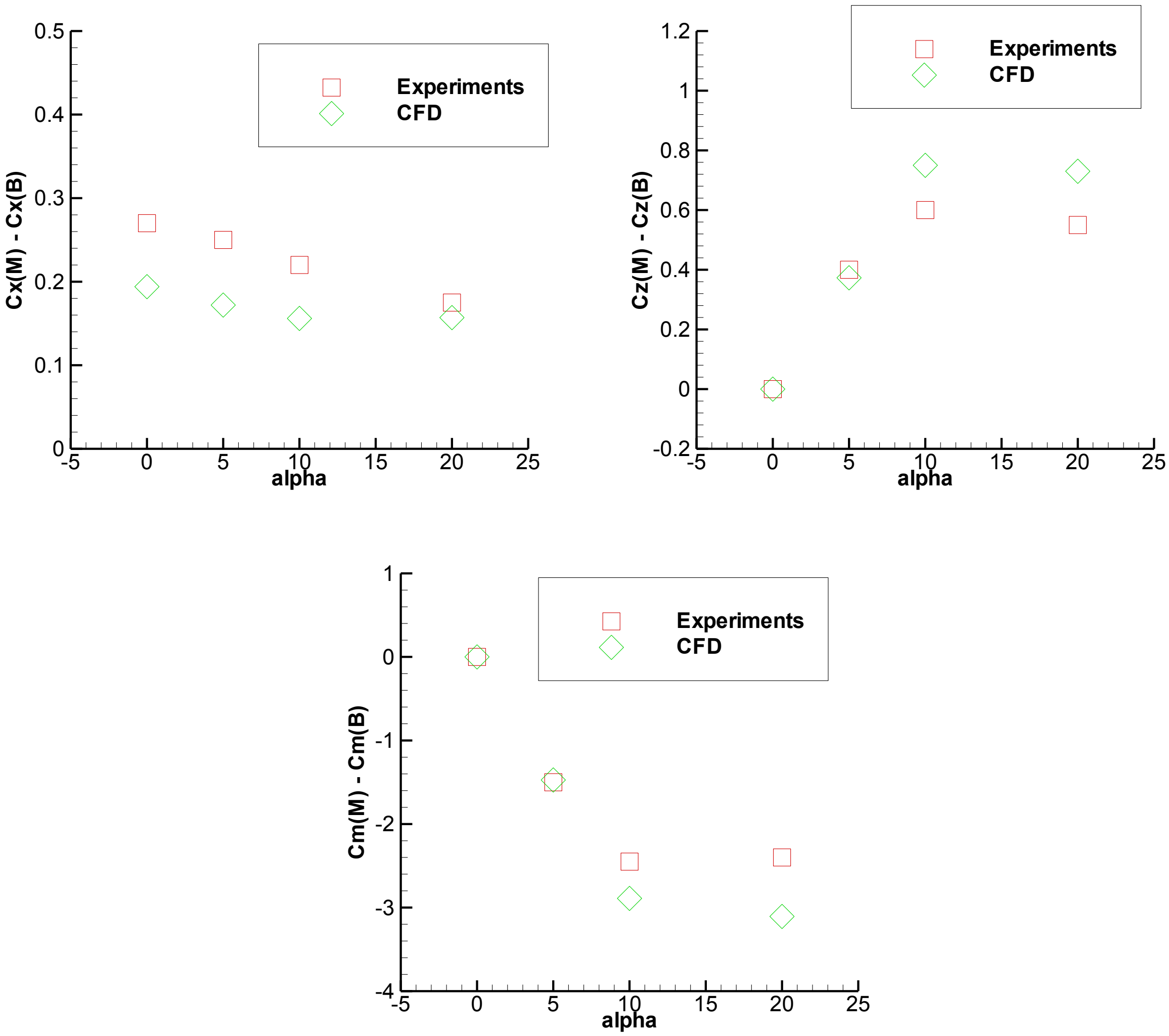

**Figure 29 : Evolutions en fonction de l'angle d'attaque de l'écart entre les coefficients de force axiale, Cx, de force normale, Cz et de moment de roulis du missile et du corps seul à Mach 1.8.**

La seconde phase a été d'évaluer les capacités du disque d'action, couplé à la base de données, en présence d'un angle d'attaque. À cause des limitations de la base de données pour les nombres de Mach élevés, les calculs avec angle d'attaque ont été effectués à Mach 1.8. Les écarts obtenus numériquement pour les coefficients de forces normales et axiales et pour le moment de tangage et des angles de 5, 10 et 20 degrés sont comparés aux résultats expérimentaux [Esch, 2000] sur la Figure 29. Un accord modéré avec l'expérience est observé pour la force axiale avec une sous-estimation de l'écart expérimental comme pour la prédiction de la traînée sans angle d'attaque (voir Figure 28). Pour la force normale et le moment de tangage l'accord entre le calcul et la mesure expérimentale est très bon pour 5º. Par contre la différence calcul/expérience est importante pour les angles d'attaque supérieurs.

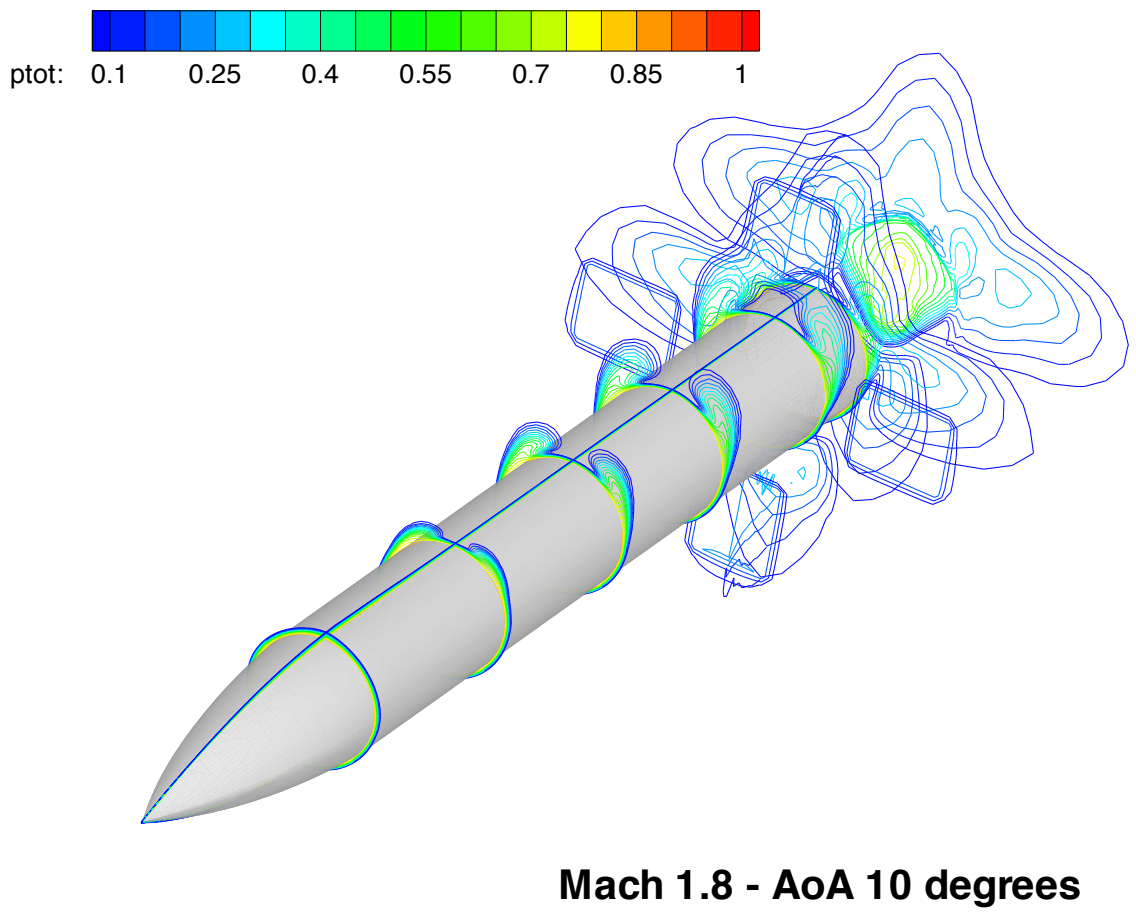

**Figure 30 : Distribution de pression totale le long du missile pour un Mach de 1.8 et 10º d'angle d'attaque.**

La différence mise en évidence pour le moment de tangage qui est calculée au point neutre du véhicule (position du centre de pression sans angle d'attaque) est due à sa sensibilité vis à vis de la force normale. Cette différence avec l'expérience pour les angles d'attaque de 10 et 20º provient de l'interaction entre les ailes et le corps du missile. L'expérience d'Esch [Esch, 2000] a mis en évidence la présence d'un décollement suivi par le développement de tourbillons à partir d'un angle d'attaque de 8º. L'analyse fine de l'écoulement pour 10º d'angle d'attaque et notamment la distribution de pression totale pour plusieurs plans transversaux, visible sur la Figure 30, a mis en évidence le développement d'un écoulement tourbillonnaire le long du corps du missile un impact sur une des ailes. La présence de ces tourbillons entraîne localement des conditions d'écoulement différentes. Le nombre de Mach en amont de l'aile est transsonique avec des angles locaux de tangage et de roulis différents du reste de l'écoulement. Or, la base de données est limitée à Mach 1.8 et n'est pas valide pour les régimes transsoniques. Par conséquent, elle n'est pas adaptée dans le cas d'un décollement ce qui conduit à des valeurs erronées des coefficients aérodynamiques interpolés. Le développement d'une base de données plus complète ou d'une méthode plus fiable pour la prédiction des coefficients aérodynamiques des ailes en nid d'abeilles est nécessaire pour améliorer les performances du disque d'action pour les écoulements avec un angle d'attaque important. Néanmoins les simulations numériques ont établi le potentiel de l'approche pour prédire les performances aérodynamiques globales d'un véhicule muni de ce type d'aile et ce pour un coût de calcul seulement 50 % supérieur à celui nécessaire pour le calcul d'un corps seul. Il est à noter que sans le disque d'action ce coût serait cinq à six fois plus élevé.

### 2.5.3 Couplage avec le module semi empirique

Le disque d'action a été intégré dans le code TAU capable d'utiliser des maillages non structurés. Dans le but d'améliorer les performances du disque d'action, celui-ci a été couplé à un module numérique basé sur la théorie semi empirique des ailes en nid d'abeilles [Belotserkovsky, 1987]. En utilisant cette théorie, les coefficients de force de l'aile en nid d'abeilles ne sont plus interpolés dans une base de données mais calculés à l'aide de relations semi empiriques. Les résultats expérimentaux obtenus au DLR [Esch, 1997, 1999, 2000] lors de campagnes d'essais ont été utilisés pour améliorer et modifier si nécessaire la théorie semi empirique.

Le fonctionnement du module est décrit brièvement ci-après, plus de détails sont disponibles dans [J4]. Ce module a été validé à l'aide de mesures expérimentales de Mach 0.3 a 6 et ce jusqu'à des angles d'attaque de 90º. Les coefficients aérodynamiques de l'aile sont calculées avec le module en fonction des caractéristiques locales de l'écoulement (nombre de Mach, angles locaux d'attaque et de roulis) et de la géométrie de l'aile. Différentes formules semi empiriques sont utilisées pour le calcul des coefficients aérodynamiques, elles sont fonction du régime de l'écoulement. Quatre régimes d'écoulement ont été identifiés, séparés par trois nombres de Mach critiques, $Ma_{cr1}$, $Ma_{cr2}$ et $Ma_{cr3}$. Le régime subsonique est limité par $Ma_{cr1}$ ($Ma \leq Ma_{cr1}$), qui correspond au nombre de Mach pour lequel un écoulement sonique est atteint dans une des cellules de l'aile. Le régime transsonique débute à $Ma_{cr1}$ et se poursuit jusqu'à $Ma_{cr2}$ pour lequel l'écoulement devient entièrement supersonique dans la cellule de calcul de l'aile. Au régime transsonique correspond une restructuration de l'écoulement avec la formation de chocs forts et de zones localement supersoniques. Le régime supersonique commence à partir de $Ma_{cr2}$ pour lequel l'onde de choc est attachée au bord d'attaque de l'aile. Au delà un dernier nombre de Mach critique, $Ma_{cr3}$, est atteint pour lequel les ondes de choc et d'expansion n'ont plus d'impact sur les plans de la cellule.

## 2.5.4 Résultats numériques

L'outil numérique obtenu a été testé en premier lieu sur une aile isolée et ce pour une large plage de nombres de Mach et d'angles attaque [O2]. Les calculs ont mis en évidence un très bon accord avec l'expérience. Cet accord est quasiment parfait pour la prédiction de la force normale, la force axiale est cependant légèrement sous estimée. Cela est dû au fait que la technique a été développée pour des ailes en nid d'abeilles avec une géométrie uniforme. Ici l'aile représentée sur la Figure 24 présente une géométrie non uniforme au niveau des cellules situées de part et d'autre du support de l'aile. Ces cellules plus petites induisent une force axiale un peu plus importante qu'elle ne le serait pour une aile uniforme. Ce point explique la sous estimation de la traînée qui reste cependant faible car inférieure à 8%.

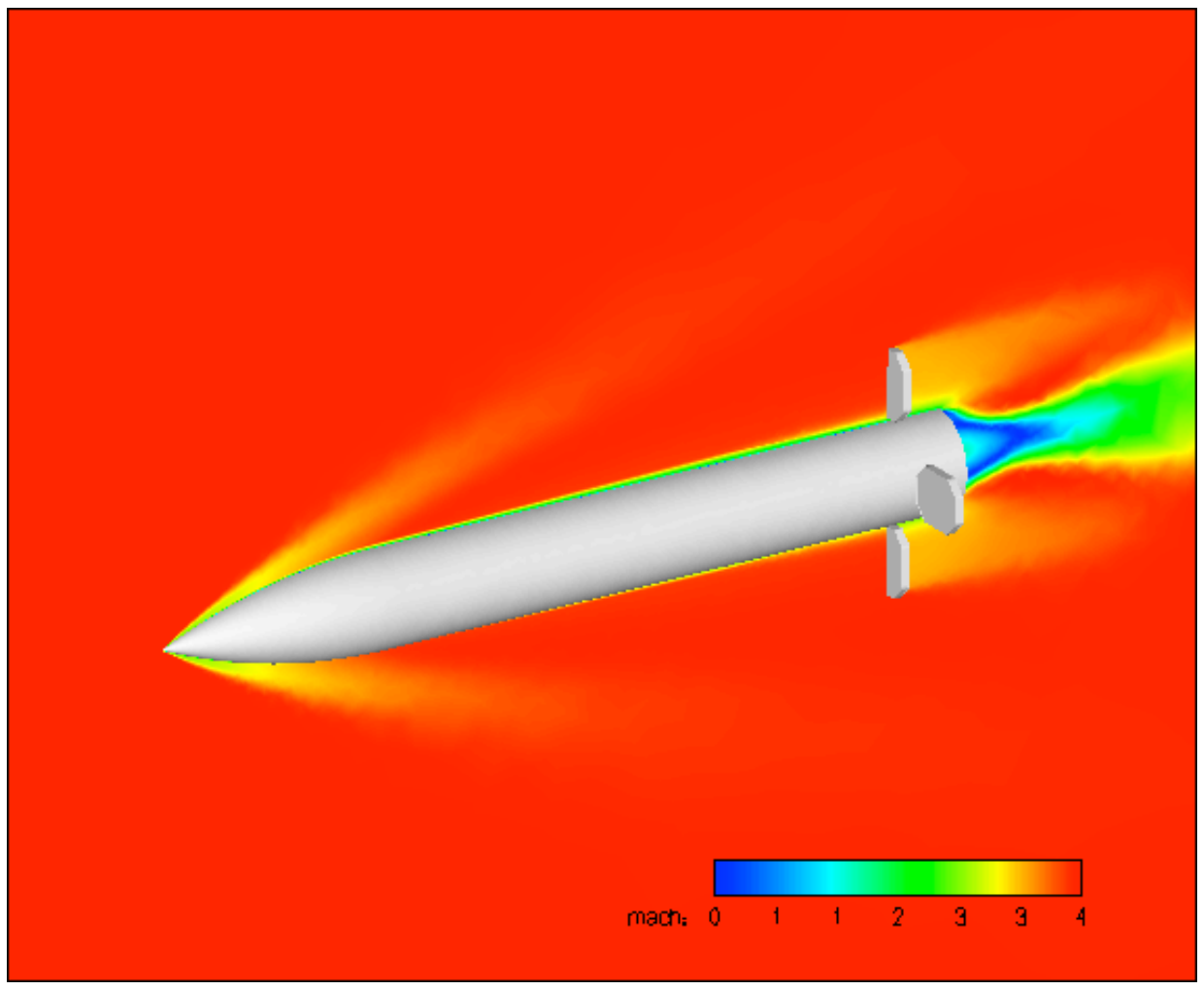

**Figure 31 : Distribution du nombre de Mach dans le plan de symétrie du missile à Mach 4.**

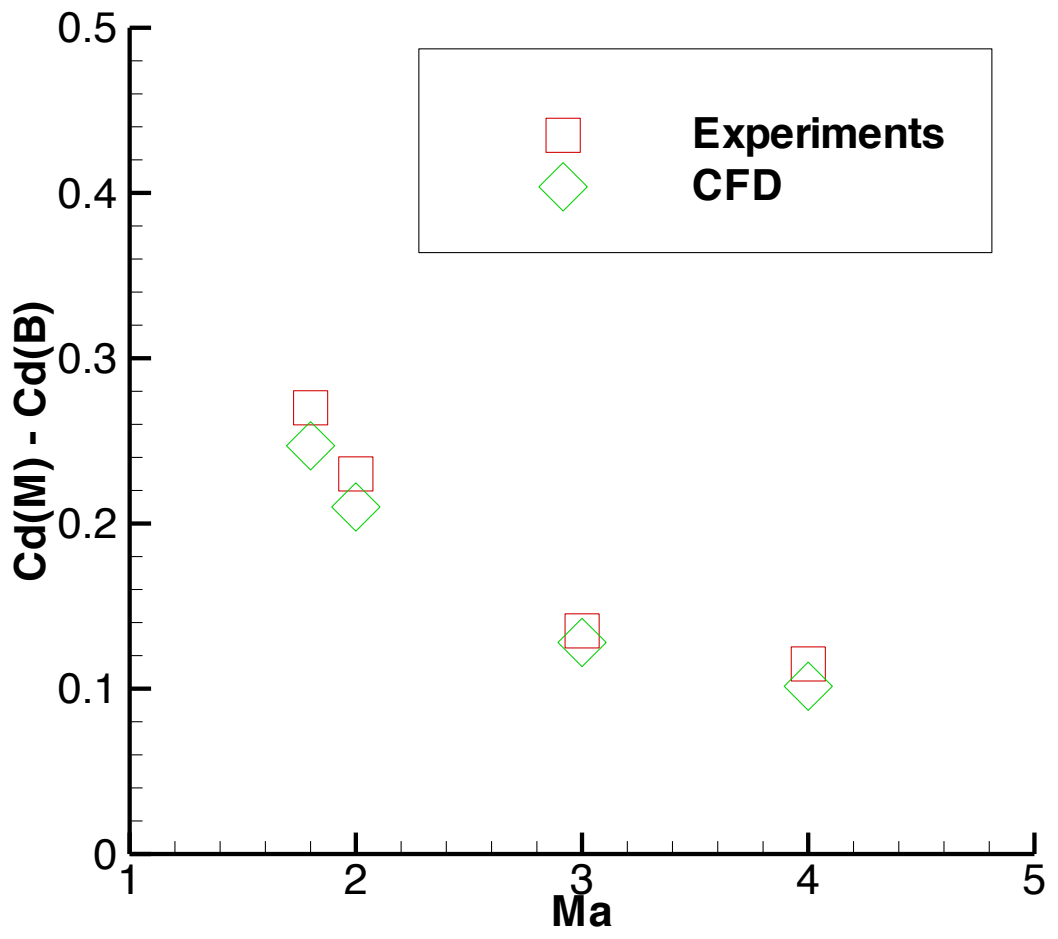

**Figure 32 : Différences entre les traînées du missile et du corps seul obtenues numériquement et expérimentalement avec TAU couplé au module semi empirique pour les nombres de Mach 1.8, 2, 3 et 4.**

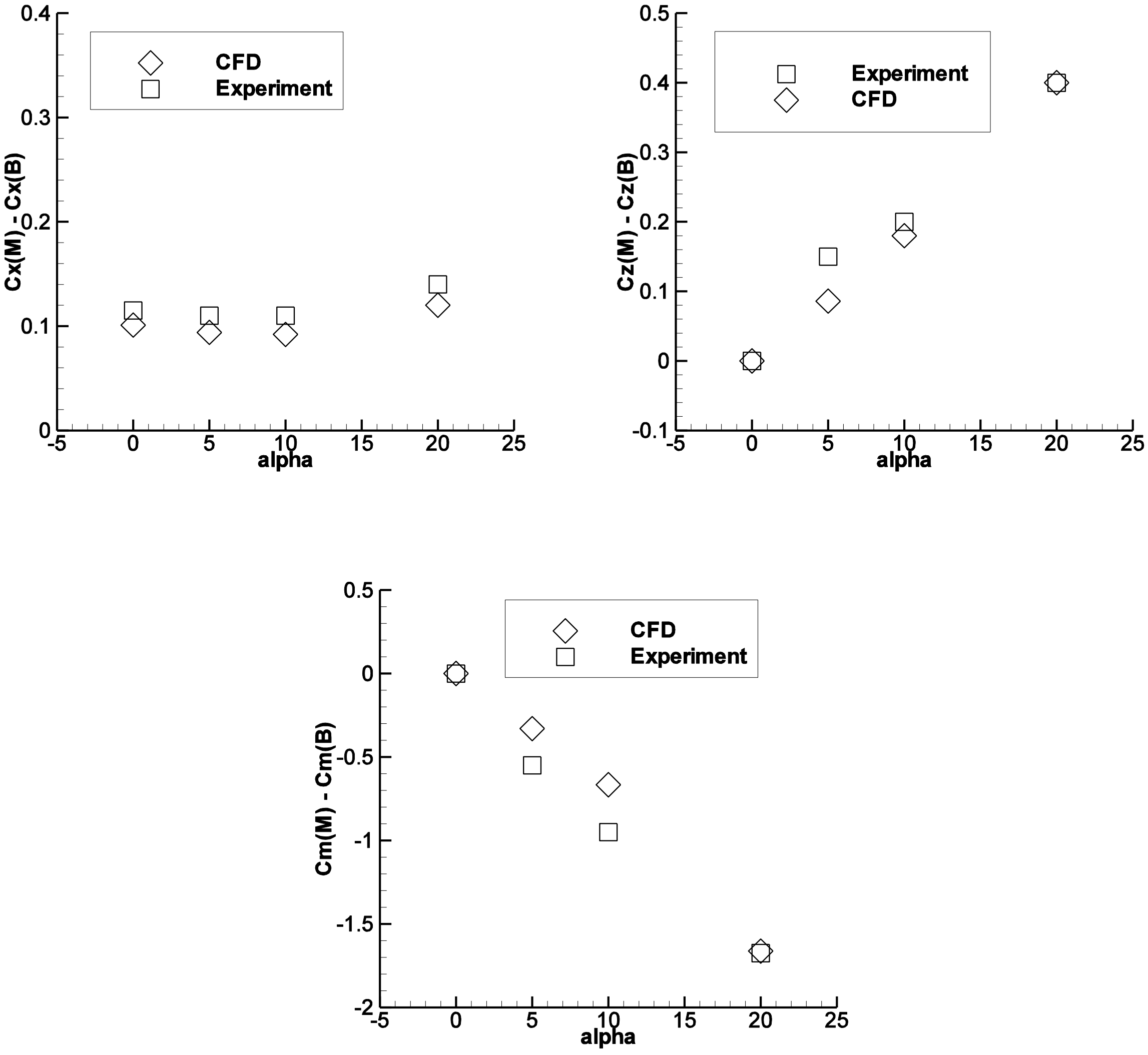

**Figure 33 : Evolutions en fonction de l'angle d'attaque de l'écart entre les coefficients de force axiale, Cx, de force normale, Cz, et de moment de roulis du missile, Cm, et du corps seul à Mach 4.**

Comme précédemment le missile complet et le corps seul ont été calculés sans angle d'attaque pour des nombres de Mach de 1.8 à 4. La distribution de nombre de Mach autour du missile à Mach 4 est représentée sur la Figure 31. Les résultats ont été comparés aux mesures expérimentales d'Esch [Esch, 2000]. Les différences de traînées entre le missile et le corps seul sont reportées sur la Figure 32. Cette figure met en évidence un bon accord entre le calcul et l'expérience quant à la prédiction de l'effet des ailes en nid d'abeille sur la traînée : au moins 90% de l'écart est prédit par les simulations. Cet effet est légèrement sous-estimé par le calcul mais ceci est induit, comme pour les calculs effectués avec une aile isolée, par la géométrie non uniforme de l'aile. La comparaison de la Figure 28 avec la Figure 32 met en évidence les progrès effectués pour la prédiction de la traînée par l'utilisation du module semi empirique par rapport à la base de données utilisée initialement.

L'extension des capacités de l'outil numérique a permis de simuler le missile complet et le corps seul en présence d'un angle d'attaque pour un nombre de Mach de 4. La Figure 33 montre les variations des coefficients aérodynamiques obtenus numériquement et expérimentalement. Un bon accord est observé entre le calcul et l'expérience que cela soit pour le coefficient de force axiale, normale ou pour celui relatif au moment de tangage. La comparaison avec la Figure 29, même si les calculs précédents avaient été menés à Mach 1.8, illustre les progrès effectués pour les calculs avec angle d'attaque. Ici l'outil numérique permet finalement de reproduire l'évolution des coefficients aérodynamiques et ce avec une marge de 10 % par rapport aux résultats expérimentaux. Ce, pour un coût de calcul inférieur d'un facteur

# 2.6 Conclusions

L'objectif de ce travail était de développer un outil numérique adapté au design de missiles supersoniques équipés d'ailes en nid d'abeilles. Les prédictions de coefficients aérodynamiques mettent en évidence les progrès au niveau de l'extension des capacités du code de calcul effectués quand celui-ci est couplé à un module semi empirique au lieu d'une base de données expérimentales. Ceci est valable pour une large plage de nombres de Mach et d'angles d'attaque. L'ensemble des calculs faits avec le module basé sur la théorie des ailes en nid d'abeilles montre que les coefficients aérodynamiques prédits restent dans une marge d'erreur inférieure à 10% par rapport aux valeurs expérimentales. L'utilisation du module permet de s'affranchir des limitations de la base de données et de prendre mieux en compte les effets dus à l'interaction entre les ailes et le missile.

Le second point est l'avantage de l'approche développée, en terme de coût de calcul, quand elle est comparée à un calcul complet. Par rapport à un calcul complet, l'effort numérique est divisé d'un facteur six. Un autre avantage de la méthode est qu'aucun effort additionnel de génération de maillage n'est nécessaire pour tester une aile de structure interne différente tant que la configuration reste inchangée (en termes de nombre d'ailes, de taille et de position). De plus la nouvelle solution peut être obtenue pour un faible coût de calcul. Comparé à une simulation sans disque d'action, cela représente un gain de plusieurs semaines pour l'analyse de l'impact d'ailes en nid d'abeilles sur les performances d'un missile. Ceci démontre l'intérêt de l'approche et son utilité pour le design de véhicules équipés d'ailes en nid d'abeilles.

Ces travaux réalisés à l'Institut d'Aérodynamique et de Technologie des Ecoulement du DLR de Brunswick ont fait l'objet des communications [J3-4], [O2], [C9-10], [A5,A7] et [R3-4] référées en Annexes.

# 3 Complexité numérique : Études liées à la prédiction des débris

## 3.1 Introduction

Depuis quelques années de plus en plus d'études portent sur les débris engendrés par les activités spatiales. La plupart des études rapportées dans la littérature relatives à ce sujet concerne soit les lanceurs tels que Saturn [Neher, 1971], Ariane 5 [Marraffa, 1996, Mazoué, 1996] ou la Space Shuttle [Kofski, 1992 ; Pike, 1990] soit la destruction lors de leur rentrée atmosphérique de véhicules en fin de vie [Koppenwallner, 2004]  Les travaux relatifs aux débris sont menés dans deux directions différentes, pour d'une part prédire l'impact au sol de la retombée d'astronefs et pour d'autre part évaluer la production de débris stables en orbite polluant les orbites de satellites en exploitation. Le premier point concerne directement la sécurité des populations au sol. Le second aspect de la production de débris prend de plus en plus d'ampleur puisque le satellite Cerise a vu son antenne endommagée par l'impact d'un débris de faible taille. Les activités spatiales sont et dorénavant contrôlées par des normes internationales et les études pour la prédiction des débris de plus en plus courantes.

Dans ce chapitre, nous présenterons deux études, effectuées à l'ESTEC, ayant trait à la prédiction de débris. La première concerne la prédiction de l'explosion de l'ATV ("Automated Transfer Vehicle") qui avait pour objectif de raffiner l'approche basée sur des modèles d'ingénieur utilisée dans le cadre du projet. Son originalité est de coupler des calculs numériques à une analyse de l'explosion pour voir si une explosion prématurée est possible en cas de fuite de propergols et de fissure de la structure du véhicule. La seconde étude concerne la prédiction de changement de phase et de possibilité de blocage d'une conduite lors de la passivation des réservoirs d'hydrazine du SCA (Système de Contrôle d'Attitude) d'Ariane 5 à la fin du lancement. Pour cette étude les parties correspondant à la modélisation du comportement thermodynamique de l'hydrazine ainsi que le développement de l'approche numérique ont été menées à l'ESTEC tandis que les mesures expérimentales ont été effectuées à l'ONERA et à EADS-Brême.

Sur le plan numérique, les phénomènes à étudier sont complexes. Dans le premier cas il s'agit de coupler des calculs CFD à une analyse d'explosion. Comme la rentrée d'un astronef est un processus dynamique, il a fallut vérifier la validité de l'approche utilisée qui repose sur des calculs numériques stationnaires. L'aspect dynamique de la rentrée n'est pas pris en compte et elle est considérée comme une succession d'états stationnaires. D'autre part la simulation de l'écoulement autour et à l'intérieur de l'ATV dans le cas d'une perforation exige du solveur numérique de pouvoir simuler un écoulement hypersonique avec un nombre de Mach supérieur à 20 tout en demeurant capable de simuler le remplissage (dans la phase non convergée du calcul) puis l'écoulement à l'intérieur du véhicule qui est à faible nombre de Mach. Dans la deuxième application, il s'agit de la passivation du SCA (Système de Contrôle d'Attitude) d'Ariane 5 et donc de prédire la dépressurisation d'hydrazine liquide le long d'une conduite avec une pression finale proche de celle du vide. L'écoulement est à très bas nombre de Mach, inférieur à 0.01, et avant son éjection l'hydrazine atteint sa pression de vapeur saturante et commence à se gazéifier. Il s'agit donc pour le solveur de prédire le changement de phase d'un liquide compressible à un nombre de Mach très bas. Le principal obstacle est que la plupart des schémas numériques sont basés sur la densité et deviennent instables à bas nombre de Mach [Wesseling, 2001]. Ici le problème a été résolu en utilisant un schéma basé sur la pression [Bijl, 1996, 1998) qui a pour avantage d'être uniformément stable et ce quel que soit le nombre de Mach.

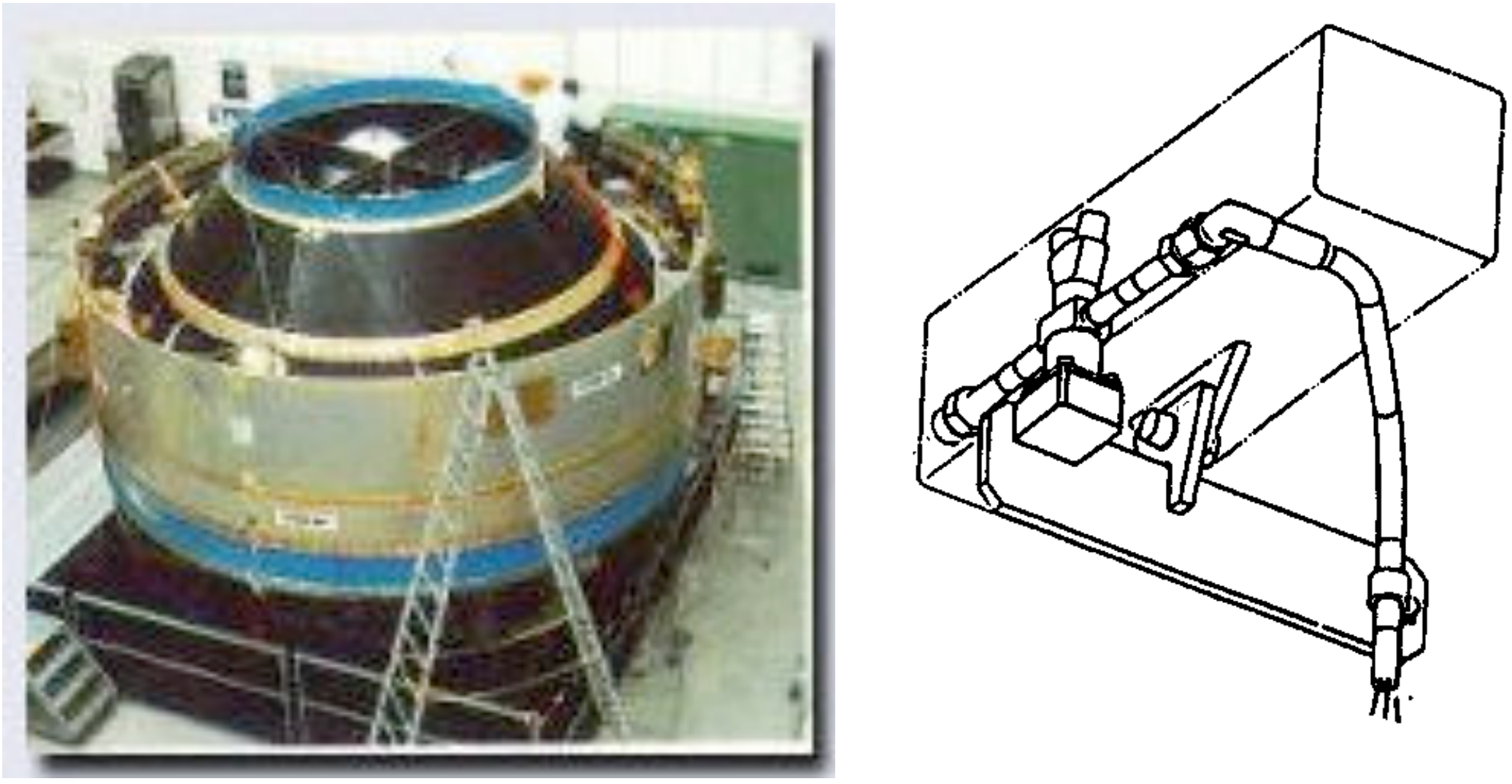

**Figure 34 : Á gauche, case d'équipements d'Ariane 5. Á droite, fin de la ligne servant à la passivation des réservoirs d'hydrazine.**

# 3.2 Passivation d'un réservoir d'hydrazine

## 3.2.1 Problème physique et modélisation

En fin du tir 503 d'Ariane 5, dans le but d'éviter une potentielle explosion due au rayonnement solaire, il a été décidé de procéder à la passivation des réservoirs d'hydrazine du SCA d'Ariane 5 représenté sur la Figure 34. La passivation de propergols stockables a déjà été étudié dans le cadre du projet Ariane 5 avec notamment la prédiction de condensation dans les tuyères [Marraffa, 1996, Mazoué, 1996]. Ici, il s'agit de passiver des réservoirs d'hydrazine liquide dans l'espace. Le but de l'étude est de valider le processus de passivation à travers une conduite et une grille et notamment de préciser le risque d'obstruction de la grille. Risque qui avait d'ailleurs été mis en évidence par Schwartzing [Schwartzing, 1998] pour des pressions d'éjection basses et en utilisant de l'eau au lieu d'hydrazine, les deux fluides ayant des propriétés thermodynamiques proches.

Sur le plan expérimental, un montage a été créé à l'ONERA [Foucaud, 1998] pour étudier le comportement de l'hydrazine liquide durant la passivation d'un réservoir. Le montage expérimental reproduisait les mêmes pertes de charge que celui de la case d'équipement avec une fin de ligne exactement identique. Le principal but de la campagne expérimentale était de préciser le mécanisme de flashing [Zeigerson-Kartz, 1998] en sortie et de mesurer le diamètre des gouttes engendrées par le processus de passivation. Le mécanisme de flashing d'un jet liquide dans le vide est un phénomène mal connu. Il correspond au breakup d'un jet liquide qui est en train de s'auto-vaporiser sous l'action d'un pression très basse [Fuchs, 1979 ; Ishimoto, 2000]. La plupart des études consacrées à ce phénomène l'ont été pour des jets d'eau liquide dans l'espace [Gayle, 1964 ; Mikatarian, 1966 ; Pike, 1990, Fuchs, 1979 ; Kofski, 1992 et Muntz, 1987]. Il est à noter que les techniques de mesures sont peu performantes pour les liquides, mais ces dernières sont encore plus médiocres pour le cas de l'injection cryogénique [Ingebo, 1992]. Cela est dû au fait que pour les fluides cryogéniques les températures de surface sont souvent proches du point d'ébullition du liquide [Ingebo, 1992].

D'habitude, dans les études numériques, notamment pour les problèmes de cavitation [Van der Heul, 1998 ; Reboud, 1994 ; Schnerr, 1995 ; Kubota, 1992 ; Lemonnier, 1988 ; Merkle, 1998] le liquide est supposé être parfait est donc incompressible. Le liquide est supposé changer de phase brutalement avec un changement de phase modéliser à travers un polynôme [Van der Heul, 1998]. Cependant, dans certaines études récentes la compressibilité du liquide commence à être prise en compte [Saurel, 1999]. Toutefois, l'ensemble des travaux cités ci-dessus ont été effectués pour de l'eau. Il n'a pas été trouvé dans la littérature de travaux expérimentaux ou numériques menés pour de l'hydrazine éjectée dans du vide. Ici l'originalité du travail est de simuler la dépressurisation puis le changement de phase de l'hydrazine en prenant en compte la compressibilité du liquide. Pour cela un modèle thermodynamique pour l'hydrazine a été développé à l'ESTEC [Giordano, 2001, 2002], validé à partir des données expérimentales accessibles dans la littérature. Sur le plan numérique ce modèle nécessite quelques précautions de par sa nature non linéaire et du fait qu'il est très raide au niveau du changement de phase.

### 3.2.2 Complexité numérique

Les écoulements à prédire ont pour principale caractéristique un nombre de Mach très bas, inférieur à 0.03. Aussi la méthode numérique doit être adaptée en premier lieu aux bas nombres de Mach. Cela restreint les schémas numériques qui peuvent être candidats, en effet la plupart des méthodes standard pour les écoulements compressibles ne peuvent être utilisée. Elles sont basée sur la densité et deviennent instables pour les bas nombres de Mach [Wesseling, 2001]. Plusieurs alternatives sont possibles :

- L'adaptation des méthodes compressibles pour les bas nombres de Mach par l'utilisation d'un préconditionnement [Choi, 1993, Edwards, 1998 ; Merkle, 1998]. Cependant, la limite d'application du schéma n'est que repoussée et la précision temporelle perdue.
- Les méthodes asymptotiques basées sur l'expansion en série par rapport au nombre de Mach. Pour assurer la précision, l'écoulement doit posséder un nombre de Mach uniforme.
- Les schémas incompressibles étendus au domaine compressible. Ainsi, le schéma incompressible d'Harlow et Welch [Harlow, 1965] a été adapté aux écoulements compressibles par Harlow et Amsden [Harlow, 1971]. Bijl et Wesseling [Bijl, 1996, 1998], parmi d'autres, ont développé cette méthode plus avant. Le résultat est un schéma unifié pour les écoulements instationnaires compressibles et incompressibles avec une précision et une efficacité uniforme par rapport au nombre de Mach.

C'est cette dernière alternative qui a été sélectionnée ici, elle permet en outre l'utilisation d'équations d'état non convexes [Van der Heul, 1998] telle que celle proposée par Giordano et De Serio [Giordano, 2002]. Il est à noter qu'avec le schéma retenu, la méthode incompressible d'Harlow et Welch [Harlow, 1967] est recouvrée quand le nombre de Mach tend vers zéro.

### 3.2.3 Schéma numérique

Dans un premier temps la distribution de pression le long de la ligne de décharge a été étudiée à l'aide d'une approche ingénieur basée sur les pertes de charge le long d'une conduite. Cette approche a été validée à l'aide de résultats expérimentaux obtenus à ASTRIUM pour de l'eau. La comparaison entre les débits calculés et ceux mesurés est reportée sur la Figure 35. Dans un deuxième temps la même approche a été utilisée pour l'expérience de l'ONERA et de l'hydrazine. Ces premiers résultats ont mis en évidence un risque de changement de phase de l'hydrazine dans la partie finale de la conduite. La partie finale est constituée d'une tuyauterie droite finie par une grille percée de trois trous, elle est représentée sur la Figure 34. Pour l'étude numérique, dans un souci de minimiser la complexité et les coûts de calcul de l'étude, la grille, percée de trois trous, est remplacée par un seul orifice d'éjection de diamètre équivalent (voir Figure 36). Cela a permis l'utilisation d'une approche quasi monodimensionnelle dans la première partie de l'étude. Un schéma numérique a alors été développé pour la simulation du comportement thermodynamique de l'hydrazine liquide, à l'aide du modèle développé par Giordano et De Serio (2002), avant son éjection dans l'espace. Ce modèle dons les équations sont rappelées dans (J2) et (C7) permet de décrire les phases liquide et gazeuse de l'hydrazine, changement de phase et compressibilité du liquide inclus.

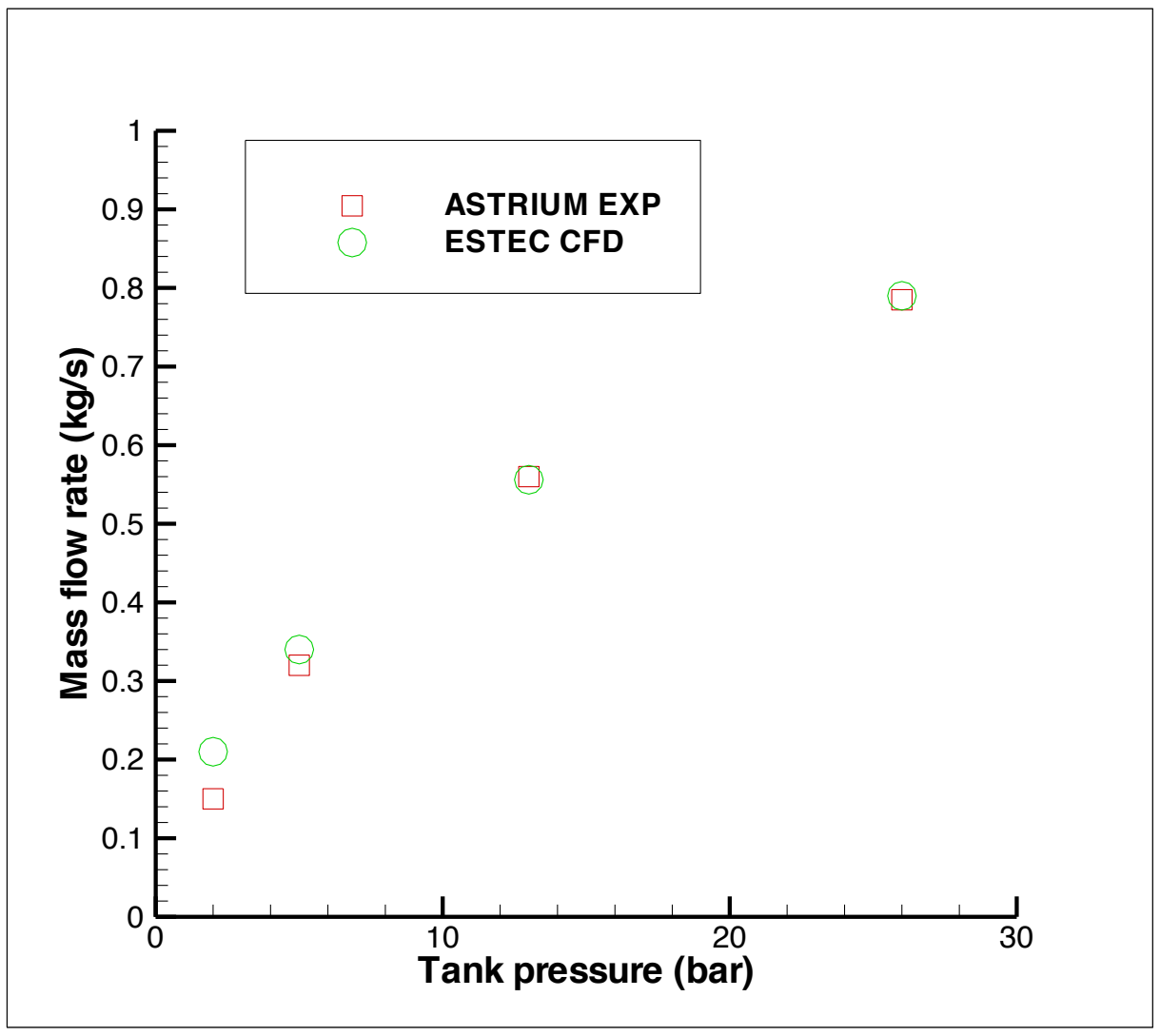

**Figure 35 : Validation de l'approche ingénieur par rapport aux mesures effectuées à ASTRIUM-Brême pour le débit dans la conduite.**

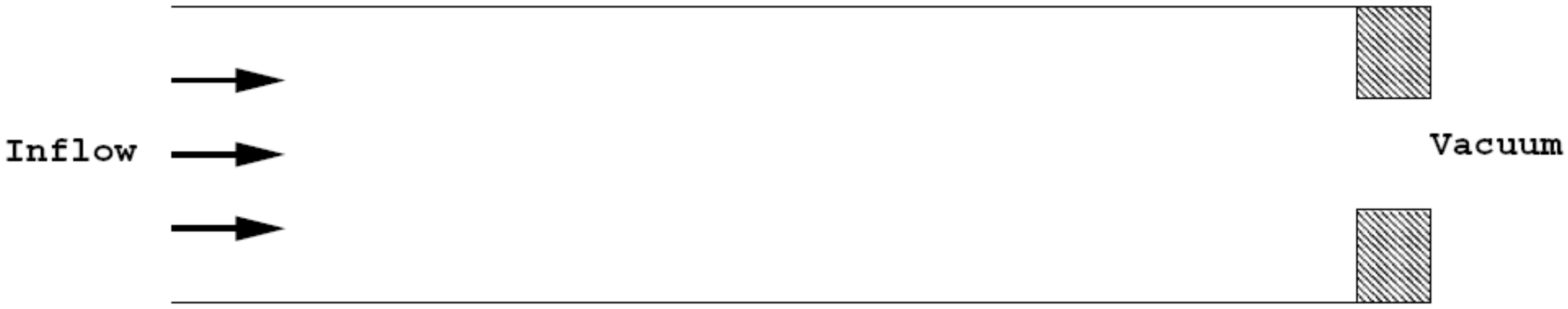

**Figure 36 : Modèle de la partie finale de la ligne de décharge utilisée dans le SCA.**

Le système d'équations à résoudre comprend outre les deux équations d'état du modèle thermodynamique, celles de conservation de la masse, de la quantité de mouvement et de l'énergie. Le schéma numérique utilise un maillage décalé, la quantité de mouvement et la vitesse sont calculées entre les nœuds, les autres variables aux nœuds. La discrétisation des équations quasi monodimensionnelles est la suivante :

$$\rho_j^{n+1} - \rho_j^n + \frac{\lambda_j}{A_j}(Am^{n+1})\Big|_{j-1/2}^{j+1/2} = 0 \qquad (3.1)$$

$$m_{j+1/2}^{n+1} - m_{j+1/2}^n + \frac{\lambda_{j+1/2}}{A_{j+1/2}}(Am^n U^n)\Big|_j^{j+1} + \lambda_{j+1/2} P^{n+1}\Big|_j^{j+1} + \Delta t\left(\frac{2C_f}{D_{j+1/2}}(Um)_{j+1/2}^n - \rho_{j+1/2}^n g\right) = 0 \qquad (3.2)$$

$$e_j^{n+1} - e_j^n + \lambda_j U_j^n e^n\Big|_{j-1/2}^{j+1/2} + \frac{\lambda_j P_j^n}{A_j \rho_j^n}(AU^n)\Big|_{j-1/2}^{j+1/2} + \lambda_{j+1/2} P^{n+1}\Big|_j^{j+1} + - \Delta t\frac{2C_f}{D_j}\left(U_j^n\right)^3 = 0 \qquad (3.3)$$

Où $\rho$ est la densité, $A$ l'aire de la conduite, $m$ la quantité de mouvement, $\lambda$ le rapport entre le pas de temps $\Delta t$ et le pas d'espace $\Delta t$, $U$ la vitesse, $P$ la pression, $e$ l'énergie interne et $D$ le diamètre de la conduite. $Cf$ est le coefficient de frottement, calculé à partir de la corrélation de Jain [Jain, 1976] pour un écoulement monophasique. Pour les écoulements diphasiques $Cf$ est donné par :

$$Cf = Cf_{sp}\,\Phi^2 \qquad (3.4)$$

Où $Cf_{sp}$ est le coefficient de friction pour un écoulement liquide tandis que $\Phi$ est déduit de la relation de Lockart-Martinelli modifiée par Richardson [Richardson, 1958] :

$$\Phi^2 = (1-\alpha)^{-1.75} \qquad (3.5)$$

Où $\alpha$ est la fraction de vide.

Les conditions limites associées au système discrétisé ci-dessus sont décrites en (J2) et (C6). Le schéma obtenu est précis au premier ordre en espace et en temps. La résolution de ce système s'effectue avec les étapes suivantes :

- (1) $e^{n+1}$ est obtenu par la résolution de (3.3) ;
- (2) Pas de prédiction dans l'équation (3.2) : $m^*$ est obtenu en utilisant $P^n$ et non $P^{n+1}$ ;
- (3) Pas de correction-prédiction : En utilisant les équations (3.1) et (3.2) un système linéaire est obtenu pour $\delta P$. $P^{n+1}$ est calculée ;
- (4) $m^{n+1}$ est obtenu ;
- (5) $T^{n+1}$ est calculée à l'aide de la méthode de Newton-Raphson ;

- (6) $\rho^{n+1}$ est calculée à partir de $e^{n+1}$ et $T^{n+1}$ ;
- (7) $U^{n+1}$ peut alors être déterminée.

### 3.2.4 Effets diphasiques

Pour la prise en compte des effets diphasiques l'hypothèse d'une faible fraction de gaz a été assumée étant donné que la pression de vapeur saturante devrait être atteinte près de la sortie. La transition vers un écoulement de type *churn flow* a lieu lorsque la fraction gazeuse se situe entre 0.08 et 0.14 [Minemura, 1998]. Dans ce cas là les effets mécaniques, notamment de tension de surface, entre les deux phases peuvent être négligés. Cette hypothèse est vérifiée tant que la fraction gazeuse est inférieure à 5% [Bilicki, 1996, 1998]. L'hypothèse d'un écoulement à bulles semble a priori valable et les calculs diphasiques ont été effectués avec un modèle de mélange.

Parmi les modèles de type mélange, un des plus populaire est le modèle à relaxation homogène [Bilicki, 1996, 1998]. Dans ce modèle, les effets de vaporisation et de condensation sont supposés être significatifs. Le liquide est supposé être sursaturé tandis que la vapeur est à l'état d'équilibre à la température de saturation. Le terme de relaxation permet de prendre en compte le déséquilibre thermodynamique et donc le retard à la nucléation et à la croissance des bulles. Une étude détaillée [Veneau, 1995 ; Hervieu, 1996] de l'ensemble de ces modèles a montré sa supériorité pour la prédiction de la dépressurisation brutale de propane liquide. En conséquence ce modèle a été retenu pour les calculs diphasiques. Une équation de transport pour la fraction sèche $\chi$ est alors résolue :

$$\frac{\partial \chi}{\partial t} + U\frac{\partial \chi}{\partial x} + \frac{\chi - \overline{\chi}}{\theta_x} = 0 \qquad (3.6)$$

Où $\theta_x$ est le terme de relaxation et $\overline{\chi}$ la fraction sèche à l'équilibre. Au plan numérique le terme de relaxation de l'équation (3.6) est résolu de façon implicite. Il correspond à la transition du déséquilibre vers l'équilibre. Durant cette transition la croissance des bulles à lieu. La croissance initiale dépend fortement des interactions mécaniques interfaciales telles que l'accélération et les forces de pression et de tension de surface. Durant cette étape les phénomènes thermiques comme le transfert de chaleur et le changement de phase sont négligeables. Lorsque le diamètre de la bulle augmente sa croissance devient principalement dépendante de la chaleur transférée qui est utilisée pour vaporiser le liquide à sa surface. Par conséquent, une prédiction réaliste des écoulements de type *flashing* nécessite de prendre en compte le déséquilibre thermodynamique entre les phases. Ici, en l'absence de données expérimentales pour l'hydrazine permettant d'évaluer le terme de relaxation, la valeur de 2 valable pour l'eau [Bilicki, 1998] a été retenue. Comme l'eau et l'hydrazine ont des propriétés thermodynamiques proches ce choix semble a priori raisonnable.

### 3.2.5 Approche quasi 1D

A l'aide du schéma numérique décrit au §3.2.3 et du modèle thermodynamique pour l'hydrazine proposé par Giordano et De Serio [Giordano, 2002], la dépressurisation de l'hydrazine le long de la conduite a été simulée pour plusieurs pressions du réservoir (25, 12 et 6 bars). Les calculs ont d'abord été effectués en tenant compte de la seule décompression de l'hydrazine liquide et ce sans changement de phase. Dans un deuxième temps le modèle thermodynamique développé pour ce projet a été utilisé dans son ensemble couplé avec le modèle à relaxation homogène.

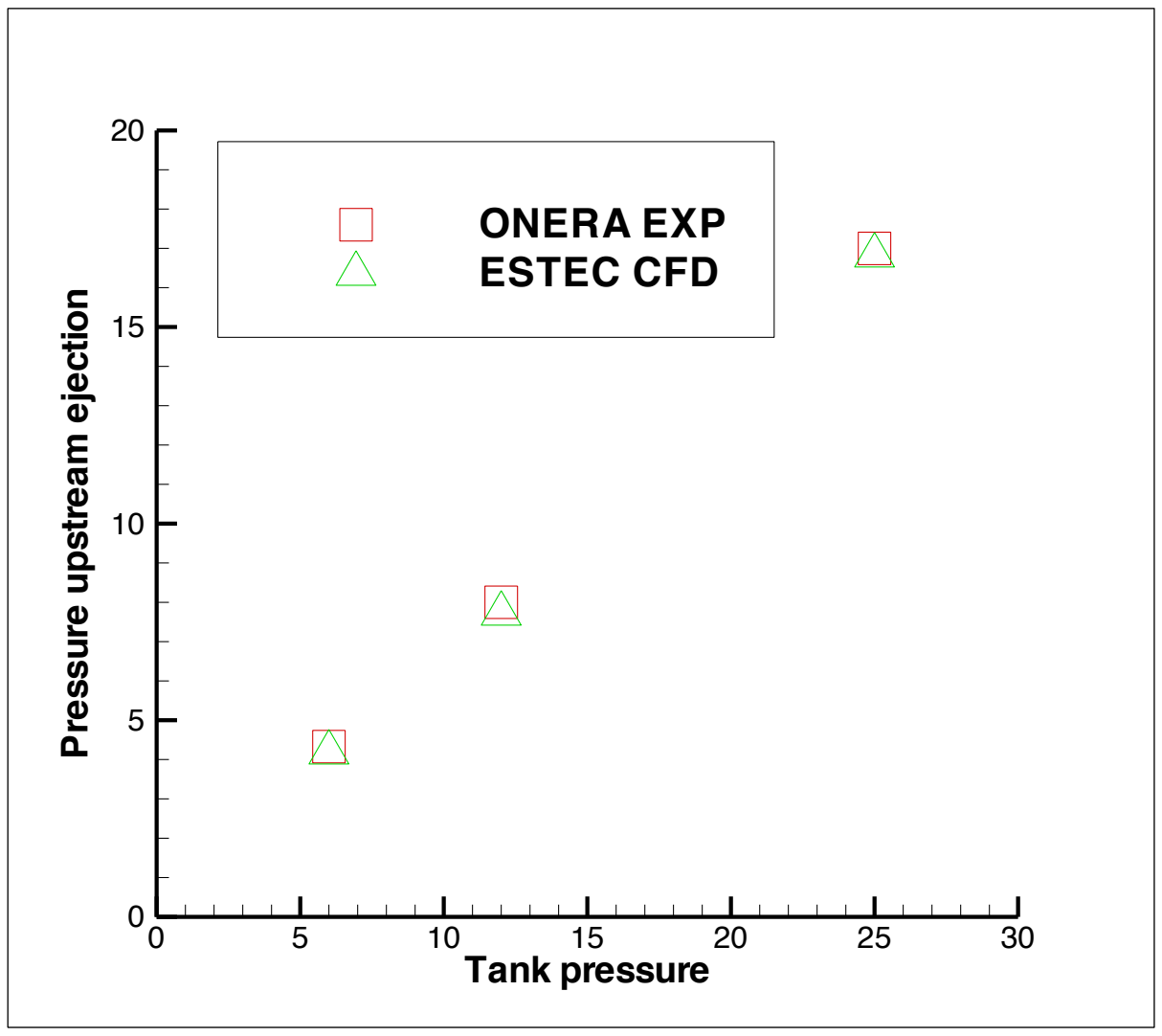

**Figure 37 : Pression calculées et mesurées en amont de l'éjection pour l'expérience de l'ONERA.**

Les calculs quasi 1D ont été validés avec les mesures expérimentales de l'ONERA effectuées pour une configuration très proche (la seule différence concerne la grille de sortie avec un seul orifice d'éjection ici au lieu de trois dans l'expérience). La comparaison entre les pressions calculées et mesurées en amont de l'éjection est montrée sur la Figure 37. Elle met en évidence un très bon accord entre le calcul et l'expérience et ce pour les trois différentes pressions du réservoir, sélectionnées pour la campagne expérimentale. Ces différents calculs ont permis de valider le code quasi 1D développé pour cette étude. Pour ce premier jeu de simulations, seule la décompression de la phase liquide a été prise en compte sans tenir compte d'un écoulement diphasique.

Pour les pressions du réservoir de 25 et 12 bars les prédictions numériques ont montré que la pression de vapeur saturante de l'hydrazine n'était pas atteinte le long de la conduite. Ce n'est pas le cas par contre pour une pression du réservoir de 6 bars. Pour ce cas, la distribution de pression le long de la partie finale de la conduite, représentée sur la Figure 38, montre que la pression de vapeur saturante est atteinte au niveau de la grille d'éjection.

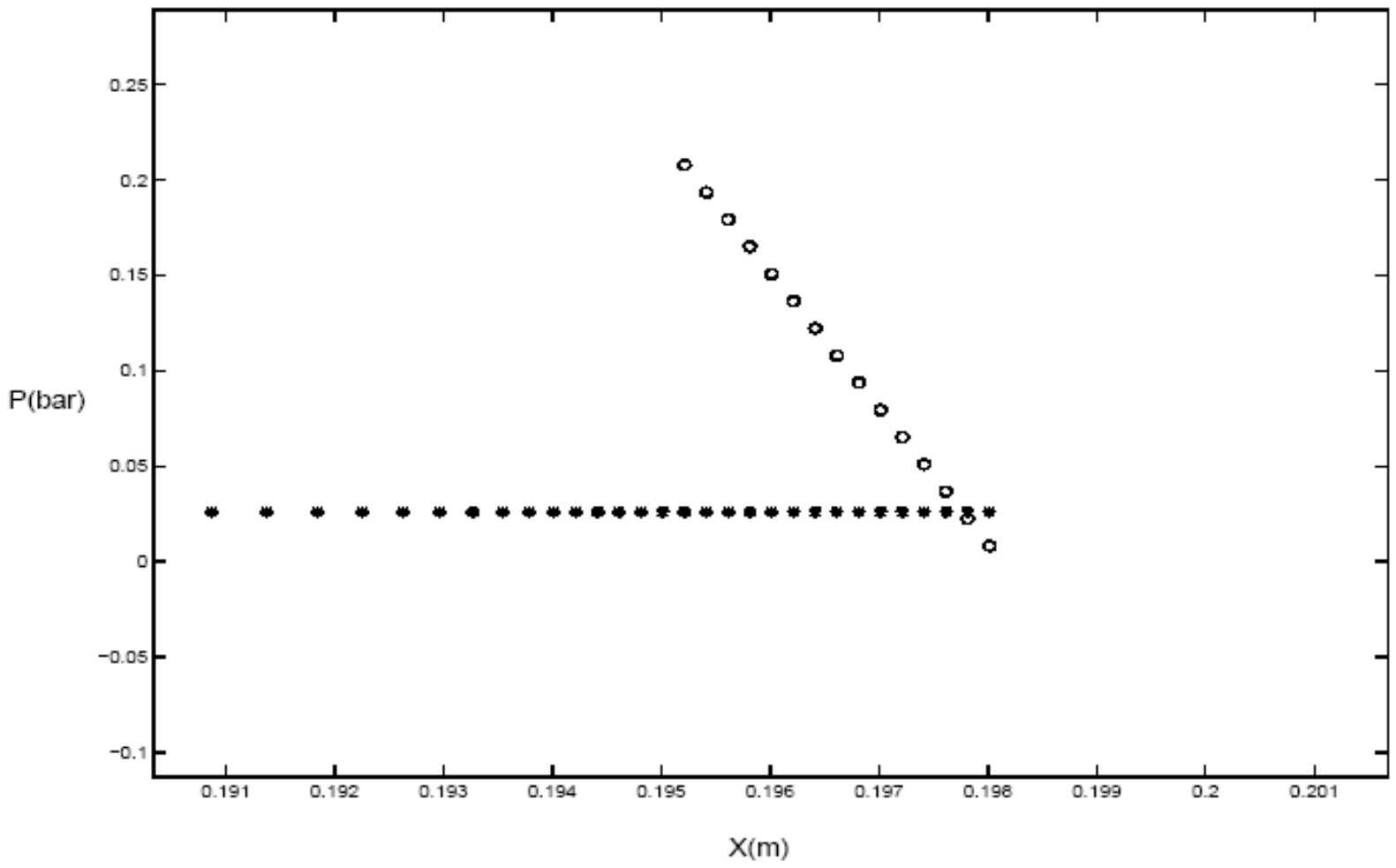

**Figure 38 : Distribution de pression le long de la partie finale de la conduite ; la ligne horizontale (∗) représente la pression de vapeur saturante.**

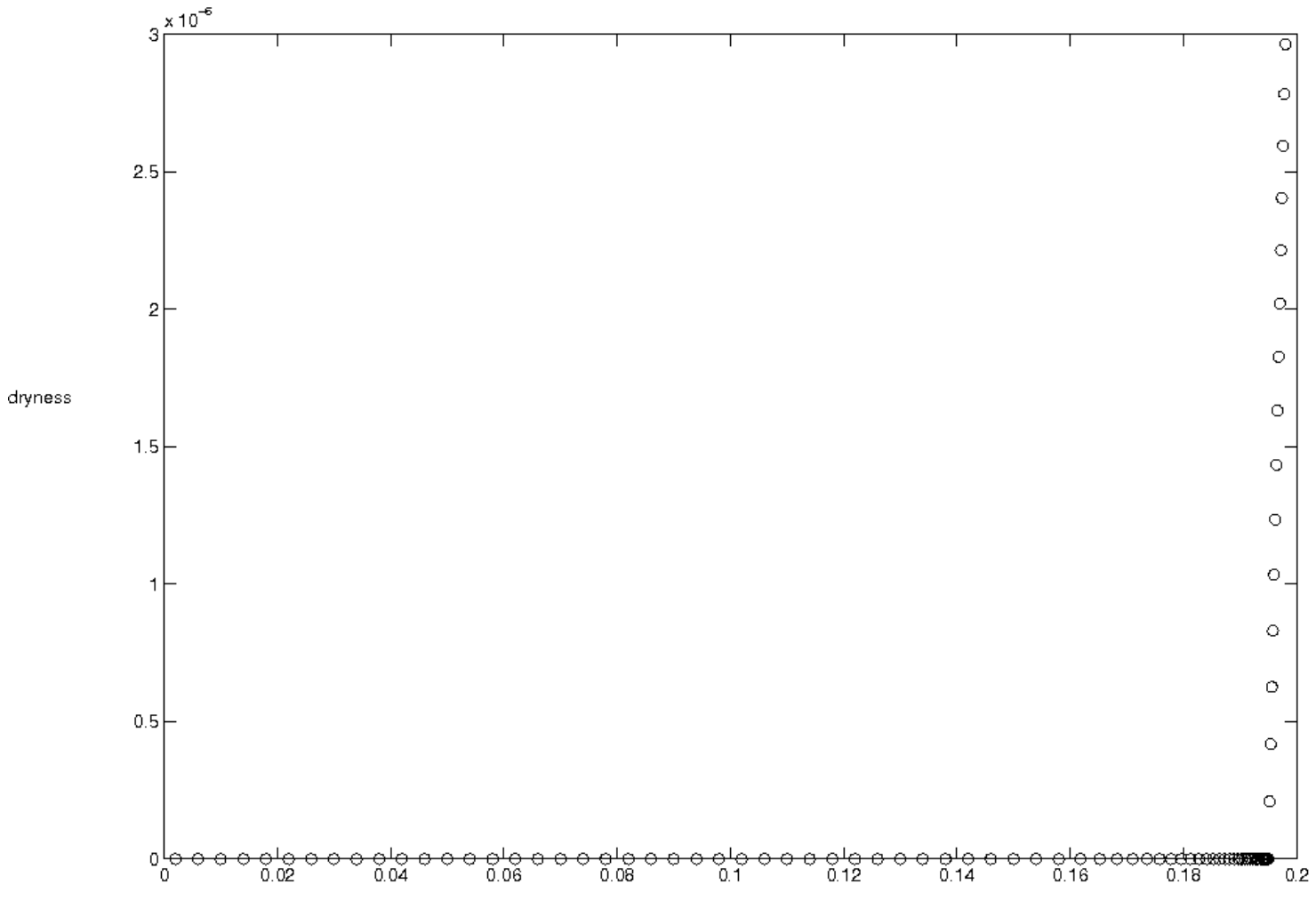

**Figure 39 : Distribution de la fraction sèche à la fin de la conduite.**

La distribution de fraction gazeuse le long de la conduite est représentée sur la Figure 39. Elle fait apparaître une faible quantité de vapeur à la sortie. La fraction sèche est de $3\ 10^{-6}$ à la sortie. La fraction gazeuse $\alpha$ est inférieure à 0.08, elle est faible mais à la limite de validité du modèle de mélange ici utilisé. Ce résultat est à rapprocher de l'expérience de l'ONERA [Foucaud, 1998] qui pour la même pression du réservoir met en évidence la présence de quelques bulles à la sortie mais avec un niveau de vaporisation très faible. Ici, il semble que le calcul surestime le niveau de vaporisation : cela pourrait être dû à un niveau de convergence insuffisant du calcul qui est très instable avec le modèle thermodynamique utilisé. Une autre possibilité pour améliorer les prédictions serait l'utilisation d'un modèle plus avancé [Downar-Zapolski, 1996] dans le quel le temps de relaxation n'est plus une constante mais une fonction. En l'absence d'une mesure expérimentale quantitative il est cependant difficile de conclure. Toutefois, le calcul comme l'expérience ne mettent pas en évidence un changement drastique des l'état thermodynamique de l'hydrazine dans la conduite et ce pour la plage de pression de fonctionnement prescrite. Ce qui exclu par conséquent le risque d'obstruction de la conduite.

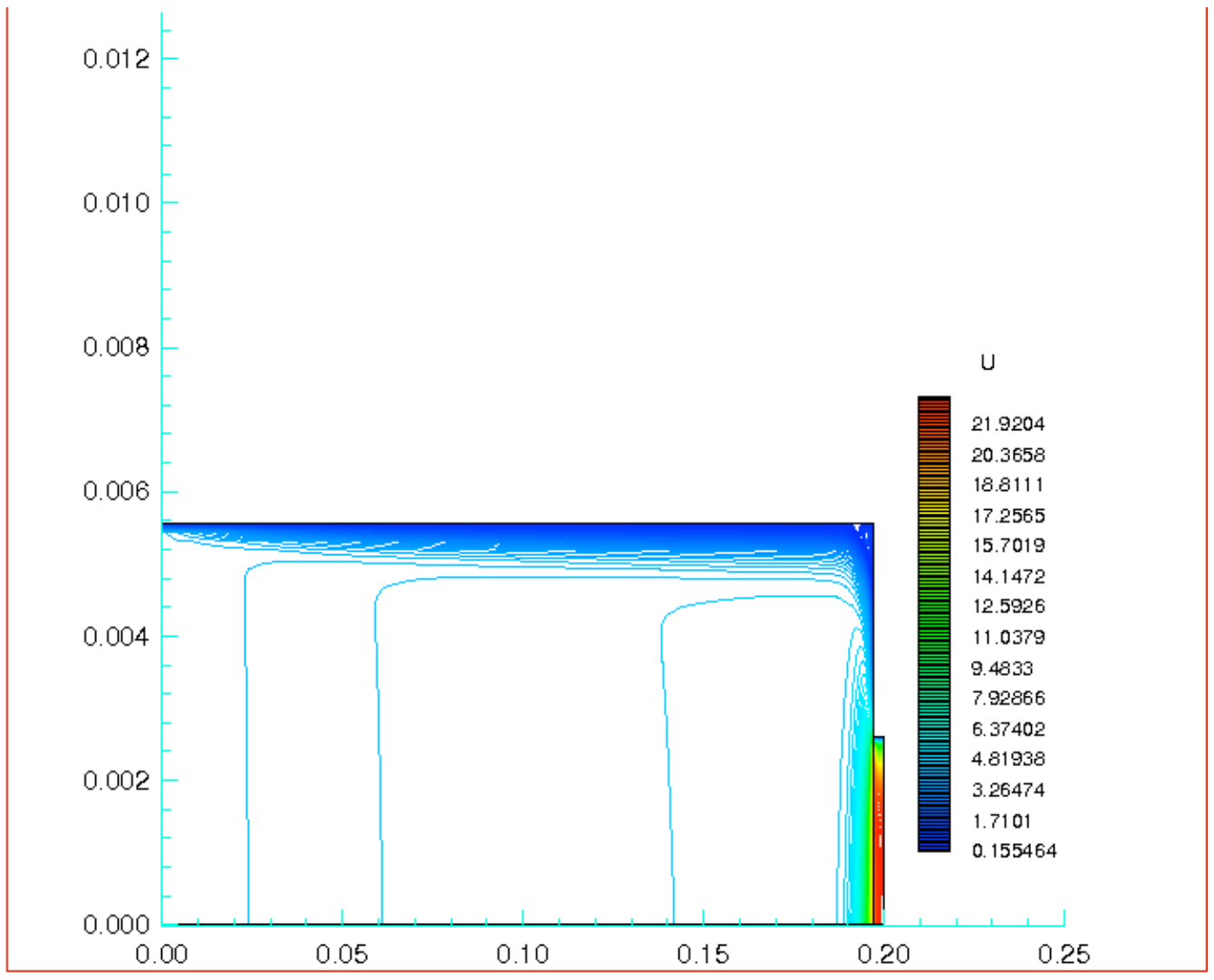

**Figure 40 : Distribution de vitesse en fin de conduite pour une pression de réservoir de 6 bars.**

### 3.2.6 Calculs axisymétriques

Dans le but de prédire plus finement l'écoulement interne des calculs axisymétriques ont été effectué avec le code DEFT [Segal, 2000] de l'Université de Delft. Ce code est basé sur la même méthode numérique de correction de pression que celle décrite précédemment. Le schéma est par contre précis à l'ordre 2 en espace de par l'utilisation d'une méthode TVD. Au vu des résultats monodimensionnel, les calculs ont été effectués pour une pression du réservoir de 6 bars avec une approche monophasique et un écoulement laminaire. La distribution de vitesse est représentée sur la Figure 40, elle ne met pas en évidence une zone de recirculation de large amplitude située au dessus de l'éjection. Cette région de séparation est prédite mais son extension est faible. La Figure 41 montre la distribution de pression dans la finale du montage. Elle ne met pas en évidence de région de basses pressions en amont de la grille d'éjection. Ces résultats corroborent les expériences menées à l'ONERA qui n'ont pas montré de niveau élevé de vaporisation avant la sortie dans la chambre à vide.

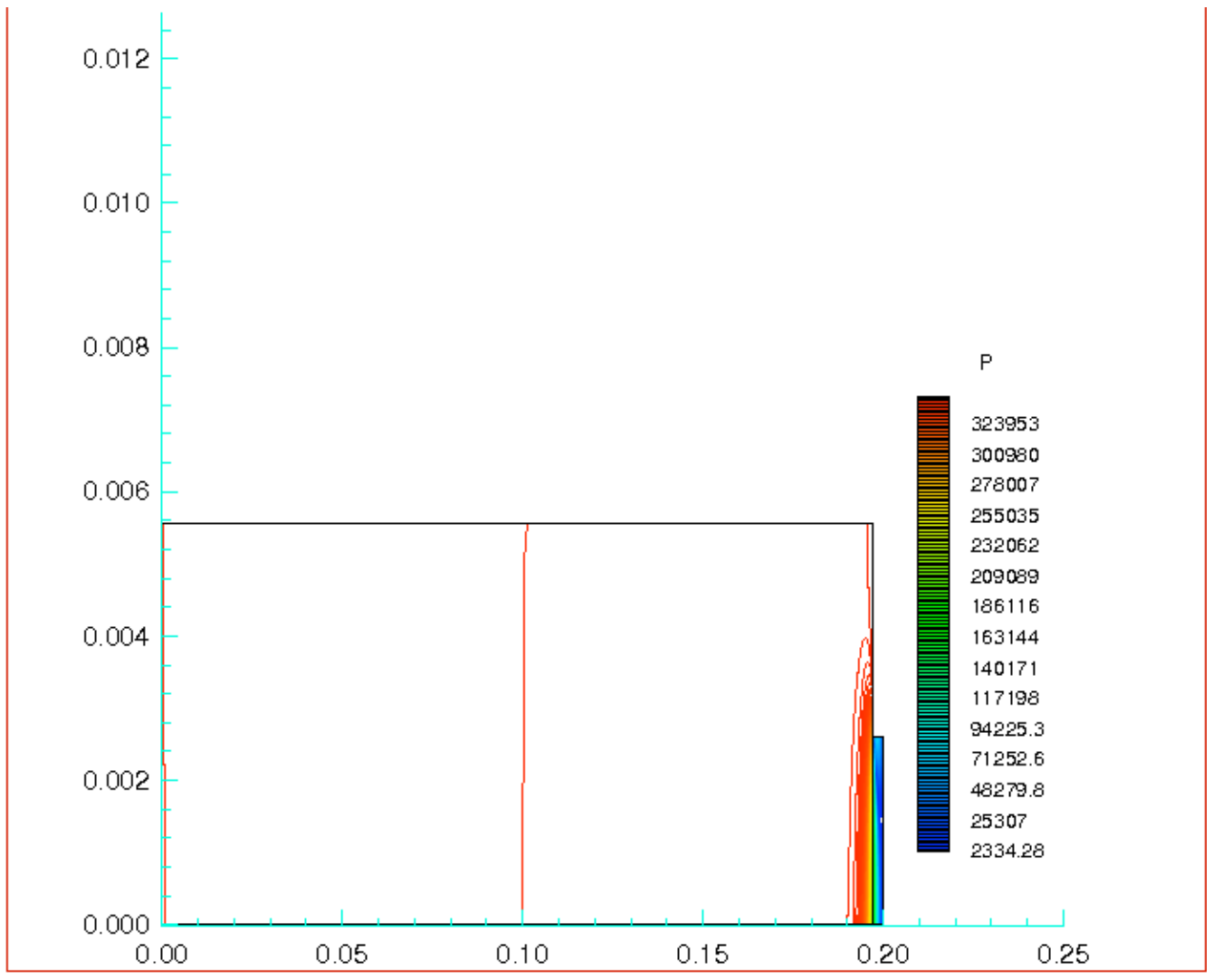

**Figure 41 : Distribution de pression en fin de conduite pour une pression de réservoir de 6 bars.**

# 3.3 Prédiction de l'explosion de l'ATV

La seconde application présentée dans ce chapitre est une analyse du risque d'explosion de l'ATV ("Automated Transfer Vehicle") lors de sa rentrée atmosphérique à la fin de sa mission. L'ATV est un cargo de ravitaillement pour la station spatiale (ISS), il est représenté sur la Figure 42. En fin de mission, il effectue une rentrée balistique dans l'atmosphère terrestre, la prédiction de sa destruction le long de sa trajectoire de rentrée est un élément important en vue d'assurer la sécurité des populations au sol. La destruction prématurée du véhicule, due à une explosion entraînée par la présence de résidus de carburant dans les réservoirs à bord, a fait l'objet de cette étude.

### 3.3.1 Approche du problème

Une première étude [Fritsche, 2001] a permis de modéliser à l'aide d'un outil d'ingénierie [Koppenwallner, 2004] la rentrée de l'ATV avec une fragmentation principale du véhicule prédite pour une altitude de 75 km. Ici, le but a été de préciser le potentiel d'explosion du véhicule dans le cas d'une fuite de carburant et en présence d'une fissure. La rentrée de l'ATV est considérée dans son cas nominal, c'est-à-dire sans rotation du véhicule.

Une simulation complète de la rentrée du véhicule en prenant en compte les aspects dynamiques est trop coûteuse du point de vue du calcul. Pour contourner la complexité numérique due à cet aspect dynamique avec un changement des conditions atmosphériques (pression, densité et température) le long de la trajectoire la rentrée a été assimilée à une succession d'états stationnaires. L'évaluation du temps de remplissage du cargo à travers une fissure [R5, JS2] a confirmé cette hypothèse. Le temps de remplissage a été évalué en assimilant la fissure au col d'une tuyère de Laval ce qui permet d'estimer le débit à travers la fissure et donc le temps de remplissage. Cette estimation au premier ordre a montré que le temps nécessaire au remplissage du véhicule correspond à une variation d'altitude de l'ordre de 200 m. Par conséquent, la rentrée de l'ATV peut être considérée comme une succession d'états stationnaires.

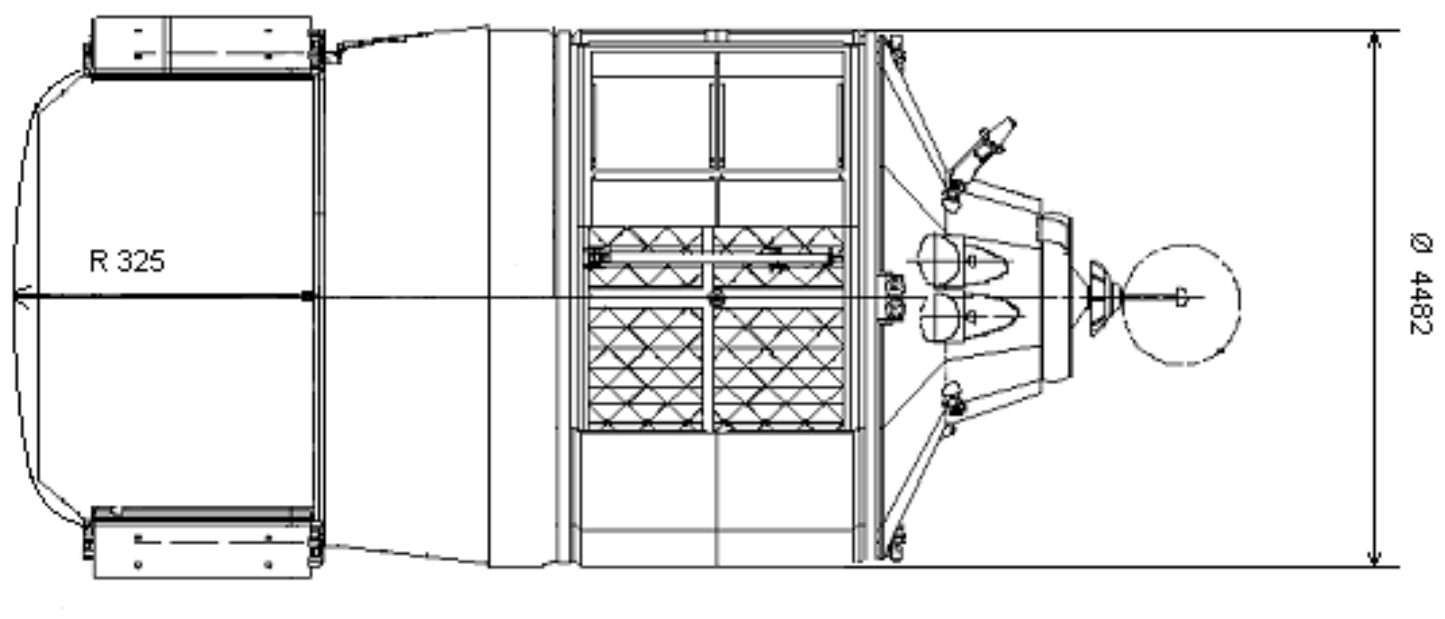

**Figure 42 : Schéma de l'ATV.**

Les carburants à bord du véhicule sont de 3 types : MMH (mono-méthyl-hydrazine), UDMH (di-méthyl-hydrazine) et NTO (tétraoxyde d'azote). Les réactions chimiques susceptibles d'entraîner l'explosion de l'ATV sont les suivantes :

$$MMH + O_2 \qquad (3.7)$$

$$MMH + N_2O_4 / NO_2 \qquad (3.8)$$

$$UDMH + N_2O_4 / NO_2 \qquad (3.9)$$

Pour analyser le potentiel d'explosion du véhicule pour ces trois réactions chimiques, nous avons procédé à un couplage faible (au sens numérique) entre le calcul CFD qui permet de préciser les conditions thermodynamiques à l'intérieur de l'ATV et une analyse d'explosion pour ces réactions.

### 3.3.2 Simulation de la rentrée

La destruction principale du véhicule est prédite pour une altitude de 75 km. L'écoulement autour de l'ATV et à l'intérieur de celui-ci dans le cas de la présence d'une fissure a été simulé pour les altitudes de 75 et 80 km. Les calculs ont été effectués avec le code TAU [Gerhold, 1997] du DLR, un modèle thermochimique à 5 espèces ($N_2$, $O_2$, O, N et NO) et 17 équations [Gupta, 1990 ; Park, 1990] pour prendre en compte les effets hors équilibre. Pour l'étude la paroi est considérée comme isotherme à 800 K et complètement catalytique étant donné que la structure est en aluminium. Les calculs ont été effectués pour une configuration axisymétrique et un véhicule intact dans un premier temps. L'indépendance des résultats par rapport au maillage a été vérifiée à l'aide du module adaptatif du code de la même façon que dans le Chapitre 2. Le maillage final pour la géométrie fissurée est représenté sur la Figure 43.

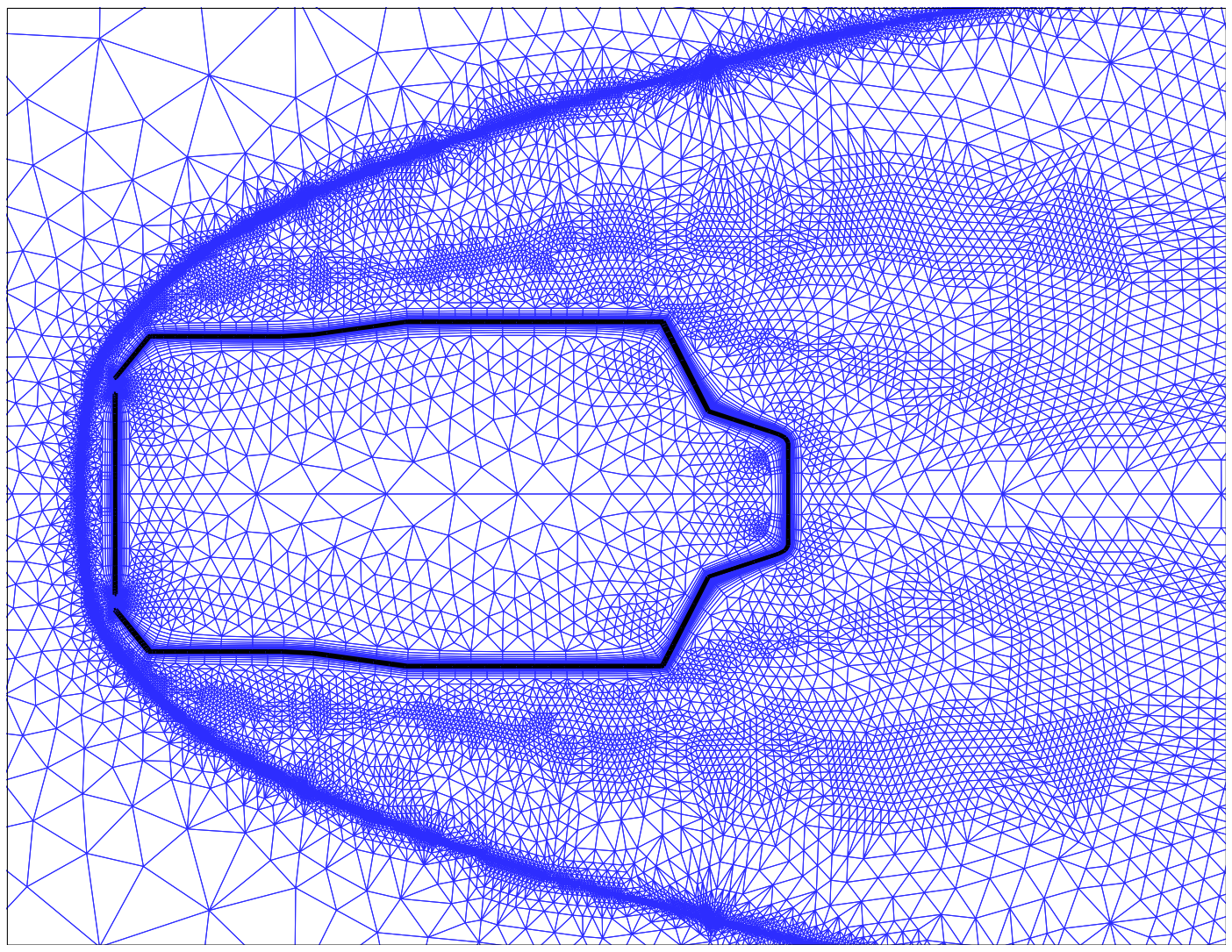

**Figure 43 : Maillage final obtenu pour l'ATV.**

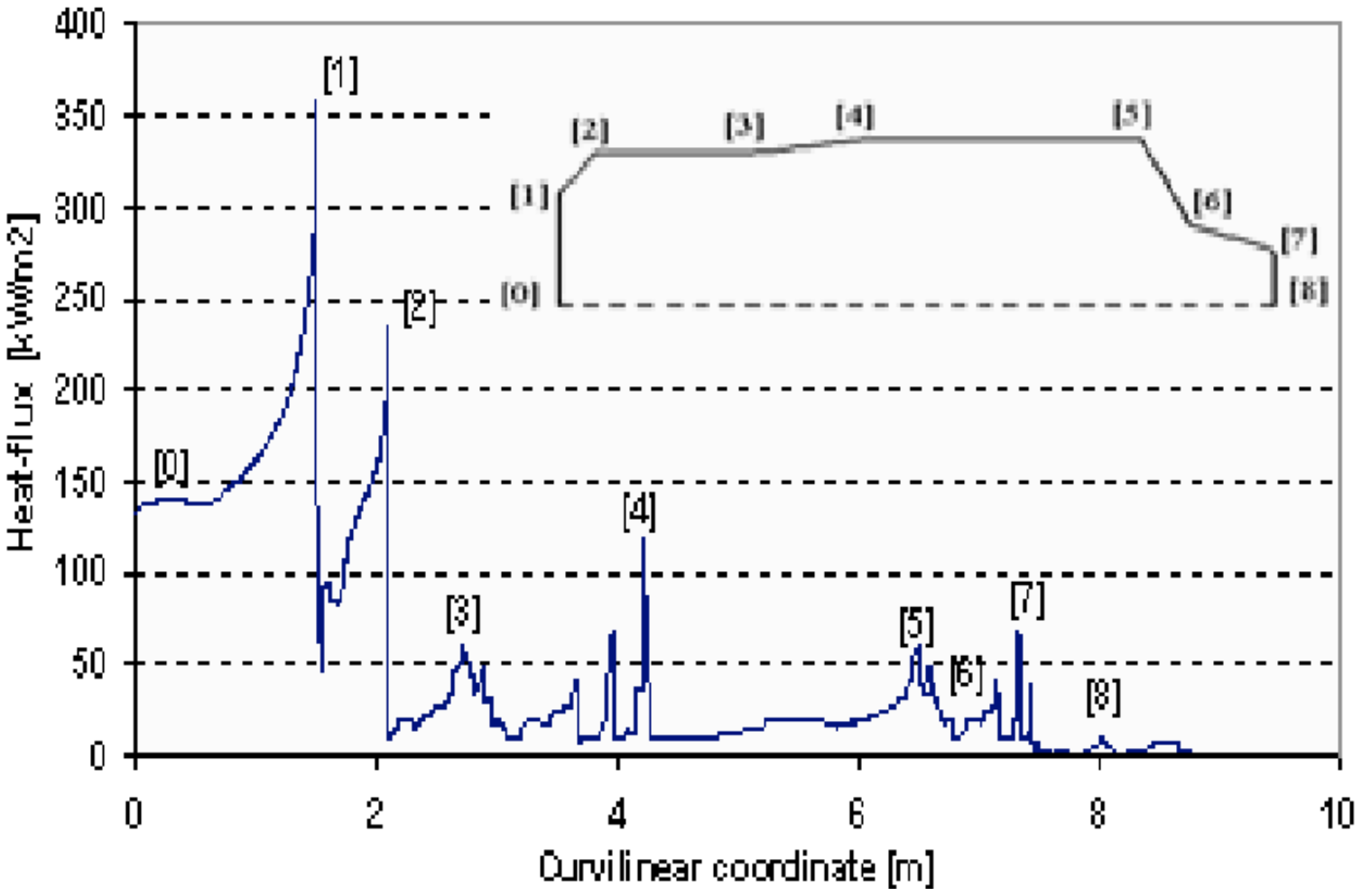

**Figure 44 : Distribution de flux de chaleur le long de la surface.**

Les résultats obtenus pour le véhicule intact ont permis de déterminer l'emplacement le plus probable de la fissure. La distribution de flux de chaleur est montrée sur la Figure 44. Le pic de flux (360 $kW/m^2$) est atteint non pas au point d'arrêt mais au niveau du coin supérieur du véhicule [1], cela est du à la présence de la paroi verticale perpendiculaire à l'écoulement, visible sur la Figure 44. L'endroit le plus probable pour l'apparition d'une fissure dans la structure est donc au voisinage de ce point [1]. Pour l'évaluation du potentiel d'explosion de l'ATV, des calculs ont donc été menés pour une telle configuration.

Le point délicat du calcul est de converger la simulation pour un écoulement extérieur hypersonique tandis que l'écoulement interne devient progressivement quasiment incompressible. L'écoulement a été considéré comme convergé lorsque la pression à l'intérieur du véhicule est égale à celle de la couche de choc. Cela est représenté sur la Figure 45 où le tracé des lignes de courant montre qu'il n'y quasiment plus de débit à travers la fissure. Il ne reste plus alors qu'un écoulement interne résiduel.

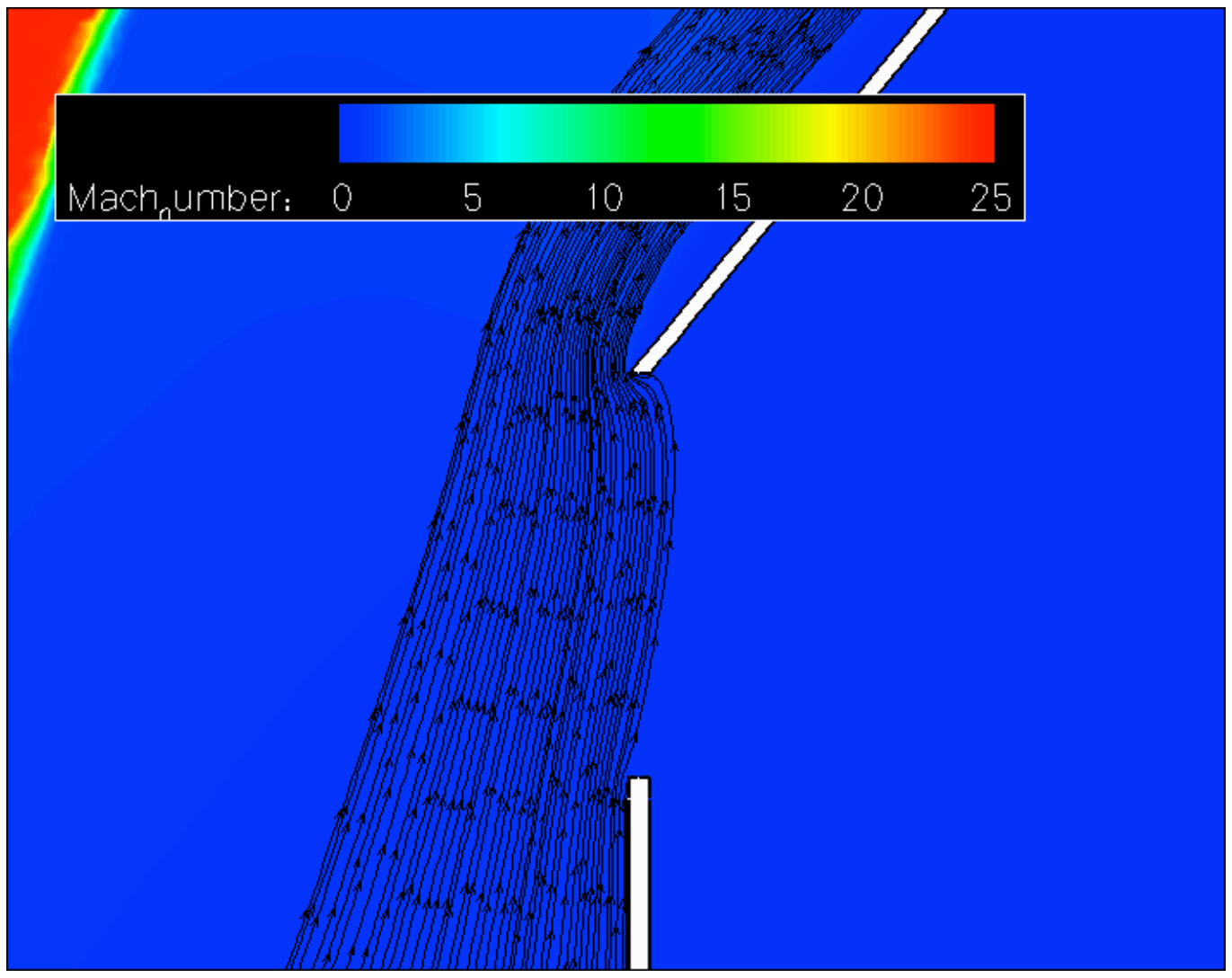

**Figure 45 : Zoom de l'écoulement au voisinage de la fissure avec tracé des lignes de courant.**

Les simulations numériques ont permis d'évaluer les pressions partielles d'oxygène atomique, élément le plus réactif, et d'oxygène diatomique à l'intérieur de l'ATV. La détermination de ces deux quantités est nécessaire pour évaluer le potentiel d'explosion du véhicule à travers les relations (3.7), (3.8) et (3.9). Les simulations numériques ont mis en évidence une faible pression partielle d'oxygène atomique, inférieure à 50 Pa, à l'intérieur du véhicule (voir Figure 46). La pression partielle d'oxygène diatomique est représentée sur la Figure 47. Cette dernière met en évidence une pression élevée d'oxygène diatomique dans l'ATV. Cela est dû à la température de l'ordre 1300 K quasiment uniforme dans le véhicule. Ce niveau de température est beaucoup plus bas que la température de 2500 K nécessaire pour dissocier l'oxygène diatomique. En conséquence lorsqu'il pénètre dans le véhicule l'oxygène atomique de la couche de choc à une forte tendance à se recombiner en oxygène moléculaire. La pression partielle résiduelle d'oxygène atomique est due à la présence d'un point chaud situé à l'intérieur de l'ATV et au dessus de la fissure. Ce point chaud est créé par le mouvement résiduel du gaz à l'intérieur du véhicule qui continue d'entraîner du gaz chaud en provenance de la couche de choc à travers la fissure.

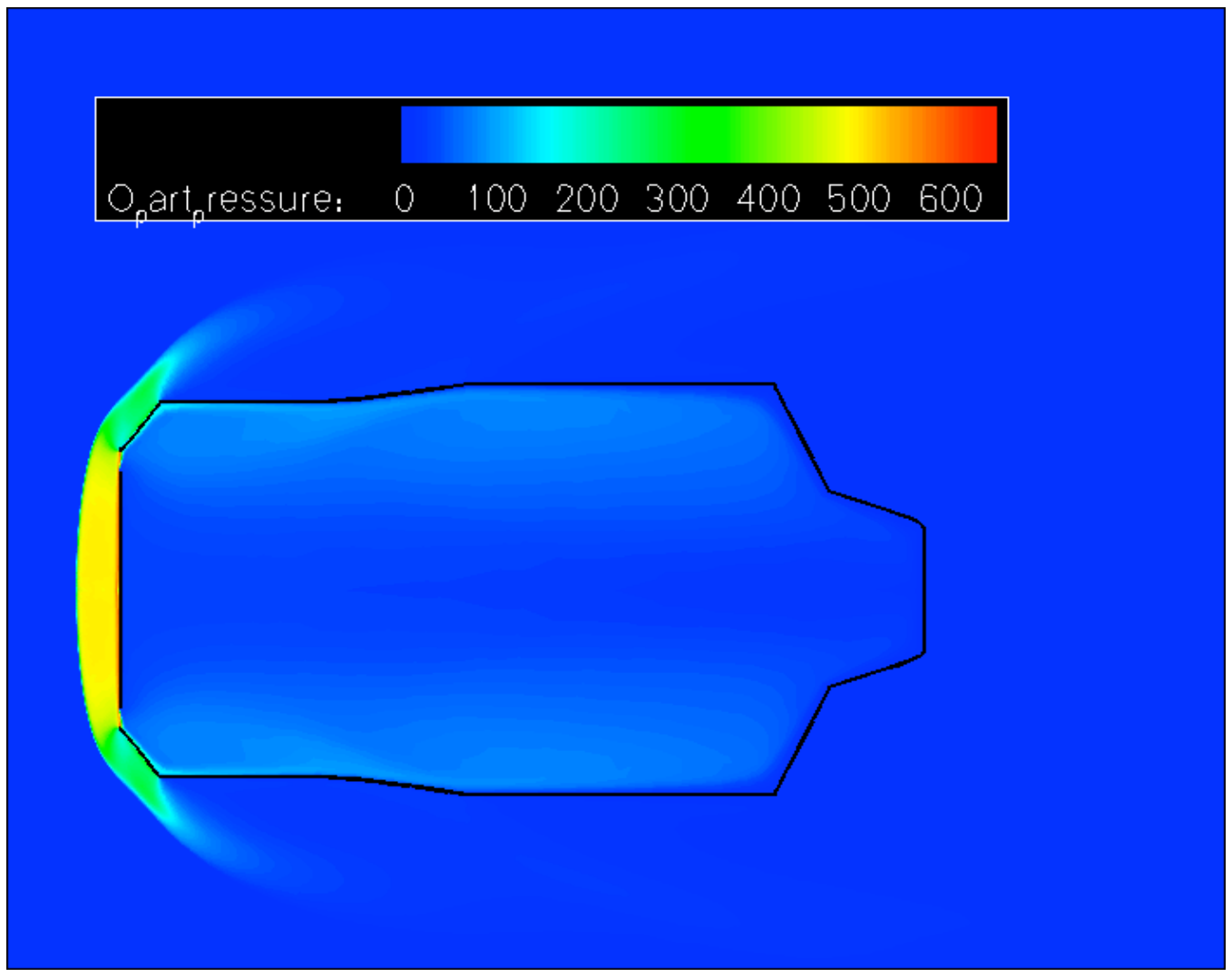

**Figure 46 : Pression partielle d'oxygène atomique pour une altitude de 75 km.**

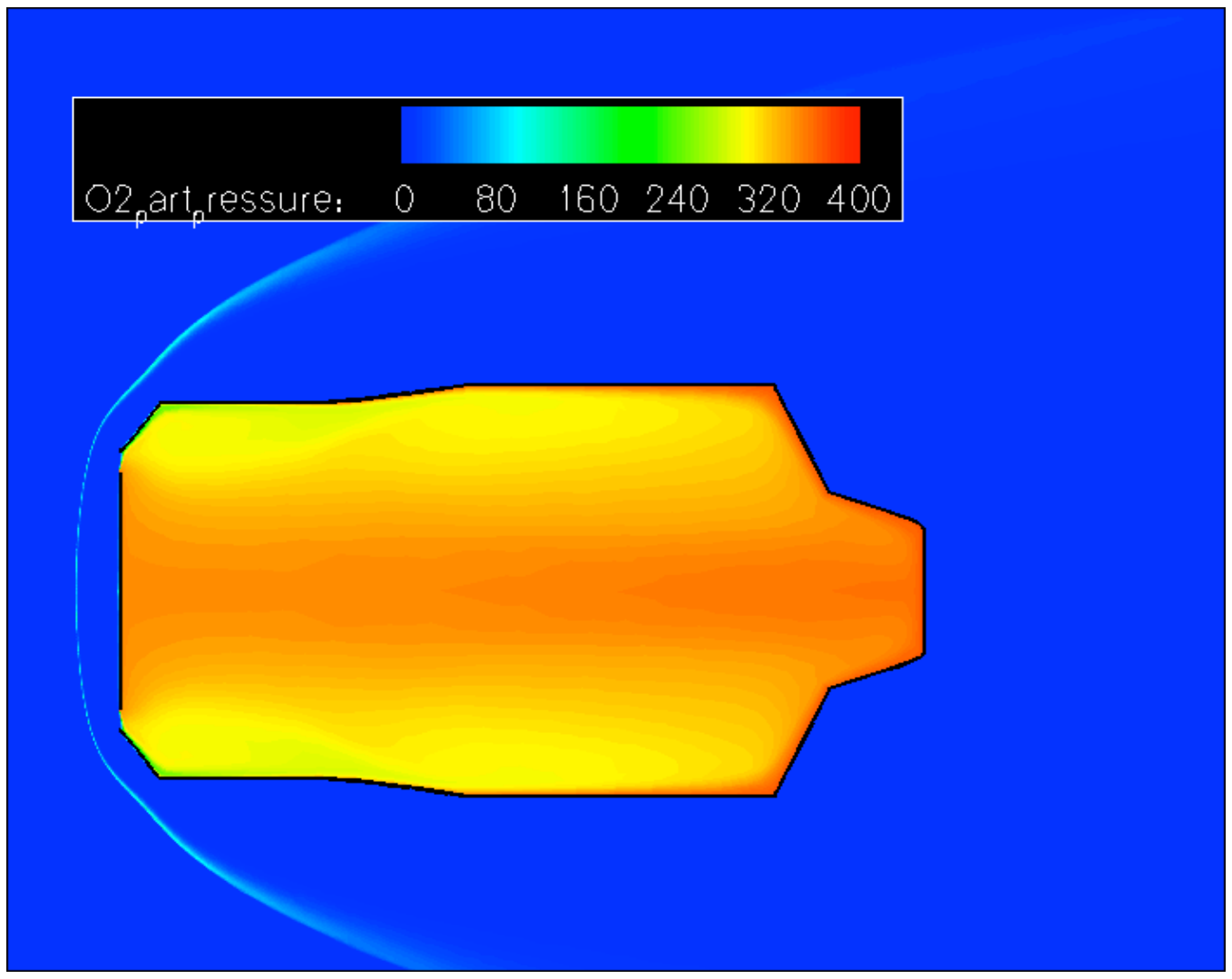

**Figure 47 : Pression partielle d'oxygène diatomique pour une altitude de 75 km.**

### 3.3.3 Estimation de la probabilité d'explosion

Dans le but d'évaluer le potentiel d'explosion du mélange à l'intérieur du véhicule les pressions minimales d'auto-allumage des réactions (3.7), (3.8) et (3.9) ont été déterminées. Cela a été effectué à l'aide de données accessibles dans la littérature [Gray, 1974]. L'interpolation de ces données permet de déterminer les pressions d'auto-allumage en fonction de la température. Pour la réaction (3.7) entre le MMH et l'oxygène moléculaire nous obtenons :

$$p_{i,\min} = T^2 e^{5115,5/T - 13,425} \qquad (3.10)$$

Pour la réaction (3.8) entre le MMH et le NTO,

$$p_{i,\min} = 70T^2 e^{2100/T-15,8} \qquad (3.11)$$

Enfin, pour la réaction (3.9) entre l'UDMH et le NTO,

$$p_{i,\min} = 133{,}3T^2 e^{1811{,}6/T-16{,}3136} \qquad (3.11)$$

Les délais d'auto-allumage des réactions ont été évalués à partir de différents éléments de la littérature. Que ce soit pour la réaction (3.7) [Gray, 1974] ou pour les réactions (3.8) et (3.9) [Seamans, 1967 ; Catoire1995], ces délais sont inférieurs à la seconde et donc du même ordre que le temps de remplissage de l'ATV.

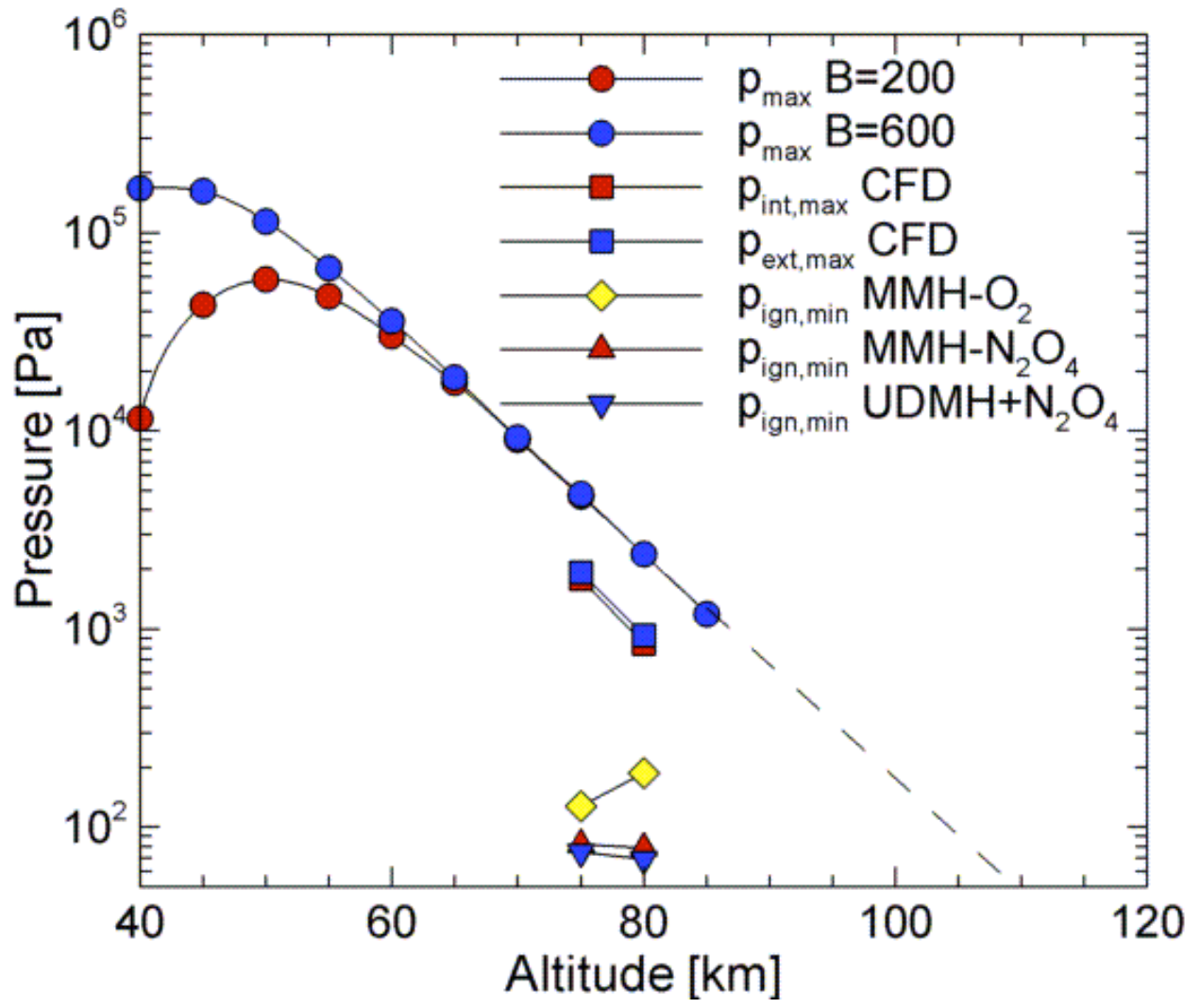

**Figure 48 : Pressions le long de la trajectoire de rentrée pour déterminées à l'aide d'une approche ingénieur pour deux coefficients balistiques (B=200 et 600), et à l'aide du calcul CFD, et pressions d'auto-allumage des différentes réactions.**

Les résultats obtenus avec l'approche développée ont été comparés aux seuils d'auto-allumage des différentes réactions. Les niveaux de pression prédits pour les altitudes de 75 et 80 km sont reportés sur la Figure 48. Les trajectoires obtenues à l'aide de l'approche ingénieur pour des coefficients balistiques de 200 et 600 sont elles aussi reportées sur la figure. Pour les altitudes considérées le coefficient balistique ($B = S\, C_D /m$ ; où S est la surface de référence, $C_D$ la traînée et m la masse) n'a pas d'influence sur la pression externe. Les comparaisons avec le calcul CFD mettent en évidence une surestimation de la pression par la méthode ingénieur. Sur la même figure, les pressions minimales d'auto-allumage des différentes réactions ($p_{ign,min}$)sont elles aussi reportées. La pression dans le véhicule est d'un ordre de magnitude plus élevée que celle d'auto-allumage. Donc, une fuite de MMH ou d'UDMH dans le cas d'une perforation de la structure a les plus grandes chances d'entraîner une explosion prématurée de l'ATV. Ce risque est amplifié par la présence d'oxygène atomique à l'intérieur de véhicule mais aussi dans le cas d'une fuite simultanée de NTO.

# 3.4 Conclusion

Les études liées à la production de débris dans l'espace mettent en jeu des processus variés et complexes : aspects dynamique de la rentrée pour la fin de vie d'un astronef, écoulement multiphasiques et décompression jusqu'à la pression du vide pour la passivation d'un réservoir pour les deux exemples abordés. Les deux applications présentées dans ce chapitre ont servi à illustrer les études liées aux problèmes de débris qui deviennent récurrentes avec le durcissement des normes relatives aux débris atterrissant sur le sol de notre planète.

Dans la première étude, les simulations numériques ont servi en premier lieu de support pour la préparation des campagnes expérimentales. Avec celles-ci, elles ont permis la validation du processus de passivation pour la plage de fonctionnement prescrite par le CNES: à savoir une pression de réservoir comprise entre 6 et 25 bars. La passivation des réservoirs d'Ariane 5 durant le vol 503, pour lequel cette étude a été menée, s'est déroulée sans problème en accord avec les prédictions numériques et avec les essais effectués au sol. Sur le plan numérique l'aspect la plus attractif a été le développement d'une méthode permettant de prédire un changement de phase et donc un écoulement à densité variable pour des nombres de Mach très bas pour lesquels les méthodes usuelles utilisées sans préconditionnement sont défaillantes.

La seconde partie est un premier pas pour traiter les problèmes liés aux rentrées d'astronefs en fin de vie. Le fait de considérer la rentrée de l'ATV comme une succession d'états stationnaires semble justifié par l'estimation du temps de remplissage après ouverture de la fissure. L'analyse a fait l'impasse sur des points tels que le couplage fluide/structure de l'écoulement qu'il est nécessaire de prendre en compte pour l'apparition de fissures et notamment de leur taille. Il aurait été aussi intéressant de connaître le flux de chaleur auquel pourrait être soumis les réservoirs de carburant. Toutefois, l'analyse a permis d'établir que les outils de calcul sont maintenant proches du point de maturité où ils pourront résoudre ce type de problème.

Ces travaux réalisés à l'ESA-ESTEC ont fait l'objet des communications [JS2], [J2], [C7-8,C16] et [R5] référées en Annexes. La deuxième partie portant sur la rentrée de l'ATV a été effectuée dans le cadre du stage jeune ingénieur de D. Boutamine [Y5].

# 4 Projets et perspectives : La rentrée atmosphérique

## 4.1 Introduction

La rentrée atmosphérique, dernière application abordée dans ce document, met en jeu une large palette de phénomènes à prendre en compte lors des calculs des bases de données aérodynamiques des capsules de rentrée ou pour estimer les flux de chaleur au niveau des protections thermiques. Lors d'une rentrée atmosphérique une capsule subit divers types d'écoulements. A haute altitude,l'écoulement est d'abord raréfié avec un grand nombre de Knudsen, lorsque le véhicule atteint les couches plus denses de l'atmosphère l'écoulement devient transitionnel avec un régime appelé "merged layer regime" ou la couche limite n'est pas distincte de l'onde de choc. Enfin, lorsque le nombre de Knudsen ($Kn = \lambda/L$ ; où $\lambda$ est le libre parcours moyen et L une grandeur caractéristique) est inférieur à 0.1, l'écoulement peut alors être considéré comme continu. Au niveau des équations cela se traduit par des systèmes différents à résoudre : équations de Boltzmann en régime raréfié résolues à l'aide de simulations de type Monte-Carlo [Bird, 1994], de Burnett (entre autres) en régime transitionnel [Agarwal, 2001 ; Dubroca, 2002 ; Mieussens, 1999] et de Navier-Stokes en régime continu.

Au niveau de l'écoulement différents phénomènes dépendant de la vitesse d'entrée et de la composition de l'atmosphère sont en prendre en compte. Ce sont notamment les effets hors équilibres chimiques (dissociation, ionisation) et thermiques ainsi que le rayonnement du plasma entourant la capsule [Marraffa, 1998-1 ; Mazoué, 2005]. Pour certaines rentrées (Vénus, Jupiter, Titan) la composante radiative du flux de chaleur est plus importante que la composante convective [Park, 1999]. Le flux de chaleur convectif est drastiquement augmenté dans le cas d'un écoulement turbulent [Charbonnier, 2003], l'analyse préalable de la transition vers la turbulence est nécessaire au design des protections thermiques. L'ablation des protections thermiques [Duffa, 1996] est aussi un point important car ce phénomène réduit considérablement la température de surface mais est à pour effet une certaine déstabilisation de la couche limite génératrice de turbulence.

Un point particulier est l’aspect stabilité de la capsule : la stabilité statique peut être perturbée en début de rentrée si la ligne sonique interagit avec le bord de fuite de la capsule [Gnoffo, 1995]. L’instabilité dynamique [Mitcheltree, 1999 ; Winchenbach, 2002] affecte l’ensemble des capsules de rentrée pour les bas nombres de Mach supersoniques. L’origine de ce type d’instabilité semble provenir d’un hystérésis de pression provenant de [Jaffe, 1970 ; Karatekin, 2002] du point de séparation au niveau du bord de fuite de la capsule.

Il est difficile de dupliquer les conditions de vol à l’aide des moyens expérimentaux existants. L’extrapolation en vol est généralement fait à l’aide d’outils numériques. Leur validation est effectuée à l’aide des données expérimentales existantes. La reconstruction d’essais effectués dans les souffleries à haute enthalpie ou dans des tubes à chocs nécessitent une connaissance pré-établie du moyen d’essais et des phénomènes s’y déroulant : écoulements en déséquilibre ou figés, présence de particules parois partiellement catalytiques. Ici, l’outil numérique intervient en aval pour sa validation [Park, 1997] mais aussi en amont. Il permet d’évaluer les capacités d’un moyen d’essais pour un environnement non testé et de cerner les problèmes liés aux tests dans des atmosphères exotiques.

Dans ce chapitre nous mettrons en avant plusieurs thèmes de recherche pour lesquels des travaux ont été réalisés ou sont en cours. Les perspectives ont été divisées en trois parties. La première traite des travaux déjà effectués ou en cours liés à l’ablation, la radiation, la catalyse, le black-out et la transition vers la turbulence pour des entrées terrestres et/ou martiennes. L’analyse des moyens d’essais est abordée dans la seconde section. Enfin la dernière partie reprend les activités sur l’optimisation des formes de capsule pour les rentrées atmosphériques et les manœuvres d’aérocapture.

Les travaux présentés ici ont été réalisés à l’ESTEC, certains d’entre eux se poursuivent actuellement dans le cadre de la société Ingénierie et Systèmes Avancés créée en Avril 2006. Les études réalisées ont fait l’objet des communications [JS1], [J5], [C17, C15-11, C6], [A4-3], [R2] et [S4] référées en Annexes. Plusieurs stages de fin d’étude ou de jeunes ingénieurs ont été menés dans le cadre de ces activités [Y5-Y1].

## 4.2 Trajectoires et optimisation de formes

Un des premiers aspects de la rentrée est la trajectoire. Elle peut être raide et associée à des flux thermiques élevés ou avec un angle de rentrée faible et des flux plus faibles mais s’exerçant pendant une durée plus longue. Le flux de chaleur est aussi fonction de la géométrie de la capsule et plus particulièrement du rayon du nez. Un des objectifs d'une étude de trajectoire est d'optimiser la géométrie de la capsule en fonction de la mission. Une étude en ce sens a été effectuée à l'ESTEC (Y2) en coopération avec l'Université Polytechnique de Milan pour optimiser à l'aide d'algorithmes génétiques la géométrie d'une capsule pour une manœuvre d'aérocapture martienne. Les résultats obtenus ont permis de démontrer la capacité de l'algorithme développé à travers l'obtention de géométries de capsules différentes suivant le type de contrainte imposé : coefficient balistique, flux de chaleur maximal, rapport entre la portance et la traînée et vitesse d'injection.

L'application de méthodes d'optimisation est une preuve, s'il en est nécessaire, de la maturité de l'outil numérique pour les problèmes de rentrée atmosphérique et notamment pour le choix de la forme de capsule optimale. L'exécution de cette tâche, effectuée auparavant de façon artisanale et empirique, est en train de déboucher sur la création d'outils couplant la CFD et l'optimisation. Les algorithmes qu'elle met en oeuvre peuvent être génétiques ou basées sur la méthode de l'adjoint, suivant l'objectif recherché. Ces derniers développements préfigurent l'apparition d'outils numériques multi-objectifs et multidisciplinaires.

# 4.3 Phénomènes liés aux rentrées

Parmi les phénomènes physiques qui affectent les rentrées atmosphériques certains ont fait de notre part l'objet de travaux récents ou en cours. Il s'agit d'études portant sur l'ablation des protections thermiques, le rayonnement dû au plasma entourant les capsules en début de rentrée, de la catalyse qui est un facteur déterminant pour le dimensionnement des protections thermiques et de la transition vers la turbulence susceptible de se produire durant la rentrée.

## 4.3.1 Ablation

L'ablation est la récession des protections thermiques entraînée par la pyrolyse du matériau du bouclier thermique. Elle est fortement dépendante de la composition du matériau et des conditions d'entrée en termes de vitesse et de composition de l'atmosphère. Elle induit de multiples effets parmi lesquels un refroidissement de la surface du bouclier. Les gaz issus de la pyrolyse peuvent avoir plusieurs effets. Leur soufflage dans la couche limite est susceptible de bloquer une partie du flux convectif mais aussi de déstabiliser la couche limite favorisant ainsi la transition vers la turbulence [Reda, 1981]. Cependant la diminution de flux de chaleur induite par le blocage convectif pourrait être compensée par l'augmentation due à une transition vers la turbulence [Gupta, 1999]. L'ablation est aussi couplée avec la radiation, pour les rentrées sévères les matériaux du bouclier thermique à base de carbone injectent des espèces au fort potentiel d'absorption ($C_2$, $C_3$) dans la couche de choc, on parle alors de blocage radiatif [Moss, 1982; Matsuyama, 2005]. Un dernier effet de la décomposition du matériau est l'injection de fines particules dans l'écoulement, leur présence induit une augmentation du rayonnement d'environ 10% [Park, 2004].

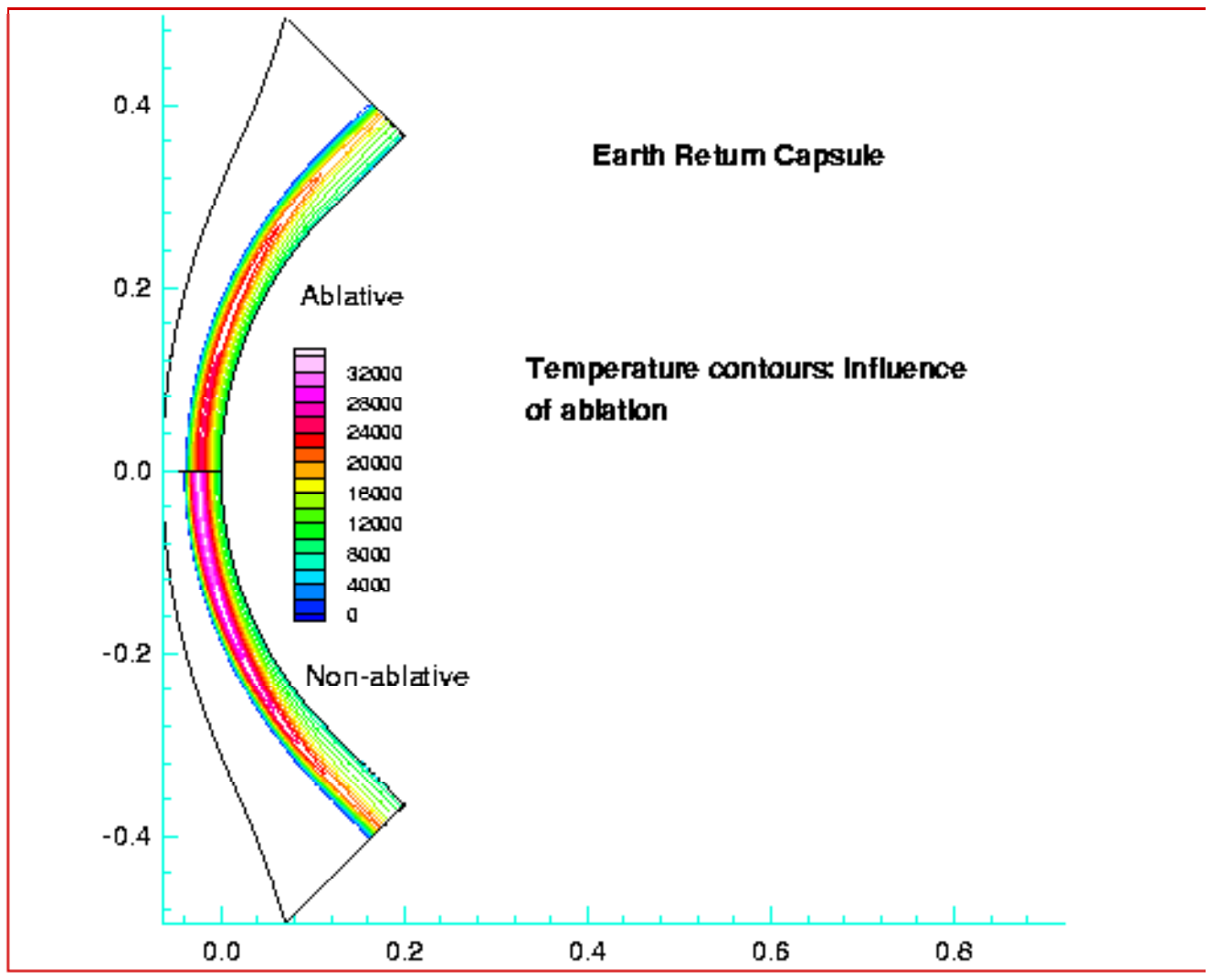

**Figure 49 : Contours de température durant la rentrée terrestre avec (moitié supérieure) et sans ablation (moitié inférieure).**

Une activité de recherche et développement en ablation a été initiées dans le cadre de mes travaux à l'ESTEC. Un premier effort a été mené dans le cadre de l'étude de faisabilité EURORETURN [Marraffa, 1998-2] d'une mission de retour d'échantillons vers Vénus menée par l'ESA. La capsule de retour d'échantillons avait une dimension de 0.4 m et une vitesse d'entrée de 14.17 km/s. Plusieurs calculs ont été mis en œuvre avec TINA [Smith, 2002] pour déterminer l'écoulement autour de la capsule. Un modèle à 24 espèces et 39 réactions, pour prendre en compte les effets hors équilibres et l'ionisation, a été utilisé pour les calculs. Le bouclier thermique était supposé être en carbone phénolique et 5 réactions modélisaient les interactions entre le bouclier et la couche de choc. La récession du bouclier n'était pas prise en compte mais un soufflage de gaz pyrolysés était imposé. Les résultats avec et sans ablation sont représentés sur le Figure 49. Ils illustrent le fort refroidissement de la couche de choc entraîné par la prise en compte de l'ablation.

Ce premier effort est en passe d'être continué dans le cadre d'une consultance de l'ESA sur le blocage convectif lors des rentrées terrestres. Les buts de l'étude seront de définir une stratégie pour le calcul du blocage et notamment du débit de soufflage des gaz pyrolysés dans la couche limite à partir de la composition du matériau et d'évaluer le taux de blocage pour une rentrée terrestre rapide.

Cette étude s'inscrit dans un support à l'Ablation Working Group mis en place par l'ESTEC en 2005 par H. Ritter, L. Marraffa et moi-même et réunissant les instituts de recherche et les industriels ayant des activités dans ce domaine. Un premier workshop a été organisé à l'ESTEC en Octobre 2005 suivi d'une seconde réunion organisé en mai 2005 avec le *3rd Workshop on Thermal Protection Systems and Hot Structures*. Plusieurs cas tests seront définis dans le cadre de ce groupe de travail avec entre autres objectifs la comparaison de résultats code à code.

### 4.3.2 Black-out

Le black-out des communications durant les rentrées atmosphériques est dû à la densité électronique élevée autour de la capsule durant la phase la plus sévère de la rentrée. La période de black-out le est toujours délicate car l'absence de communications entraîne la perte totale des données en cas d'échec. De plus dans l'objectif de conservation des données de vol cette phase doit être prédite correctement pour stocker les données durant cette phase et pour prévoir quand les communications seront reprises. Cette phase peut durer de quelques dizaines de secondes à plusieurs minutes suivant le type de rentrée. Il était de l'ordre de 30 secondes pour Mars Pathfinder et durait de 4 à 10 minutes pour les vols Apollo. Il est à noter que cette interruption de communications est toujours une phase de vol délicate lors des vols habités.

La durée du black-out dépend de plusieurs facteurs : la sévérité de la rentrée, la bande de fréquence utilisée par les antennes à bord et la composition du bouclier de protection thermique. Ainsi, la présence de métaux alcalins dans le matériau du bouclier contribue à la production d'électrons libres responsables du black-out [Evans, 1974].

Une étude préliminaire d'estimation du black-out en préparation de la mission de l'IRDT a été menée à l'ESTEC à l'aide d'une approche de type ingénieur. Connaissant la fréquence de fonctionnement de l'antenne de communication il est possible de déterminer la densité électronique produisant la même fréquence au sein du plasma entourant la capsule qui est la fréquence de coupure. Le black-out partiel est l'atténuation des communications n'est pas considéré ici. La fréquence d'un plasma est donnée par la relation :

$$f_p = \frac{1}{2\pi}\sqrt{\frac{q^2 n_e}{\varepsilon_0 m_e}} \qquad (4.1)$$

Où $q$ est la charge de l'électron, $m_e$ sa masse, $\varepsilon_0$ la permittivité du vide, $n_e$ la densité électronique et $f_p$, en Hz, la fréquence du plasma. La densité critique électronique correspondant à la fréquence de coupure $f_{link}$ est donnée par :

$$n_{e,crit} = \frac{f_{link}^{\ 2}}{80.64\ .\ 10^6} \qquad (4.2)$$

L'antenne ARTS de l'IRDT est intégrée dans le bouclier de protection thermique, elle est située à une faible distance du point d'arrêt. Pour une étude préliminaire, nous pouvons considérer que la densité d'électrons au voisinage de l'antenne est proche de celle au point d'arrêt. Cette dernière a été déterminée le long de la trajectoire à l'aide du code PMSSR [Smith, 1995] et un modèle thermochimique à 16 espèces pour l'atmosphère terrestre. Les densités obtenues et les fréquences critiques pour plusieurs bandes de communications ont été reportées sur la Figure 50.

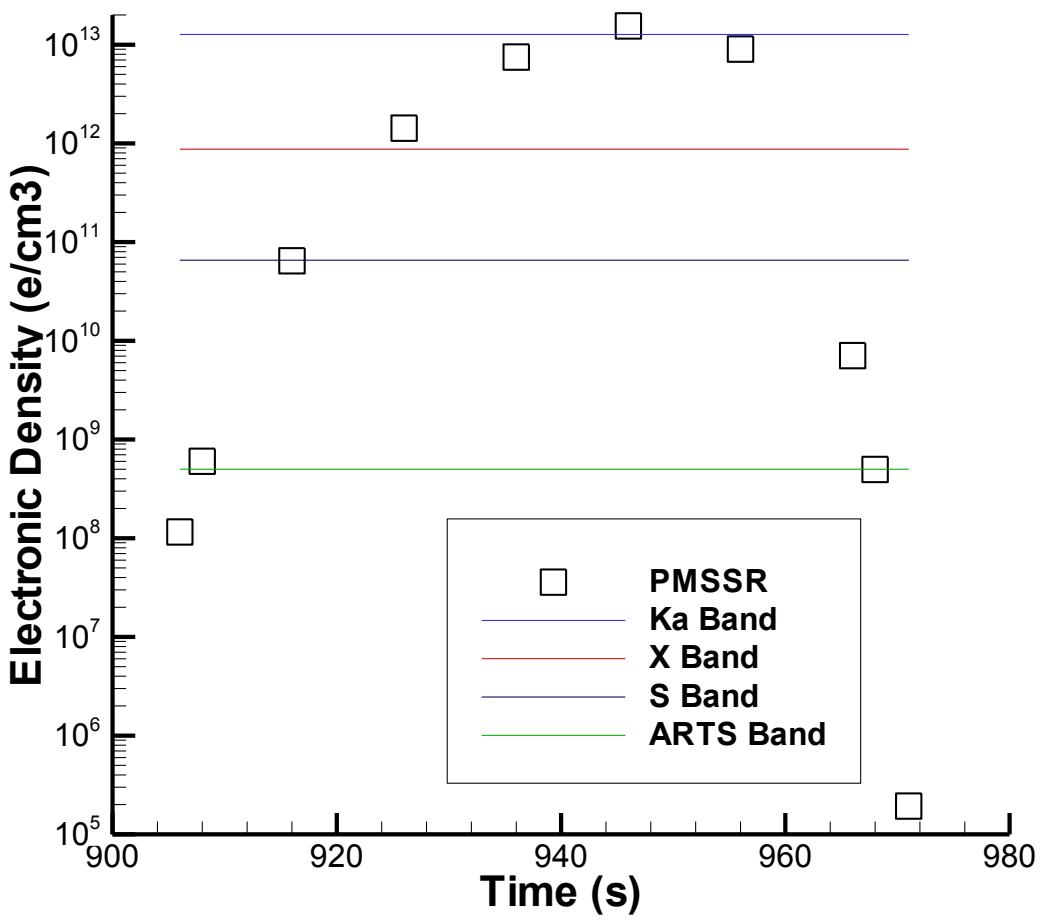

**Figure 50: Prédiction du black-out lors de la mission de l'IRDT.**

Les prédictions donnent une durée de black-out de 60 s en accord avec celle prédite par le Babakin Space Center. Cet accord suffisant pour une phase préliminaire fait l'impasse sur de plusieurs points. L'approche axisymétrique utilisée est suffisante car les comparaisons avec des calculs tridimensionnels ont montré des différences mineures [Greendyke, 1992]. Par contre le black-out est sensible à la composition du bouclier qui n'est pas prise en compte ici. Le phénomène d'avalanche électronique caractéristique du début de la rentrée est fortement dépendant du modèle thermochimique. Dans le cas présent des comparaisons pour plusieurs modèles seraient utiles. Enfin cette étude a fait l'impasse sur des phénomènes tels que la diffraction, le couplage avec le rayonnement (qui est cependant faible pour ce type de rentrée) et la propagation d'ondes électromagnétique au sein du plasma.

La mission de l'IRDT n'a pas été nominale et à ce jour la capsule n'a pas été récupérée. Les données pour le black-out sont cependant disponibles et une analyse à l'aide d'outils plus avancés s'avère nécessaire pour évaluer la validité de la démarche utilisée ici.

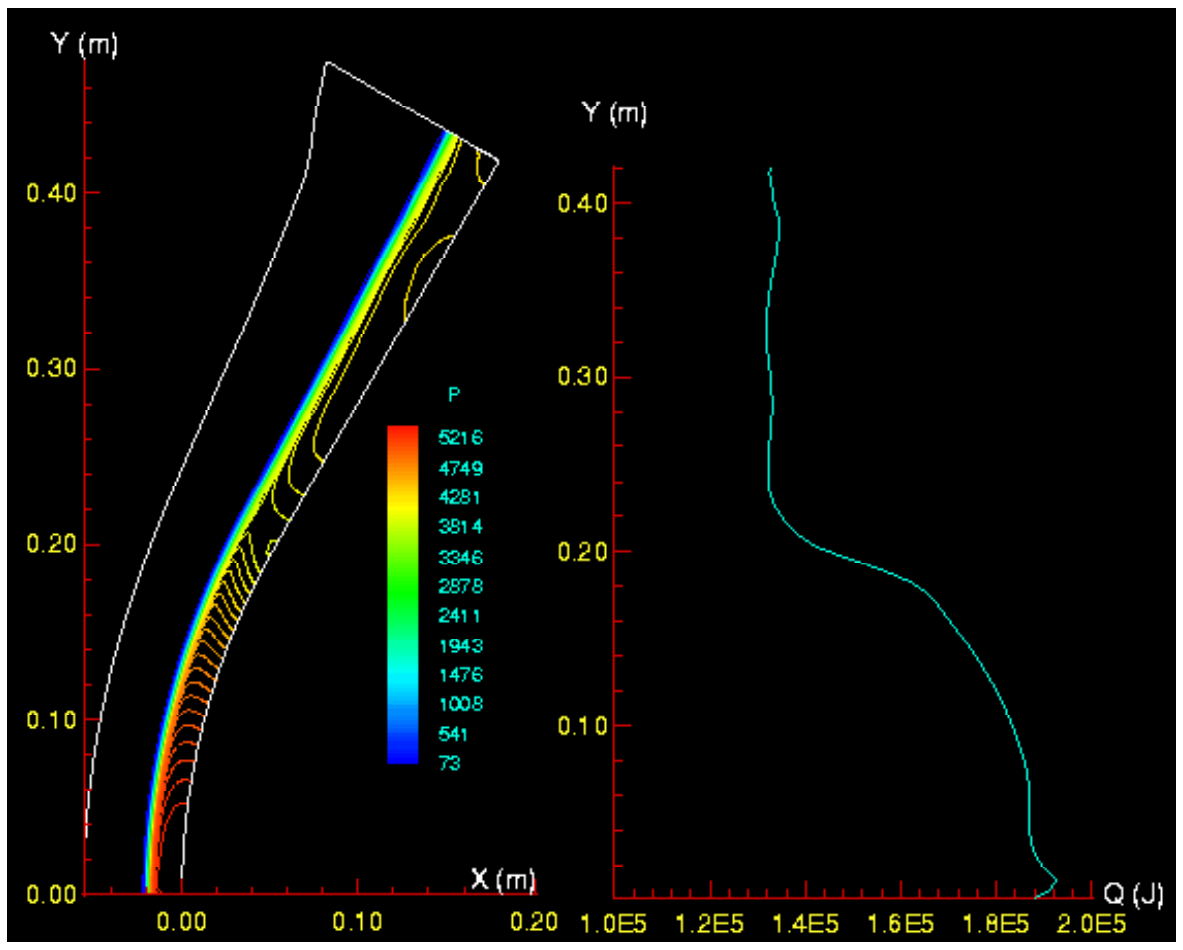

**Figure 51 : Champ de pression et distribution de flux de chaleur à la surface de la sonde InterMarsnet calculés par le code TINA.**

### 4.3.3 Catalyse

La catalyse qui est la recombinaison des espèces à la paroi du bouclier thermique est un paramètre dimensionnant pour celui-ci. La recombinaison des espèces dissociées s'accompagne de réactions exothermiques qui ont pour effet de majorer le flux de chaleur. La catalyse peut être totale ou partielle et dépend d'une part de la composition de l'atmosphère et d'autre part de celle du matériau de la protection thermique. Ce phénomène qui est toujours mal compris a suscité de nombreux travaux liés aux sondes d'exploration martienne [Kolesnikov, 1999 ; Rini, 2004]. Ainsi, les calculs de flux thermiques pour les capsules martiennes sont généralement effectués en supposant une recombinaison totale du $CO_2$ à la paroi. C'était ainsi le cas pour l'étude InterMarsnet dont les résultats obtenus avec le code TINA en terme de champ de pression et de flux convectif sont représentés sur la Figure 51. Le $CO_2$ est une molécule qui se dissocie facilement et ses éléments se recombinent aisément. Cette recombinaison à la paroi est fortement exothermique. Cela est illustré sur la Figure 52 qui représente les flux à la surface de la sonde MSRO calculés avec TINA avec les effets hors-équilibre thermique et une paroi catalytique (Qtot-Cat), une paroi non catalytique (Qtot Non-Cat), sans déséquilibre thermique et catalyse (Qtot-Eq) ainsi que le calcul de référence (RTech). Les résultats mettent en évidence un facteur multiplicatif de l'ordre de 5 pour le flux de chaleur au point d'arrêt.

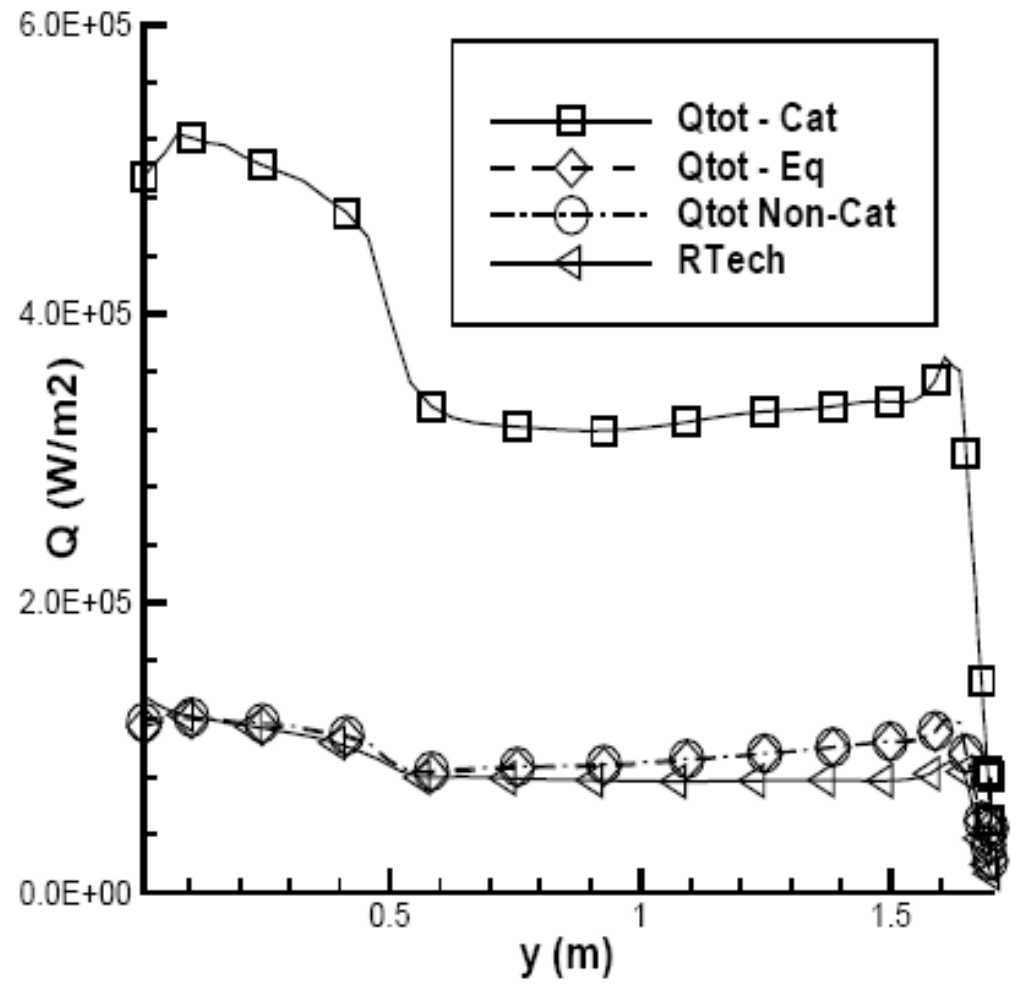

**Figure 52 : Calculs de la rentrée de MSRO avec et sans catalyse. Qtot-Eq est sans hors équilibre thermique, RTech est le calcul de référence.**

Si le facteur dimensionnant pour la protection thermique, de la catalyse est bien connu, le phénomène en lui-même est toujours mal compris. C'est particulièrement le cas pour sa relation avec l'ablation. Il dépend de la présence d'irrégularités à la surface du matériau. Dans le cas d'une rentrée martienne, ces dernières sont des sites propices à la recombinaison entre les atomes d'oxygène et de carbone. La catalyse considérée globalement dépend de phénomènes physiques dont l'échelle est mésoscropiques. Des travaux en ce sens sont actuellement en cours au laboratoire TREFLE [Perron, 2005]. Des futures activités de recherche dans ce sens pourraient par conséquent s'inscrire dans le cadre d'une collaboration entre ce laboratoire et la société ISA.

### 4.3.4 Rayonnement

Lors des rentrées atmosphériques en fonction de la composition de l'atmosphère et de la vitesse d'entrée, les espèces présentes dans la couche de choc peuvent être excitées. La désexcitation des espèces entraîne leur retour à un état stable. La transition vers un état est accompagnée d'émission ou d'absorption d'énergie, ce en accord avec la loi de Planck :

$$A + h\nu = A^* \qquad (4.3)$$

$$AB + h\nu = AB^* \qquad (4.4)$$

où la transition entre deux états des espèces A et AB nécessite ici l'absorption d'une quantité d'énergie $h\nu$, ou $h$ est la constante de Planck et $\nu$ la fréquence du rayonnement.

Le rayonnement ainsi émis (ou absorbé) peut s'avérer être important. Ainsi pour la rentrée d'Huygen [Mazoué, 2005] le flux radiatif était plus important que le flux convectif. C'est aussi le cas pour les entrées vénusiennes ainsi que pour les planètes géantes. La détermination du flux radiatif nécessite des calculs couplés CFD/Rayonnement qui sont très coûteux. Le calcul usuel se fait par couplage 1D au niveau de la ligne d'arrêt. Les seuls calculs documentés faits à ce jour en Europe avec un couplage 2D ont été effectués dans le cadre de la mission Huygens [Mazoué, 2005].

Dans le cadre de *l'Ablation Working Group* et du Workshop tenu à Porquerolles en 2004, un premier pas a été fait vers l'étude du rayonnement et ce dans le cas d'un cas test proposé par le CNES pour une rentrée martienne. Des calculs CFD ont été effectués à l'aide de plusieurs approches et modèles. Ensuite un couplage 1D entre ces calculs et PARADE [Smith, 2003] a été effectué pour évaluer le flux radiatif. Pour ce travail la molécule CO a été introduite dans PARADE et les résultats en termes de spectre et d'émission totale ont été validés par rapport à ceux prédits par SESAM [Lino da Silva, 2004]. Les comparaisons ont mis en évidence un bon accord que ce soit pour les bandes CO Ängstrom ou CO $4^{ème}$ positif. Les résultats ont confirmé que la contribution de la bande CO Ängstrom est négligeable.

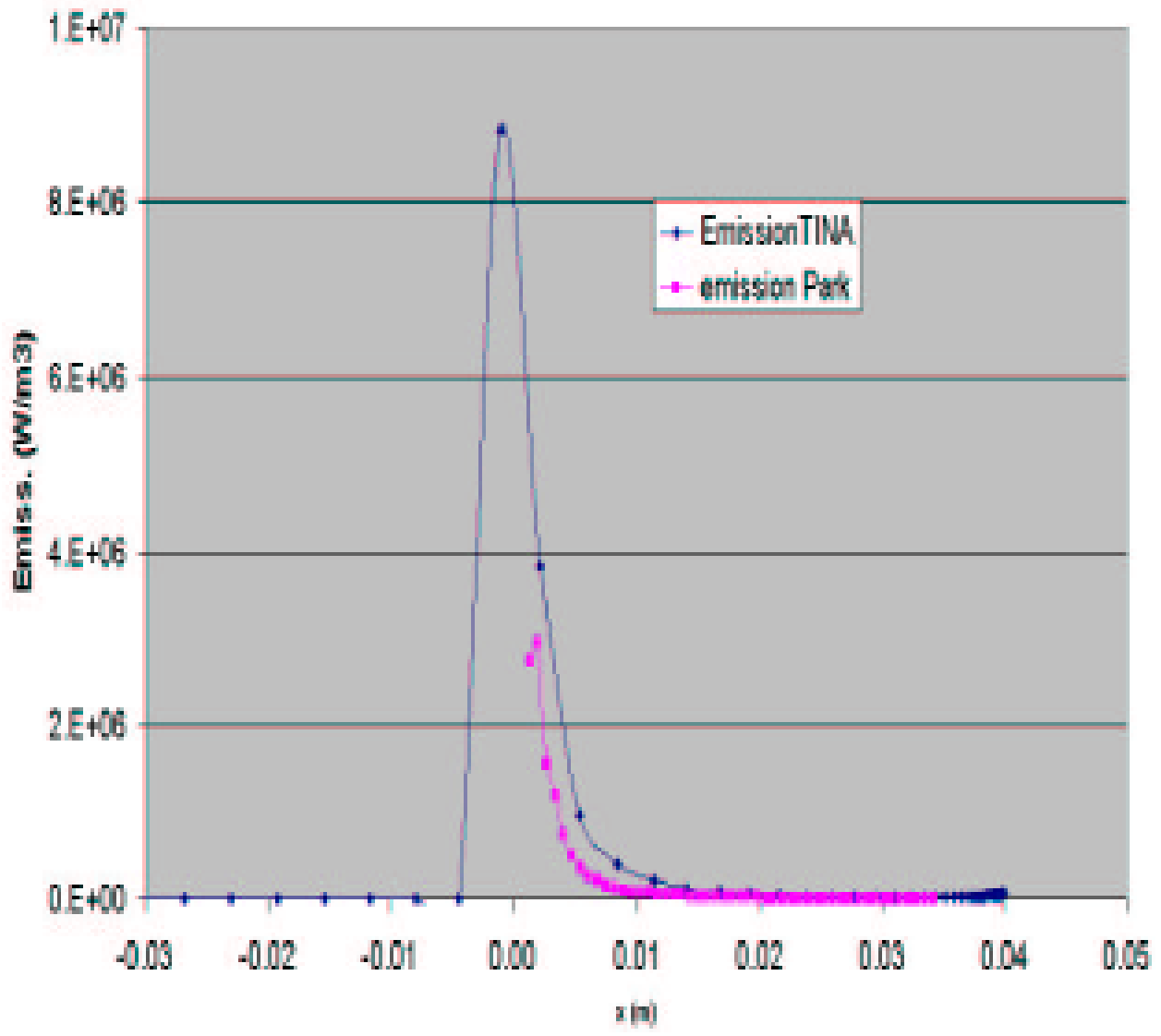

**Figure 53 : Emission totale le long de la ligne d'arrêt prédite à l'aide des modèles de Park et de TINA.**

Pour le calcul du rayonnement trois modèles thermochimiques ont été retenus : Deux avec 5 espèces (CO, $CO_2$, O, $O_2$ et C) [Park, 1990 ; Smith, 2002] et un troisième [Kay, 1993] avec 9 espèces (CO, $CO_2$, O, $O_2$, C, $N_2$, CN, NO et N), ou la présence d'azote dans l'atmosphère de Mars est prise en compte.

Les deux modèles à 5 espèces mènent à des profils similaires de température vibrationnelle et de distribution de densité de CO dans la couche de choc. Par contre la couche de choc prédite à l'aide du modèle issu de TINA [Smith, 2002] est légèrement plus large et plus chaude. Cela produit un pic plus élevé d'émission de rayonnement visible sur la Figure 53.

Les flux radiatifs prédits au point d'arrêt à l'aide des différents modèles et le calcul de référence pour le cas test du Workshop sont reportés dans le Tableau 4. Le flux radiatif prédit à l'aide du modèle de Park est voisin du résultat de référence qui a été obtenu avec le même modèle. La différence entre les niveaux de flux prédits à l'aide des modèles de Park et de TINA est difficile à expliquer et nécessiterait un effort supplémentaire. La principale conclusion est que la prise en compte de l'azote présent à un faible niveau (2,7%) dans l'atmosphère martienne augmente de façon significative le niveau de flux radiatif prédit.

| *Modèle* | *Kay* | *Park* | *TINA* | *Référence* |
|---|---|---|---|---|
| *Nombre d'espèces* | 9 | 5 | 5 | 5 |
| *Flux radiatif (kW/m2)* | 91,6 | 3,2 | 12,6 | 3,8 |

**Tableau 4 : Flux radiatif au point d'arrêt prédits à l'aide des différents modèles et calcul de référence.**

### 4.3.5 Transition vers la turbulence

La transition vers la turbulence est un aspect important durant la rentrée. Un écoulement turbulent génère des flux convectifs beaucoup plus élevés : jusqu'à 60% de flux supplémentaire dans le cas d'une rentrée martienne [Charbonnier, 2003]. La transition est un phénomène mal connu en soi. Elle semble trouver son origine dans la bifurcation des ondes d'instabilité qui peuvent être amplifiées on non suivant les conditions d'écoulement. Quand les instabilités sont suffisamment amplifiées une bifurcation tridimensionnelle peut apparaître menant à la transition [Barkley, 2002]. Le seuil de transition est dominé par l'amplification des premier et second modes d'instabilité [Malik, 2003]. Le premier mode est une extension pour les vitesses élevées de l'instabilité de Tollmien-Schlichting et domine la transition pour les régimes supersoniques. Le second mode est associé à une instabilité non visqueuse et domine la transition pour les régimes hypersoniques [Malik, 2003].

La transition est généralement associée au décollement laminaire. C'est le cas pour les écoulements comportant des décrochements dus à la géométrie. Les décrochements ont pour effet de générer de la vorticité associée à une déstabilisation de la couche limite. L'existence de régions ou la vorticité est élevée est illustrée par la Figure 54 qui représente le champ de vorticité le long de l'IRDT durant sa rentrée. L'IRDT présente deux décrochements (pour des raisons de design industriel) : le premier à l'interface du nez et du cône, le second entre le cône et la partie déployable de la protection thermique. Ici la présence de deux décrochements génère trois zones décollées et par conséquent une forte déstabilisation de la couche limite.

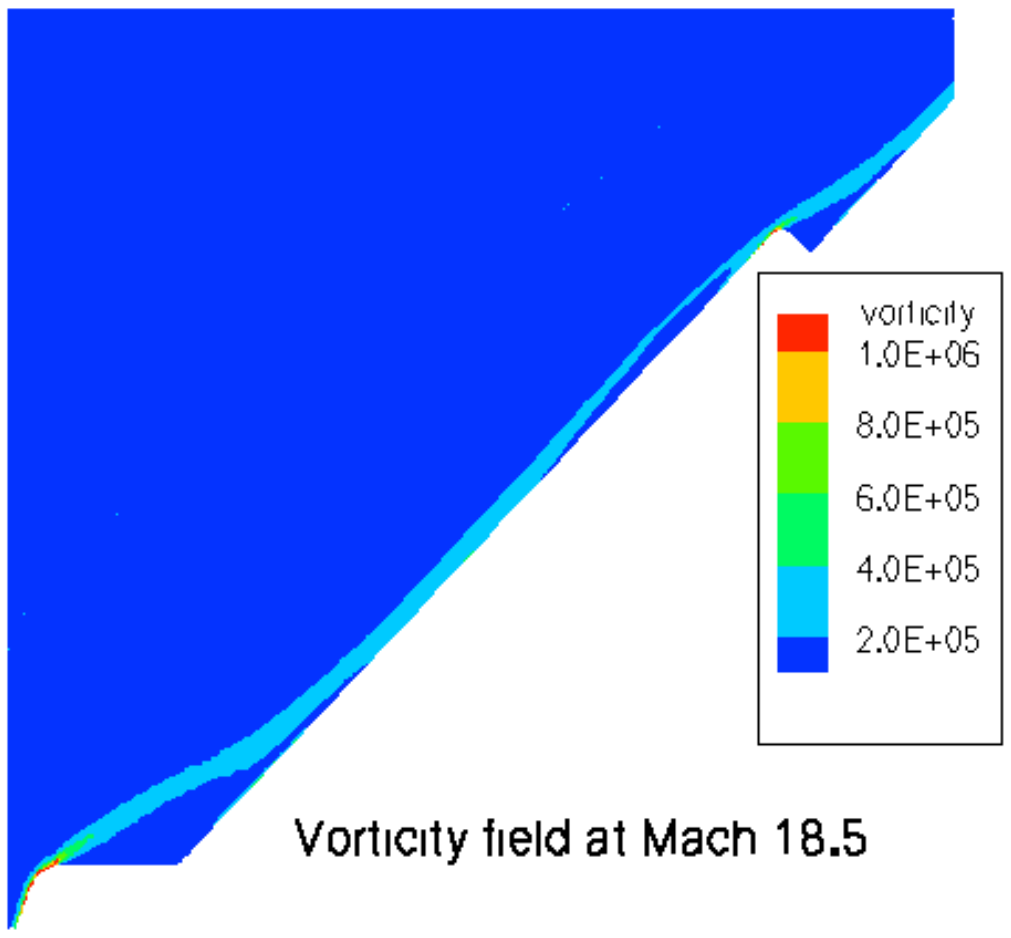

**Figure 54: Champ de vorticité le long de l'IRDT simulé avec TAU dans des conditions hors équilibres.**

Pour les rentrées atmosphériques la transition est provoquée par la présence de microrogosités sur la surface formées durant la rentrée qui perturbent la couche limite. Quand l'altitude diminue, le nombre de Reynolds augmente et des conditions d'écoulement capables d'amplifiées les instabilités peuvent être atteintes. Le problème est alors de quantifier ces perturbations en fonction de la rugosité de la surface. Ceci est généralement effectué à l'aide de critères d'ingénieur généralement basés sur un nombre de Reynolds critique calculé par rapport à la rugosité. Un grand nombre de critères plus ou moins empiriques existent [Parnaby, 2003 ; Anderson, 1974 ; Reda, 1981]. Parmi ceux-ci le critère PANT proposé par Anderson (1974) est un des plus populaire, il est donné par la relation suivante :

$$\mathrm{Re}_\theta \left[ \frac{kT_e}{\theta T_w} \right]^{0.7} \geq 255 \qquad (4.5)$$

Où $k$ est la rugosité de la surface et $T_e$ et $T_w$ les températures au bord de la couche limite et à la paroi. $\theta$ est l'épaisseur de la couche limite et $\mathrm{Re}_\theta$, le nombre de Reynolds basé sur celle-ci.

La corrélation dérivée par Reda (1981) à partir du critère PANT est :

$$\mathrm{Re}_k = \left[ \frac{\rho_e U_e k}{\mu_e} \right]_{TR} \cong 106 \qquad (4.6)$$

Où $\rho_e$, $U_e$ et $\mu_e$ sont respectivement la densité, la vitesse et la viscosité dynamique au bord extérieur de la couche limite. $Re_k$ est le nombre de Reynolds base sur la rugosité, sa valeur critique est de 106 avec une incertitude de 20%.

Une revue des corrélations existantes a été effectuée au début des années quatre-vingts [Reda, 1981] et des comparaisons effectuées à l'aide de résultats expérimentaux et de données de vol. La conclusion de cette revue est d'utiliser $Re_\theta$ comme critère de transition pour une surface lisse. Sa valeur critique varie entre 140 et 250 dans la littérature. Pour une surface rugueuse la corrélation (4.2) a été retenue. Une revue plus récente [Parnaby, 2003] a conclu que l'effet de la rugosité de la surface a un effet significatif sur la transition durant la rentrée avec un matériau ablatif. Cet effet est plus important que celui suggéré par le critère de PANT.

Dans le cadre des études de l'ESTEC menées dans le cadre des projets IRDT et PARES, des évaluations de transition vers la turbulence le long de la trajectoire ont été effectuées à l'aide des critères retenus par Reda (1981). Pour l'IRDT les calculs ont montré que le critère pour une surface rugueuse était atteint avant le maximum de flux de chaleur pour une rugosité de 1mm (voir Figure 55). La protection thermique de l'IRDT présente deux décrochage au niveau du cône (voir Figure 54) de plus, d'après le Babakin Space Center, une récession d'environ 7 mm du matériau ablatif est prévue durant le vol. Aussi le bouclier thermique a été dimensionné en prenant le long de la trajectoire le maximum entre les prédictions de flux turbulent et laminaire.

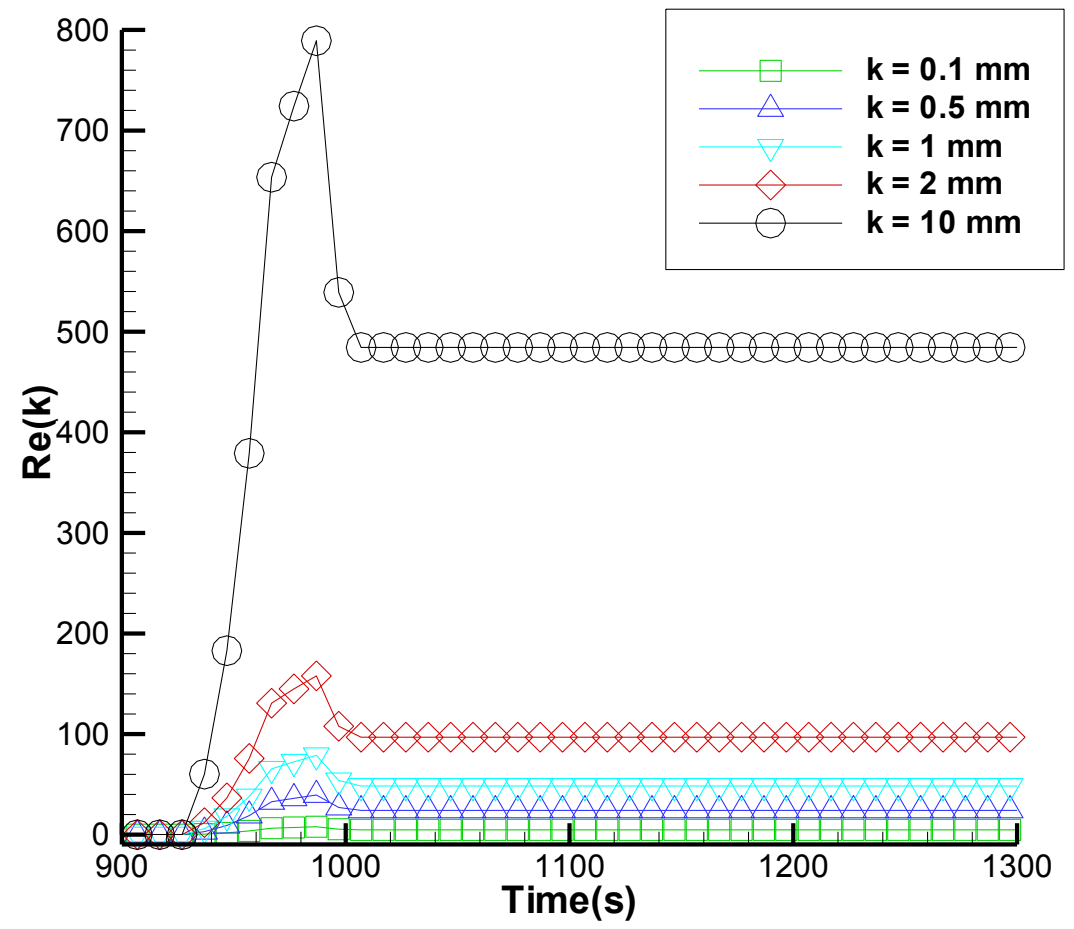

**Figure 55 : Distributions du nombre de Reynolds basé sur la rugosité $Re_k$ le long de la trajectoire pour plusieurs valeurs de k.**

Le critère de transition utilisé ici est loin d'être universel. Il a été validé à l'aide de données obtenues pour le graphite. Son application à d'autres matériaux peut être remise en cause.

Un aspect important de la transition pour des matériaux ablatifs est l'impact de l'ablation sur la transition. L'effet du soufflage du gaz issu de la pyrolyse dans la couche limite est mal connu. A priori, le soufflage devrait avoir un effet déstabilisateur sur la couche limite, c'est probablement le cas s'il est uniforme. En cas de variation du soufflage, dû par exemple à la présence de fibres carbone l'effet induit pourrait être inverse [Duffa, 2005].

Dans le domaine de la transition beaucoup reste à découvrir c'est actuellement un des points durs de la mécanique des fluides. Elle est fortement liée au décollement laminaire qui est toujours mal compris même si des efforts dans cette direction ont été récemment effectués [Rist, 2003 ; Radespiel, 2003]. Pour les rentrées atmosphériques il est difficile d'étudier ce phénomène globalement car il semble dépendant des la connaissance des phénomènes physiques intervenant au niveau mésoscopique : conditions d'écoulement du matériau vers la couche limite, récession de la surface, présence de fibres…..

# 4.4 Analyse des moyens d'essais

A priori les résultats expérimentaux servent, entre autre, à valider les codes de calcul numériques. Cependant avec la maturité croissante de l'outil numérique, les codes de calcul sont aussi utilisés pour l'analyse des moyens d'essais. Tout d'abord ils servent à la préparation des essais par l'analyse prédictive de l'écoulement dans le but d'avoir une prédiction des performances de l'installation d'essais, par l'étude des phénomènes particuliers aux moyens d'essais tels que la présence de particules de carbone dans les souffleries à haute enthalpie qui "polluent" les mesures et enfin pour la validation de nouvelles méthodes expérimentales.

Ici nous présentons quelques résultats d'une étude effectuée à l'ESTEC dans le cadre d'un stage ingénieur [Y1] dans le but d'évaluer les performances de la soufflerie du CIRA, SCIROCCO, pour les atmosphères de $CO_2$.

## 4.4.1 Modèle thermodynamique pour le $CO_2$

La première étape de l'étude a été d'établir un modèle thermodynamique pour le $CO_2$ pour étudier la composition du mélange à haute température. Pour cela trois différentes bases de données ont été utilisées : celles de CHEMKIN [Kee, 1987], de TINA [Smith, 2002] et de l'Université de la Louisiane [Esch, 1970]. Les résultats obtenus pour l'évolution du coefficient de chaleur spécifique, Cp, de l'entropie, S, et de l'enthalpie, H en fonction de la température sont reportés sur la Figure 56. Les résultats mettent en évidence un bon accord jusqu'à 5000 K, au-delà les résultats prédits par CHEMKIN divergent des autres. Cela est dû au fait qu'avec la version utilisée de CHEMKIN l'ionisation n'a pas pu être prise en compte et qu'elle devient significative au dessus de 5000 K. A partir de 15000 K, les résultats de TINA divergent de ceux de l'Université de la Louisiane. En effet, à partir de cette température les coefficients sont constants et n'évoluent plus. Le choix a finalement été fait d'utiliser la base de données de TINA et de la compléter quand nécessaire par ceux de Esch [Esch, 1970].

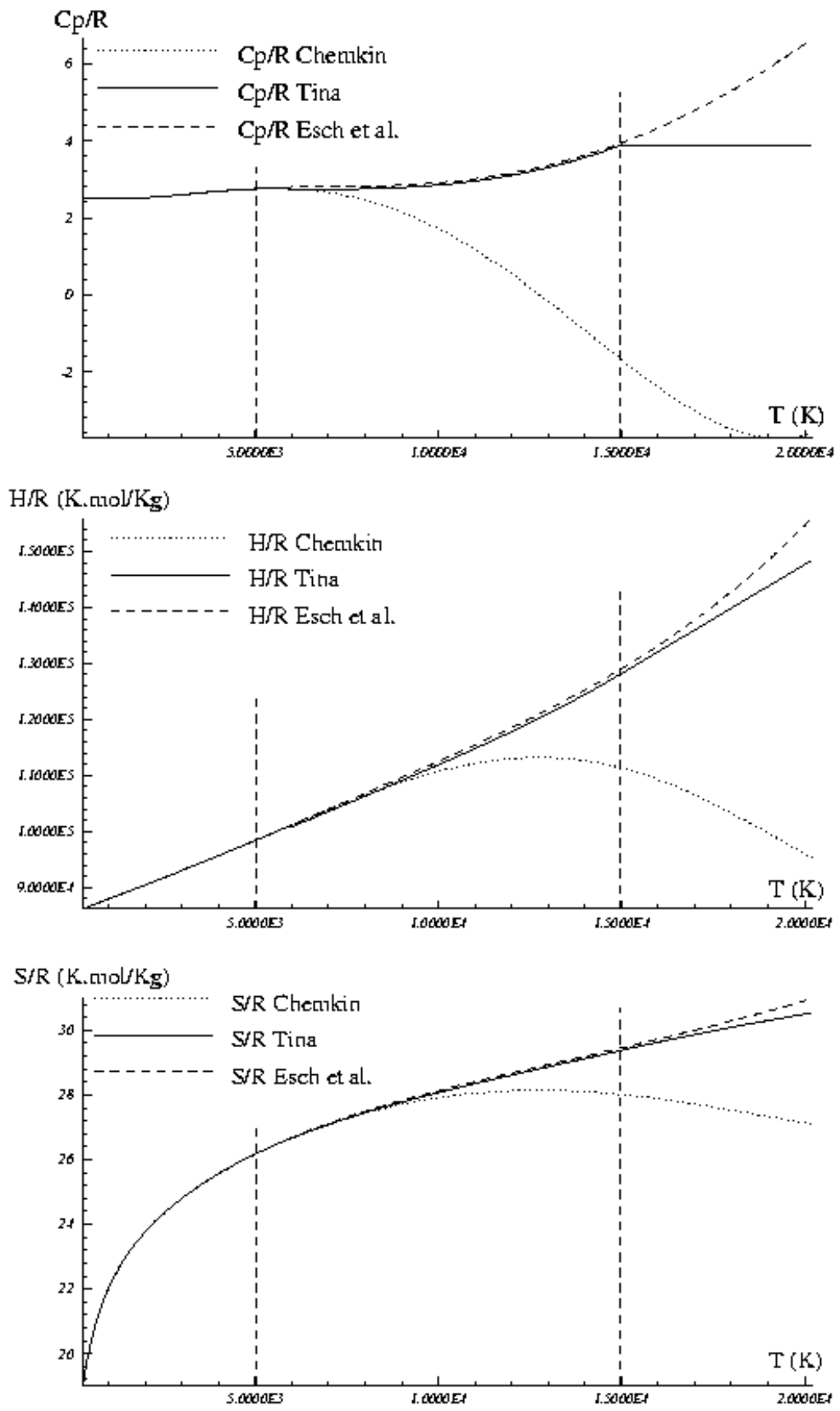

**Figure 56 : Courbes interpolées pour le carbone à partir des bases de données de CHEMKIN, TINA et de l'Université de la Louisiane [Esch, 1970]. Cp est la chaleur spécifique à pression constante, S l'entropie, H l'enthalpie et R la constante universelle des gaz.**

### 4.4.2 Approches numériques

La première étape de l'étude a été d'établir un modèle thermodynamique pour le $CO_2$ à haute température. La seconde a été de sélectionner deux approches numériques pour estimer l'enveloppe de performance de la soufflerie pour le $CO_2$. La méthode la plus rigoureuse pour prédire l'enveloppe de performance de SCIROCCO est d'effectuer des calculs hors équilibre de la soufflerie avec un code de CFD et une base thermodynamique validée. Cependant pour le coût de calcul pour ce type d'étude est assez important au vu du nombre de points de calcul nécessaire pour avoir une bonne définition de l'enveloppe. Le choix a donc été fait de développer une méthode quasi 1D à l'équilibre thermochimique pour étudier l'ensemble de l'enveloppe et d'effectuer quelques calculs à l'aide de TINA [Smith, 2002] pour avoir une comparaison code à code. La méthode quasi 1D développée consiste à résoudre les  équations de conservation de la masse, de l'enthalpie totale et de l'entropie [Shapiro, 1953] le long de la soufflerie. Le calcul de la composition du mélange a été effectué pour des conditions d'équilibre. Pour cela la méthode a été couplée à un logiciel de la NASA  développé par Gordon et Mac Bride [Gordon, 1976]. Comme le calcul a l'équilibre n'est pas réaliste ici puisque sa validité suppose un temps caractéristique des relations chimique beaucoup plus long que le temps nécessaire à l'écoulement pour traverser la tuyère, la méthode 1D a été couplée aux relations isentropiques d'un gaz parfait. Avec ce couplage en aval d'une région défini à partir d'un nombre de Mach déterminé (ici 0.7) l'écoulement est considéré comme gelé. Cette technique a été utilisée pour avoir une prédiction de l'enveloppe de performance plus réaliste qu'avec une approche complètement à l'équilibre.

### 4.4.3 Principaux résultats

Les résultats obtenus à l'aide de l'approche monodimensionnelle ont été comparés à ceux prédits par TINA. La Figure 57 montre les distributions de nombre de Mach et de densité le long de la tuyère prédites par les deux approches. La comparaison met en évidence un bon accord entre les différentes prédictions. L'accord est quasiment parfait en amont du col en aval la différence observée permet d'estimer les effets hors équilibre L'utilisation du figeage permet de se rapprocher davantage du calcul hors équilibre met l'écart demeure quand même significatif. Les comparaisons ont permis toutefois de valider l'approche quasi 1D développée ici.

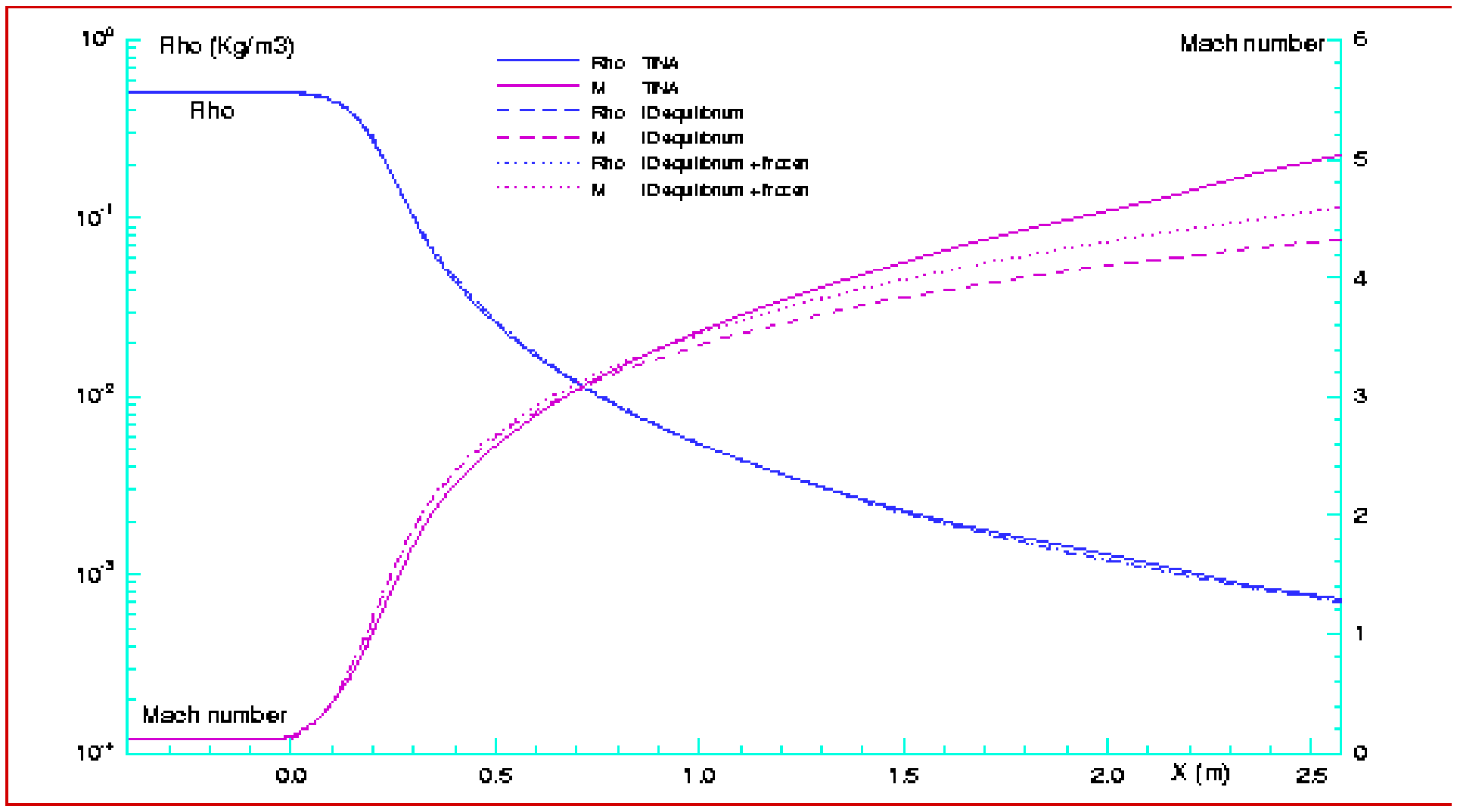

**Figure 57 : Comparaisons entre les résultats de TINA et ceux de l'approche 1D pour le nombre de Mach et la densité. Les lignes en pointillé correspondent à ceux obtenus à l'aide du figeage de l'écoulement.**

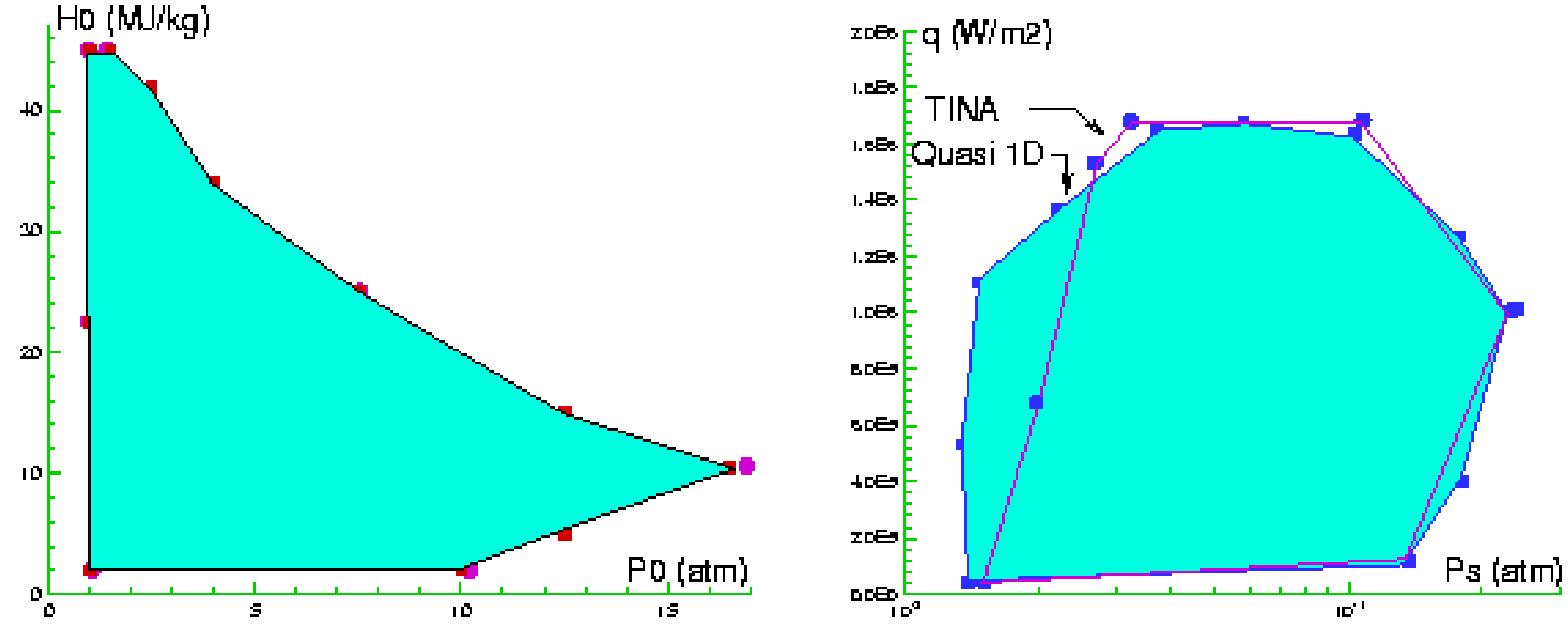

**Figure 58 : Enveloppe de performance de SCIROCCO pour la tuyère de 75 mm. A gauche conditions de réservoir pour l'air, à droite flux obtenu en fonction de la pression totale pour une atmosphère contenant 95% de CO2 et 5% de $N_2$.**

Un autre résultat intéressant est l'enveloppe de performance prédite pour une atmosphère de $CO_2$, elle est représentée sur la Figure 58. Pour estimer les performances de SCIROCCO une atmosphère contenant 95% de $CO_2$ et 5% de $N_2$ a été retenu. Elle correspond à une moyenne entre les compositions des atmosphères de Mars et de Vénus. Les conditions de réservoir connues pour l'air ont été utilisées pour cette étude préliminaire. Une étude plus approfondie nécessiterait la détermination de ces conditions pour le $CO_2$. En effet l'efficacité de l'arc électrique pour ce gaz n'est pas connue, on peu cependant s'attendre (au vu des études expérimentales menées depuis lors) à une diminution d'environ 30% de l'efficacité de l'arc électrique. Le flux convectif au point d'arrêt a été déterminé à l'aide de la relation utilisée par Rubio Garcìa et al [Rubio Garcìa, 1998] pour les mini sondes martiennes :

$$Q_{conv} = 1.156\,10^{-5}\,(H_o - H_w)\frac{\sqrt{\rho U^2}}{1.013\,10^5\,R_n} \qquad (4.7)$$

où, *Qconv* est le flux convectif, $H_0$ l'enthalpie totale, $H_w$ l'enthalpie à la paroi, $\rho$ la densité, $U$ la vitesse et $R_n$ la rayon au nez de la sonde.

La comparaison entre les enveloppes de performance prédites par les deux approches montre un bon accord sauf dans la région d'enthalpie élevée et de pression d'arrêt faible. Ceci s'explique par le fait que c'est pour cette région de l'enveloppe que les effets hors équilibre sont les plus importants.

Cette étude a montré l'intérêt de l'approche monodimensionnelle pour la détermination de l'enveloppe. Pour avoir une enveloppe de performance fine de SCIROCCO pour ce type d'atmosphère, la détermination de l'efficacité de l'arc électrique pour la soufflerie et le $CO_2$ serait nécessaire. Un autre point serait la prise en compte de l'ionisation qui n'était pas considérée dans le modèle thermodynamique utilisé avec TINA.

Les phénomènes liés à la rentrée atmosphériques abordés dans ce chapitre de Projets et Perspectives ont déjà fait l'objet de travaux réalisés à l'ESA-ESTEC. Ils ont fait l'objet des communications [JS1], [J5], [C6, C11-C18], [A3-4], [R2] et [S4] référées en Annexes. Ces travaux ont été pour certains menés dans le cadre de stage de fin d'étude [Y1-Y2] ou lors de stage à l'ESTEC de jeunes ingénieurs [Y3-4].

# 5 Conclusion

La principale conclusion de ce document est la maturité aujourd'hui atteinte par la CFD. Cela est illustré par son intérêt pour une large palette d'applications, ici aérospatiales. Au sein du secteur spatial elle est largement utilisée pour les calculs d'aérothermodynamique interne et externe avec des applications en propulsion, en rentrée atmosphérique et en thermique (calcul de boucles fluides). Elle ne sert plus seulement à effectuer des comparaisons calcul/expérience mais présente un intérêt certain non seulement pour les études paramétriques d'avant projet mais aussi pour la préparation de campagnes expérimentales et le développement de nouvelles méthodes de mesure.

Elle sert à traiter des problèmes à la physique de plus en plus complexe et les limites de son champ d'application sont petit à petit repoussées. Ces dernières se situent souvent au niveau de problèmes fortement couplés qui nécessitent la résolution non seulement des équations de Navier-Stokes mais aussi d'autres jeux d'équations : Equations de Maxwell pour le black-out, de Schrödinger pour le rayonnement, de la thermique des matériaux pour l'ablation et de la mécanique des solides pour le couplage fluide/structure.

D'autres points durs existent encore c'est notamment le cas du phénomène de breakup abordé dans le premier chapitre et pour lequel les études expérimentales sont difficiles à mener, des problèmes des films de paroi qui apparaissent dans des écoulements de combustion de type propulseurs à poudre ou dans des systèmes d'épuration du gaz naturel. Enfin, le phénomène pour lequel peu de progrès ont été effectués est la transition vers la turbulence. Pour la déterminer lors des rentrées atmosphériques de vieux critères empiriques sont utilisés de façon générale alors qu'ils ont été développés pour des conditions particulières. Un des points clés pour améliorer notre connaissance de la transition vers la turbulence semble être de commencer par mieux appréhender les problèmes de séparation laminaire. Les simulations effectuées mettent en évidence une instabilité forte de la séparation et notamment de sa région initiale. La compréhension de ce mécanisme serait un premier pas dans la résolution de la transition.

Les travaux présentés ici ont été menés dans plusieurs centres de recherche : l'IMFT, l'ESA-ESTEC et le DLR. Actuellement certains sont poursuivis dans le cadre de la société Ingénierie et Systèmes Avancés créée en Avril 2006 en partenariat avec Fluid Gravity Engineering. C'est le cas des travaux en ablation pour lesquels une étude sur le blocage est en cours et des études sur le black-out qui se poursuivent à travers une analyse de la phase de black-out lors du vol de l'IRDT.

# Références

# Annexe 1 : Curriculum Vitae

## État Civil

| | |
|---|---|
| Nom : | Reynier |
| Prénom : | Philippe |
| Date et lieu de naissance : | 7 janvier 1968 à Villeneuve sur Lot |
| Situation familiale : | Célibataire |
| Nationalité : | Française |
| Adresse personnelle : | 209, Avenue Pasteur<br>33600 Pessac |
| Adresse professionnelle : | Ingénierie et Systèmes Avancés<br>Technopole Bordeaux Technowest<br>Domaine James Watt – Tour C<br>19, Allée James Watt<br>33700 Mérignac |
| Téléphone : | 05 56 47 92 74 |
| Courriel : | Philippe.Reynier@isa-space.eu |

## Formation

**1995** Doctorat en Mécanique des Fluides, Institut National Polytechnique de Toulouse.

Thèse soutenue le 16 octobre 1995 à l'Institut National Polytechnique de Toulouse.

Jury : MM. Borghi, Chassaing, Duthoit, Ha Minh, Kuentzmann, Mojtabi, Robert et Zaleski.

Titre : Analyse physique, modélisation et simulation numérique des jets simples et des jets coaxiaux turbulents, compressibles et instationnaires.

**1991** DEA de Mécanique des Fluides, Institut National Polytechnique de Toulouse.

**1990** Maîtrise de Mécanique et 2º année de Magistère MATMECA, Université Bordeaux 1.

**1989** Licence de Mécanique et 1º année de Magistère MATMECA, Université Bordeaux 1.

**1988** Deug A "Physique-Informatique", Université Bordeaux 1.

**1985** Baccalauréat C, Lycée de Villeneuve sur Lot.

## Expérience professionnelle

**Depuis 2006 ISA, Mérignac – Direction d'une start-up**

Création et management de la société Ingénierie et Systèmes Avancés, start-up créée en partenariat avec Fluid Gravity Engineering (GB).

Activités techniques :

- Support technique pour l'Ablation Working Group' mis en place par l'ESA.
- Etude sur le blocage convectif lors des rentrées terrestres à grande vitesse.
- Reconstruction de la phase de blackout observée durant le vol de l'IRDT.

Activités d'enseignement

- Séminaire sur la rentrée atmosphérique et les phénomènes associés effectué dans le cadre de l'option de 3° année de l'ENSCPB, Energétique et Systèmes Industriels.

**2003-2006 AOES, Leyde, Pays-Bas – Ingénieur Consultant à l'ESA – ESTEC dans les Divisions Propulsion et Aérothermodynamique puis Structures et Thermique**

Activités dans le cadre de rentrées terrestres orbitales (AEOLUS, ATV, EXPERT, IRDT, PARES), du programme d'exploration de Mars, AURORA et des missions planétaires (EXOMARS HUYGENS et JEP) :

- Mise en place d'un groupe de travail européen en ablation et préparation d'un premier workshop.
- Support technique pour et les activités futures de l'ESA : Préparation dans le cadre d'AURORA du programme technologique relatif aux rentrées terrestres et martiennes et à l'aérocapture terrestre ainsi que d'appels d'offres pour des études de technologie : Phénomènes associés aux rentrées terrestres à grande vitesse et stabilité dynamique des capsules.
- Préparation des programmes de recherche et de technologie (TRP et GSTP) de l'ESA : Physique des écoulements avec particules, thermochimie du $CO_2$, ablation, tests du nez et de la charge utile relative à l'interaction choc/couche limite pour EXPERT.
- Participation aux revues de projets : EXOMARS, IRDT et PARES ; et aux études CDF (Concurrent Design Facility) : Aerocapture Working Group, Jupiter Entry Probe.
- Evaluation de l'influence d'un 'aerospike' sur la protection thermique pour l'étude JEP.
- Etudes de trajectoire et prédiction du black-out et de la transition vers la turbulence lors de rentrées terrestres (Earth Aerocapture, IRDT, PARES).
- Simulation numérique et évaluation des flux thermiques lors de rentrées dans les atmosphères de Mars et de la Terre avec prise en compte du déséquilibre thermochimique et de la catalyse à l'aide de codes structurés et non structurés.
- Etude de l'impact de l'ablation sur le comportement de la protection thermique d'HUYGENS.
- Étude de la destruction de l'ATV lors de sa rentrée en fin de mission.

- Analyse de l'aérodynamique des minidrones dans l'atmosphère martienne et participation au Jury du Concours Minidrones organisé par l'ONERA et la DGA.

**2000-2003 DLR (Centre Aérospatial Allemand), Brunswick, Allemagne – Ingénieur de recherche dans le Groupe Véhicules Spatiaux**

Prédiction et analyse des comportements aérodynamiques et aérothermodynamiques des missiles à grande vitesse dans le cadre des projets du DLR :

- Développement et application de solveurs Euler et Navier-Stokes structurés et non structurés en hypersonique pour des véhicules munis d'ailes en nid d'abeilles.
- Génération de maillages 3D structurés et non structurés pour des configurations complexes.
- Modélisation d'ailes en nid d'abeilles à l'aide d'un disque d'action couplé à une base de données expérimentales puis à une approche semi-empirique basée sur la théorie des ailes en nid d'abeilles.
- Support numérique pour la préparation des campagnes d'essais en soufflerie.
- Étude thermomécanique d'un radôme avec prise en compte du couplage fluide-structure.

**1998-2000 ESA – ESTEC (Agence Spatiale Européenne), Noordwijk, Pays-Bas – Ingénieur de recherche dans la Division Propulsion et Aérothermodynamique**

Étude de la passivation du Système de Contrôle d'Attitude d'ARIANE 5 :

- Responsable du développement du modèle numérique au sein de l'équipe en charge du projet. Coopération avec les différents partenaires : le CNES, le DLR, l'ONERA et EADS-ASTRIUM.
- Prédiction de la dépressurisation et des changements de phase lors de la vidange d'un réservoir d'hydrazine dans l'espace. Intégration dans un code de calcul d'une loi d'état thermodynamique non linéaire pour un liquide compressible (hydrazine) avec possibilité de vaporisation; et simulation numérique d'écoulements multiphasiques à faibles nombres de Mach en collaboration avec l'Université de Delft (Pays-Bas).
- Préparation et suivi d'essais à l'ONERA. Validation des résultats numériques par rapport aux mesures de l'ONERA et d'EADS-ASTRIUM.

Études aérothermodynamiques pour les missions d'exploration spatiale :

- Étude de trajectoire et dimensionnement de la capsule de retour pour une mission de retour d'échantillons vers Mercure.
- Simulations numériques des conditions d'entrée et évaluation des flux thermiques lors d'entrées dans les atmosphères de Mars et de la Terre : Prise en compte du déséquilibre thermochimique, de l'ablation et de la catalyse.
- Évaluation des capacités de la soufflerie à haute enthalpie SCIROCCO (coopération avec le CIRA, centre aérospatial italien) à la simulation d'entrées dans les atmosphères de Mars et de Vénus.

Support pour les activités futures de l'agence :

- Proposition d'un projet sur une micro-capsule de retour d'échantillons.

- Préparation d’un TRP sur les écoulements multiphasiques.

**1991-1997 Institut de Mécanique des Fluides de Toulouse - Doctorant MESR puis post-doctorant du CNES**

Analyse, modélisation et simulation numérique des jets d’injection du moteur-fusée VULCAIN pour ARIANE :

- Développement et application de solveurs Navier-Stokes à des écoulements instationnaires. Simulation numérique d’instabilités dans les jets turbulents, subsoniques, transsoniques et supersoniques.
- Modélisation de la turbulence dans les jets ronds compressibles à l’aide de modèles k-ε linéaires et non linéaires. Evaluation de l’anisotropie dans les jets coaxiaux avec un modèle au second ordre (équations de transport pour les contraintes de Reynolds).
- Validation des résultats numériques par comparaison avec les prédictions du code Thésée de la SNECMA et avec des données expérimentales de la littérature.
- Etude de l’influence du contraste de densité sur l'instabilité des jets d’injection de moteur-fusée.
- Développement d'une approche monophasique des sprays, valable dans les jets coaxiaux diphasiques lorsque les effets de tension de surface sont négligeables par rapport à ceux de quantité de mouvement.

# Annexe 2 : Publications

**Revues internationales à comité de lecture (5 + 2 soumises)**

**Ouvrages internationaux à comité de lecture (2)**

[O1] Reynier Ph. and Ha Minh H.
*Influence of density contrast on instability and mixing in coaxial jets.*
Fluid Mechanics and its Applications, Vol. 41, pp. 25-32, L. Fulachier, J. L. Lumley, F. Anselmet (Eds), Kluwer Academic Publishers, 1997 (Proceedings of the IUTAM Symposium on "Variable Density Low-Speed Turbulent Flows", Marseille, 8-10 July 1996).

**Thèse de Doctorat**

[T] Reynier Ph.
*Analyse physique, Modélisation et Simulation numérique des jets simples et des jets coaxiaux, turbulents, compressibles et instationnaires.*
Thèse de Doctorat de l'INPT, n°1062, Toulouse, Octobre 1995.

**Conférences avec Publication des actes et Comité de lecture (18)**

[C18] Reynier Ph.,
*Modelling of re-entry flows: Expertise available at ISA.*
In Proceedings of the 2nd International Workshop on Radiation and High Temperature Gases in Atmospheric Entry, Rome, 6-8 Sept., 2006.

[C17] Marraffa, L., Reynier Ph., Mazoué, F. & Reimers, C.,Boutamine D. E.,
*Some aerothermodynamics aspects of ESA entry probes.*
In Proceedings of EWHSFF 2005, East-West High Speed Flow Field Conference, Beijing, Oct. 19-22, 2005.

[C16] Schmehl R., Reynier Ph., Boutamine D. E., Steelant J. & Marraffa L.
*Computational Analysis of ATV Re-entry and Explosion Assessment.*
In Proceedings of EUCASS Conference on Aerospace Science, Moscow, July 4-7, 2005.

[C15] Hänninen P. G., Lavagna M., Reynier P. & Marraffa L.
*Evolutionary Algorithms for Multidisciplinary optimization in space: Atmospheric vehicles design.*
In Proceedings of 1st ECCOMAS Conference on Computational Methods for Coupled Problems in Science and Engineering, Santorini, Greece, May 25-28, 2005.

[C14] Reynier P. & Marraffa L.
*Aerothermodynamics investigations for Earth orbital entry vehicles.*
In Proceedings of 4th International Symposium on Atmospheric Reentry Vehicles and Systems, Arcachon, March 21-23, 2005.

[C13] Hänninen P. G., Lavagna M., Finzi A. E., Reynier P. & Marraffa L.
*Space probe Multidisciplinary Design for atmospheric phases.*

**Autres Communications (7)**

**Rapports Techniques (5)**

**Séminaires (4)**

[S2] Reynier Ph.
*Unsteady simulations of jets using turbulence models.*
Joint Research Center of the European Commission, Ispra, Italy, Nov. 10, 1997.

[S1] Reynier Ph.
*Unsteady simulations of jets using turbulence models.*
UMIST, Manchester, Oct. 17, 1997.

# Annexe 3 : Encadrement de travaux

**Encadrement d'ingénieurs débutants à l'ESA-ESTEC**

[Y5] Djamel Boutamine, Ingénieur en Génie Mécanique de l'Ecole Polytechnique d'Alger (2003),
*Computational analysis of ATV re-entry flow and explosion assessment*.
Stagiaire International, Septembre 2004-Septembre 2005. Participation 50%.

[Y4] Felipe Dengra Moya, Ingénieur de l'Universitat Politècnica de Catalunya (2001)
*Unsteady aerodynamic study of an airfoil in Martian flight conditions.*
Stagiaire International, Septembre 2002-Septembre 2004. Participation 40%.

[Y3] Quentin Morel, Ingénieur Université Technologique de Compiègne (2003),
*Structure/Flow interaction in inflatable structures.*
Stagiaire, Septembre 2003-Septembre 2004. Participation 20%.

**Encadrement de stages de DEA et équivalents à l'ESA-ESTEC**

[Y2] Petri-Giovanni Hanninen, Politechnico di Milano,
*Multidisciplinary optimization for space vehicles during aero-assisted manoeuvres: An evolutionary algorithm approach.*
Stage de Laurea d'Ingénierie Spatiale, Mars - Septembre 2004. Participation 60%.

[Y1] Vincent Marieu, École Matmeca, Universite Bordeaux 1
*Evaluation of SCIROCCO capabilities to entry simulations in $CO_2$ atmospheres.*
Stage d'Ingénieur Matmeca et DEA de Mécanique, Mars – Août 1999. Participation 60%.

# Annexe 3 : Sélection de Publications